\documentclass[a4paper,11pt,titlepage]{article}
\usepackage[T1]{fontenc}
\usepackage[latin1]{inputenc}
\usepackage{psfrag}
\usepackage[dvips]{graphicx}
\usepackage{amsmath}
\usepackage{amsfonts}
\usepackage{amssymb}  % just for the square in the end

\newcommand{\ul}{\underline}

\newcommand{\td}{\tilde}
\newcommand{\C}{^*}
\newcommand{\T}{^\mathrm{T}}
\newcommand{\TC}{^\mathrm{T*}}
\newcommand{\Tr}{\mathrm{Tr}}

\newcommand{\dol}[1]{\stackrel{\hspace{0.55mm}\ul{\ul{\hspace{1.9mm}}}}{#1}}
\newcommand{\dyadic}[1]{\stackrel{_\leftrightarrow}{#1}}
\newcommand{\vecb}[1]{\mathrm{\mathbf{#1}}}
\newcommand{\upd}{\mathrm{d}}
\newcommand{\nvec}{\hat\vecb n}
\newcommand{\svec}{\hat\vecb s}
\newcommand{\kvec}{\hat\vecb k}
\newcommand{\xvec}{\hat\vecb x}
\newcommand{\yvec}{\hat\vecb y}
\newcommand{\zvec}{\hat\vecb z}
\newcommand{\hvec}{\vecb H}
\newcommand{\rvec}{\vecb r}
\newcommand{\gvec}{\vecb g}
\newcommand{\fvec}{\vecb f}
\newcommand{\avec}{\vecb a}
\newcommand{\bvec}{\vecb b}
\newcommand{\cvec}{\vecb c}
\newcommand{\Cvec}{\vecb C}
\newcommand{\dvec}{\vecb d}
\newcommand{\rhovec}{\hat{\boldsymbol{\rho}}}
\newcommand{\chivec}{\hat{\boldsymbol{\chi}}}

\newcommand{\nuvec}{\boldsymbol{\nu}}
\newcommand{\Deltavec}{\boldsymbol{\Delta}}
\newcommand{\sigmavec}{\boldsymbol{\sigma}}
\newcommand{\Deltavecre}{\Deltavec_{\mathrm{R}}}
\newcommand{\Deltavecim}{\Deltavec_{\mathrm{I}}}
\newcommand{\cvecim}{\textrm{Im}~\cvec}
\newcommand{\vF}{v_{\mathrm{F}}}
\newcommand{\NF}{N_{\mathrm{F}}}
\newcommand{\kB}{k_{\mathrm{B}}}
\newcommand{\kF}{k_{\mathrm{F}}}
\newcommand{\iu}{\mathrm{i}}
\newcommand{\me}{\epsilon_m}
\newcommand{\nam}{\breve} % nambu matrix
\newcommand{\Sgn}{\textrm{Sgn}} % nambu matrix
\newcommand{\Det}{\textrm{Det}} % nambu matrix
\newcommand{\nc}{s}
\newcommand{\im}{\textrm{Im}}
\newcommand{\re}{\textrm{Re}}
\newcommand{\op}{\breve} % operator accent
\newcommand{\uar}{\uparrow}
\newcommand{\dar}{\downarrow}

\newcommand{\levelone}{\section}
\newcommand{\leveltwo}{\subsection}
\newcommand{\levelthree}{\subsubsection}

\begin{document}

\title{
\begin{figure}[!hb]
\begin{center}
\includegraphics[width=0.2\linewidth,angle=0]{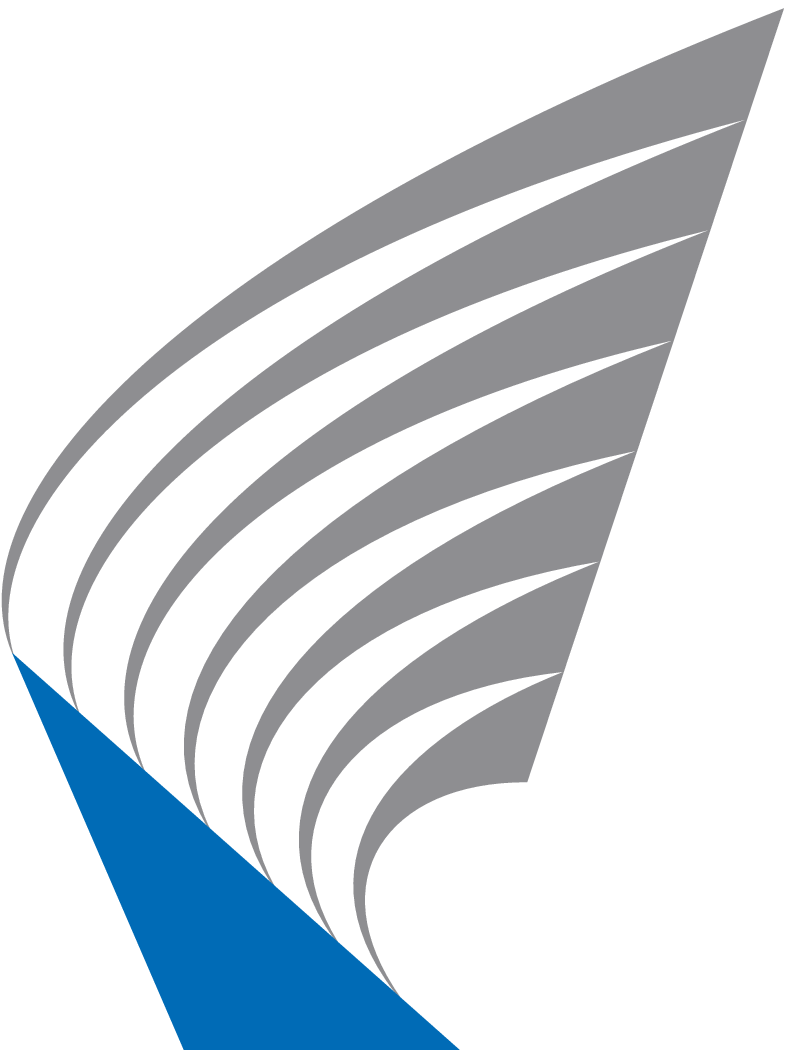}
\end{center}
\end{figure}
\Huge Master's Thesis \\ \bfseries Pinhole model for the 
Josephson $\pi$ state in $^3$He}
\author{Janne Viljas \\
\emph{Helsinki University of Technology}, \emph{Low Temperature
Laboratory}, \\ \emph{P.O. Box 2200, FIN-02015 HUT, FINLAND}, \\
\emph{E-mail: Janne.Viljas@hut.fi } }
\date{\today}
\maketitle
%\begin{abstract}
%Abstract abstraction.
%\end{abstract}

\noindent
%{\huge\bfseries Preface}\\
{\huge\bfseries Foreword}\\

The work for this Master's Thesis was done mainly within in a period
of one year in the theory group of the Low Temperature Laboratory at 
Helsinki University of Technology. 
%It is now time to present some appreciation for all the people who have
%been involved in the completion of this work.
Most of all I would like to thank my instructor, Docent Erkki Thuneberg,
without whose constant attention and support I would never have
finished it. 
I also want to thank Professor Mikko Paalanen and the
rest of the laboratory personnel for providing such an inspiring 
environment for doing individual research. Further thanks go to my
supervisor, Professor Martti Salomaa, for reading and checking the 
manuscript. 

The people who have worked a long time with me in the same room also deserve some
appreciation, not least due to interesting discussions over the morning
coffee. For this, I thank the members of our group,
Juha Kopu and Risto Hänninen. Risto's experience in the practical
aspects of the quasiclassical method was also helpful.
Doctor Adriaan Schakel aided with some general theoretical 
issues during the last few months. 

I feel that I have learned a lot of new things during
the project and I am confident I will still have use for them later on.
Practically all of the numerical work presented below
was done by me alone, whereas in some analytical quasiclassical
calculations I received a lot of help from Erkki.
I want to emphasize his important contribution to the work. 
I should also thank the Center 
for Scientific Computing (CSC) for providing fast computers and 
good software for doing all the more demanding numerics.

Finally, I would like thank my parents for all the things they did for me
during my school years and undergraduate studies. And I thank all
my friends, who ensured that some of my time was spent on
other things as well.

\vspace{10mm}

Otaniemi, April 17th 2000

\vspace{15mm}
Janne Viljas

%Last, but definitely not least, I personally feel that
%the relatively simple and well-defined pinhole calculation has worked 
%as an excellent introduction to making practical calculations with the 
%quasiclassical method. 

%\vspace{6mm}
\newpage

\noindent
{\huge\bfseries Alkusanat} \\

Tämä diplomityö on tehty pääasiassa viimeisen vuoden aikana
työskennellessäni tut\-ki\-mus\-apu\-lai\-se\-na Teknillisen Korkeakoulun
Kylmälaboratorion teoriaryhmässä.
Erityisesti haluan kiittää ohjaajaani Dosentti Erkki Thunebergia, jonka tuella ja 
avustuksella työ on viimein saatu päätökseen. Professori Mik\-ko
Paa\-la\-sel\-le sekä laboratorion muulle henkilökunnalle kiitos kuuluu
inspiroivan työilmapiirin luomisesta ja valvojalleni Professori
Martti Salomaalle työn lukemisesta ja tarkistamisesta.

Samassa työhuoneessa kanssani pitkään työskennelleille
ryhmätovereilleni Juha Kopulle ja
Risto Hänniselle annan kiitokset muun muassa mielenkiintoisista
kahvipöytäkes\-kus\-te\-luis\-ta. Riston kokemus kvasiklassisen teorian
käytännön ongelmista oli myös arvokasta. Tohtori Adriaan Schakel avusti 
joissakin yleisissä teoreettisissa ongelmissa vii\-meis\-ten
kuu\-kau\-sien aikana.

Olen oppinut projektin aikana paljon uusia asioita ja olen
vakuuttunut, että mi\-nul\-la on käyttöä oppimalleni jatkossakin. 
Käytännössä kaiken alla kuvatusta numeerisesta työstä 
tein itse, mutta esimerkiksi joissakin vaikeimmissa kvasiklassisen teorian
analyyttisissä laskuissa sain paljon apua Erkiltä. Tahdon korostaa
hänen tärkeää osuuttaan työssä. Kiitän myös  
CSC - Tieteellinen Laskenta Oy:tä tehokkaiden tietokoneiden ja hyvien
ohjel\-misto\-jen tarjoamisesta vaativimpien numeeristen ongelmien
ratkaisemiseksi.

Lopuksi haluan kiittää vanhempiani kaikesta siitä tuesta ja
kannustuksesta, jonka heil\-tä olen saanut koulunkäyntini ja opintojeni
aikana. Lausun kiitoksen myös kaikille ystävilleni, jotka ovat
pitäneet huolen siitä, että olen käyttänyt osan ajastani myös muihin
harrasteisiin.

\vspace{10mm}

Otaniemi, 17. huhtikuuta 2000

\vspace{15mm}
Janne Viljas

\newpage

\tableofcontents
\newpage

%\fontsize{12pt}{13pt}\selectfont

%$\boldsymbol{\mathrm{k}}_{c}^{bulk}$

%%%%%%%%%%%%%%%%%% Chapter 1: INTRODUCTION!
\levelone{Josephson effect}
The effects related to a weak coupling
between two phase-coherent
quantum systems are often collectively termed Josephson phenomena.
The name dates back to Josephson, who first predicted such effects
(tunneling of Cooper pairs) to
take place in a weak connection between two superconductors, a so-called
Josephson junction \cite{josephson}. In this Master's Thesis I am concerned with 
superfluids rather than superconductors, but the following main
results are at least approximately true for all cases. 
The name ``Josephson junction'' is usually reserved only for
the pure tunneling junctions in superconductors, and microbridges or
superfluid junctions etc. are put under the more general cathegory of 
``weak links'' \cite{likharev}. 

\leveltwo{Basic concepts of superfluid weak links}
Imagine a container of superfluid, divided into two parts (1 and 2) by a thin
membrane. The two condensates are described by some 
macroscopic wavefunctions, or order parameters, which have
well-defined but different phases $\phi_1$ and $\phi_2$. We assume
that there are no other relevant degrees of freedom.
The two sides are then weakly coupled by introducing some (sufficiently) 
small orifice(s) in the dividing membrane. If we denote the phase difference
between the condensate wavefunctions by $\phi=\phi_1-\phi_2$, then 
there will be a supercurrent flowing through the weak link thus
formed, given by the simple formula 
\begin{equation} \label{e.sine}
I(\phi)=I_c\sin\phi,
\end{equation}
where $I_c$ is a critical current specific to the junction.
If we apply a finite pressure difference $\Delta P$ between the two
sides keeping the temperature constant, 
there will also be a difference in the chemical potentials
$\Delta\mu=v\Delta P$, $v$ being the specific volume $V/N$. The
well-known Josephson-Anderson phase-evolution equation \cite{anderson1}
\begin{equation} \label{e.ac}
\frac{\partial\phi}{\partial t}=-\frac{\Delta\mu}{\hbar}
\end{equation}
then tells that $\phi$ will change in time. As a result of this and
the periodic form of Eq. (\ref{e.sine}), a constant $\Delta\mu$ results in an oscillating
supercurrent through the weak link.

The current-phase relation (\ref{e.sine}) is accurate only for pure
tunneling junctions. For micro\-bridge-type weak links, to which we
include the superfluid ones, there are deviations from the sine form
\cite{likharev,kulik}. The general trend is that at lower temperatures, or
stronger coupling, the sine will become slanted.
The length scale determining the effective size and thus the strength
of the coupling is always the supefluid coherence length, which grows with 
temperature and diverges at the transition.
For large enough effective aperture sizes, the current-phase relation
will become hysteretic (multivalued). A hysteretic $I(\phi)$ implies that
there is dissipation taking place at a constant rate for a constant
$\Delta\mu$, since there then exists a net dc current component. The
dissipated energy per period of $I(\phi)$ equals
$\hbar\int_0^{2\pi}\max\{I(\phi)\}\upd\phi$, taking $I(\phi)$ to be the particle current.
This holds because the net work done by the pressure difference 
$\Delta P$ in a time $\upd t$ in pushing the fluid through the orifice 
is %\cite{barone}
\begin{equation} \label{e.dene}
\upd F=v\Delta P \upd N=-\hbar\upd N\frac{\partial\phi(t)}{\partial t}
=-\hbar\frac{\partial N(t)}{\partial t}
\frac{\partial\phi(t)}{\partial t}\upd t
=\hbar I(\phi)\upd \phi,
\end{equation}
where we used Eq. (\ref{e.ac}) and defined 
$I(\phi(t))\equiv-\partial N(t)/\partial t$, see Ref. \cite{barone}. This thermodynamic
argument thus defines the free energy of the weak link 
$F(\phi)$ and relates it to the particle current $I(\phi)$ through the equation
\begin{equation}
I(\phi)=\frac{1}{\hbar}\frac{\partial F(\phi)}{\partial \phi}.
\end{equation}
The dissipation is generally explained with the concept of 
phase slips, introduded by Anderson \cite{anderson1}. A phase slip
occurs when a vortex nucleates at one wall inside the orifice,
crosses the flow under the influence of the Magnus force, and then
vanishes on the other side. As a result, the phase difference changes
(or ``slips'') by $2\pi$ and some energy is withdrawn from the flow.
This happens because it costs energy to nucleate the vortex, and when
it vanishes, its energy is dissipated as heat.
The chemical potential difference is then maintained by a constant
rate of vortex motion across the orifice: we have
\begin{equation} \label{slippi}
\begin{split}
-\langle\mu_1-\mu_2\rangle_{t}
=\hbar\left\langle\frac{\upd(\phi_1-\phi_2)}{\upd 
t}\right\rangle_{\!t} 
%&=\lim_{\tau\rightarrow\infty}\left[\frac{\hbar}{\tau}
%\int_0^\tau \upd t \frac{\upd}{\upd
%t}\left(\int_{AB}\nabla\phi\cdot\upd\vecb s \right)\right]\\
=2\pi\hbar\left\langle\frac{\upd n}{\upd t}\right\rangle_{\!t},
\end{split}
\end{equation}
where $\langle \upd n/\upd t \rangle_t$ is the average rate of phase 
slip events. In what follows, we shall not be dealing with dynamical 
effects of this kind. In the later sections we shall also specialise to weak links 
in superfluid $^3$He and concentrate on some novel peculiarities in 
the current-phase relations $I(\phi)$ found in them.

\leveltwo{Josephson effect in $^3$He and $^4$He}

As soon as the superfluid phases of $^3$He were discovered in 1972, it seemed
obvious that $^3$He should display an equivalent to the Josephson-type 
effects found in superconductors. For $^4$He, they had already been
speculated about for long \cite{anderson1}, but for both systems they
kept on defying unambiguous experimental confirmation for quite some
time. Evidence for single phase-slips 
taking place in $^4$He, which satisfied Eq. (\ref{slippi}) very well, was clear by
mid-80s \cite{avenel1}. On the other hand, ideal nondissipative
Josephson behavior in $^4$He still remains to be observed. 

With $^3$He, the earliest experiments were made at Cornell and here in
Otaniemi in the late 70s \cite{web}. It was not until 1988 that
the first succesful observation was reported from University of Paris
in Orsay, France \cite{original}.
At vapor pressure and temperatures close to $T_c$ their $^3$He-B weak link showed
nearly ideal hydrodynamic behavior, so that Eq. (\ref{e.sine}) seemed to be
well satisfied. Dissipation only started to take place at lower temperatures.
The basic reason why ideal Josephson behavior is so much more difficult to
achieve in $^4$He than in $^3$He is in the two orders-of-magnitude 
difference in their coherence lengths: in $^4$He $\xi_0$ is on the
order of $1.5$ Å whereas in $^3$He it is some $700$ Å. For ideal
behavior, the dimensions of the weak
link should be of the same size or smaller than the coherence length,
and obviously $1.5$ Å (atomic size) is technically very difficult to achieve.

While many people thought the case for $^3$He was then settled, a group at
Berkeley persistently kept on developing their own experiment. After another
decade, in 1998, they reported the discovery of a new feature in the 
current-phase relation \cite{berkeley}. (The earlier progress of their
work is described in Refs. \cite{pere_jltp95,bp_nat97,bp_sci97}.)
%The weak link they used consisted of an array of $65\times 65$
%circular holes $100$ nm in diameter, spaced $3~ \mu$m apart in a $50$ nm thick
%membrane.\footnote{Actually, there is some ambiguity about the actual
%size and shape of the holes; these are the measures we originally assumed. } 
The weak link they used consisted of a regular parallel array of $65\times65$ small
apertures, each some $100$ nm in diameter, spaced $3~\mu$m apart in a
$50$ nm thick membrane.
They found that at temperatures below about $0.5 T_c$, the
weak link could be stabilised in a local 
minimum of energy at the phase difference
$\pi$. This state was seen in the $I(\phi)$-relation as a new branch 
around $\phi=\pi$, known as the $\pi$ branch, or the $\pi$ state.
Subsequent refinements of their apparatus brought about better
resolution, and the $\pi$ state could then be seen as a continuously
evolving kink in $I(\phi)$ with decreasing temperature
\cite{bistability}. In addition, the weak link could be found in two
different states with their own distinct current-phase relations and
critical currents. Finally, it has recently been reported that behavior reminiscent of the
$\pi$ state can be seen in a single narrow slit as well as the aperture array
\cite{singleaperture}.

In the wake of this excitement, we were among the first to start doing
calculations on the $\pi$ state.
A number of previous calculations of the Josephson effect in $^3$He 
already existed, but none were general enough to be able to explain
the observations \cite{mt86,kop86,hook87,kurkijarvi,thu88,uf89,tks,salomaa}.
The new attempts include
Refs. \cite{bose,hatakenaka,spoil}, while, \emph{a priori}, the most
convincing would seem to be Yip's work \cite{yippi} and our own
\cite{viljas}. This diploma
thesis reviews our previously published results and is a continuation of
that work. Among other things, we show that Yip's simplified model
is not likely to be a correct interpretation of the $\pi$ state. Ours is
perhaps better, but not at all flawless either. It will not always be
explicitly mentioned, but in all our estimates and comparisons to
``experiment'' we will be consistenly assuming the parameters of the
experimental aperture array of Ref. \cite{bistability}.

Much work still remains to be done, both in experiment and in
theory. For example, most work thus far has been done on B-phase and 
the effects of an external magnetic field have not been studied in detail. 
The next step in increasing computational difficulty would be the self-consistent 
calculation for apertures of finite size.
Furthermore, introduction of A-phase, or even A-B interfaces at the weak link --- with
or without magnetic fields --- and so on, provide combinations for further
research for years to come. For a short recent review on the $\pi$ state,
see the article by R. Bowley in Ref. \cite{bowley}.

\leveltwo{Applications of superfluid weak links?}

%To try to give a some sort of \emph{raison d'etre} for this whole
To try and give some sort of a motivation for these investigations,
I would like to note the following.
The Josephson effect in superconductors is best known for its
important applications in measurements of magnetic fields. 
Perhaps a bit surprisingly, in addition to its purely academic
interest, the superfluid equivalent may also be of practical use in
the future. Whereas in superconducting SQUID loops the phase is
coupled to the magnetic field, in superfluid
loops with a weak link it is, somewhat analogously, coupled to
rotation. Based on this it is possible to use superfluid weak links
for very accurate measurements of absolute rotation \cite{salmelin,schwab}.
The resolution obtained thus far is not quite as good as what can be
achieved using laser interferometry (Sagnac effect), but the
techniques are improving; see Refs. \cite{hakonen,rotation}.

\newpage

%In the superfluid transition, the quasiparticles near the fermi
%surface condense into a pair-correlated state, which can be described with the
%BCS-type ground state \cite{vollhardt}
%\begin{equation}
%|\Psi_{\textrm{BCS}}\rangle=\prod_{\vecb k}\prod_{\alpha}\bigg(
%u_{\vecb k \alpha\alpha} +\sum_\beta v_{\vecb k \alpha\beta}
%a_{\vecb k\alpha}^+a_{\vecb -k\beta}^+\bigg) |\textrm{vac}\rangle.
%\end{equation}

%\begin{equation}
%\langle\Psi| a^+_{\vecb k \alpha} a_{\vecb k' \beta}|\Psi\rangle
%=\delta_{\vecb k \vecb k'}\delta_{\alpha \beta}n_{\vecb k \alpha}
%\end{equation}

%\begin{equation}
%n_{\vecb k \alpha}=\sum_\beta v_{\vecb k \alpha \beta} 
%v_{\vecb k \alpha \beta}^*
%\end{equation}

%\begin{equation}
%F_{\vecb k \alpha\beta}=v_{\vecb k \alpha \beta}u^*_{\vecb k
%\beta\beta} = -F_{-\vecb k\beta\alpha}
%\end{equation}

%\begin{equation}
%\begin{split}
%H_{\textrm{MF}}-&\mu N=\\
%&\sum_{\vecb k\alpha}\xi_{\vecb k\alpha}
%a^+_{\vecb k \alpha}a_{\vecb k \alpha}+\frac{1}{2}
%\sum_{\vecb k \alpha\beta}(\Delta^*_{\vecb k \alpha\beta}
%a_{-\vecb k \beta}a_{\vecb k \alpha}+
%a^+_{\vecb k \alpha}a^+_{-\vecb k \beta}\Delta_{\vecb k \alpha\beta})
%-\sum_{\vecb k \alpha\beta}
%\Delta^*_{\vecb k \alpha\beta}F_{\vecb k \alpha\beta}
%\end{split}
%\end{equation}

%In $^3$He, as well as superconductors, mean field and Landau theory
%works all the way to transition temperature because of the extremely
%narrow critical region \cite{goldenfeld}. (Due to large size of cooper 
%pair and large correlation length.)

\levelone{Order parameter in $^3$He}
%Before going any further I should try to give some simplified
%account of the underlying microscopic theory of the superfluid phases 
%of $^3$He. 
Let us first consider the microscopic description of the
superfluid phases of $^3$He.
This is important, because the ``unorthodox'' results
to be discussed below are direct consequences of the more complicated 
order-parameter structure of $^3$He, as compared with, say, conventional superconductors.
The discussion given here is somewhat simplified;
for a more complete introduction, see Refs. \cite{mineev,leggett,ketterson,vollhardt}, for 
example.

Liquid $^3$He is a fermion system, just as electrons in a metal, with the
atoms possessing a (nuclear) spin $\frac{1}{2}$. Unlike electrons, 
the $^3$He atoms are neutral, but they are very strongly interacting. 
In fact, the hard-core repulsive interaction makes
pair formation in the $s$-wave state impossible, so that if pairing to a 
superfluid state is to take place, higher 
angular-momentum states are a necessity. Whereas in metals the weak
attractive interaction needed for pairing is provided by phonon
exchange, in $^3$He it is the Van der Waals interaction. Strictly speaking it 
is not the bare $^3$He atoms which form the ``Cooper pairs'' in the
superfluid state: the Landau ``Fermi liquid'' theory can be used to 
reformulate the description in terms of weakly interacting
fermionic excitations, called \emph{quasiparticles} \cite{leggett}.

\leveltwo{Spin triplet and $p$ wave}

To a good approximation, the pairing state in superfluid $^3$He is
spin triplet and $p$-wave. This means that the Cooper pairs have the
total spin $s=1$ and an orbital
angular momentum $l=1$, which is consistent with the requirement of the 
antisymmetry of the total fermionic wavefunction. Possible mixing in
of other kinds of $l,s$ combinations is usually neglected. 
%Let us now briefly consider the spin and orbital characteristics of the pair state.

\levelthree{Orbital angular momentum, $l=1$}

Let $\{\xvec^l_i, i=1,2,3\}$ denote an orthonormal triad of vectors in 
real space. We use these as the quantization axes for the orbital angular
momentum:
$\op\vecb L=\xvec_1^l\op L_1+\xvec_2^l \op L_2+\xvec_3^l \op L_3$.
The operator $\op L_3$, for example, has the three eigenstates 
$|Y^1_{+1}\rangle$, $|Y^1_{-1}\rangle$ and $|Y^1_{0}\rangle$,
with the eigenvalues $1$, $-1$ and $0$, respectively ---
the functions $Y^l_m(\theta,\phi)=\langle\theta,\phi|Y^l_m\rangle$
are the (suitably normalised) spherical harmonics.
From these, we construct the following orthonormal basis states
$|L_1\rangle\equiv-|Y^1_{+1}\rangle+|Y^1_{-1}\rangle$,
$|L_2\rangle\equiv\iu(|Y^1_{+1}\rangle+|Y^1_{-1}\rangle)$,
$|L_3\rangle\equiv\sqrt{2}|Y^1_{0}\rangle$,
which have the property $\op L_i|L_i\rangle=0, i=1,2,3$. 
A general normalised $p$-wave state may now be expressed as
$|\kvec\rangle=\hat k_1|L_1\rangle+\hat k_2|L_2\rangle+\hat k_3|L_3\rangle$, 
where
$\kvec=\hat k_1\xvec_1^l+\hat k_2\xvec_2^l+\hat k_3\xvec_3^l$ is a
unit vector. The vector transformation property of the $\hat k_i$'s 
is seen as follows. 

Let the quantization axes transform
under the rotation $\dyadic{R}\!{}^l(\nvec,\theta)$ as 
$\xvec^{l'}_j=\dyadic{R}\!{}^l\cdot\xvec_j^l=\sum_i\xvec_i^lR_{ij}^l$. 
The basis states for this new set of axes
are then obtained from the old ones via
$|L'_j\rangle=\op U^l|L_j\rangle=\sum_i|L_i\rangle U^l_{ij}$,
where $\op U^l(\nvec,\theta)=\exp(-\iu\theta\nvec\cdot\op\vecb L)$,
i.e., $\nam{\vecb L}$ is the generator of rotations in the angular-momentum 
space \cite{inui}. 
For  $l=1$ and this particular basis, it turns out that 
$U_{ij}=\langle L_i|\op U^l|L_j\rangle=R_{ij}$.
%In the new real space and $L$-space bases the (unrotated) $\kvec$ and 
%$|\kvec\rangle$ become
%$\kvec=\sum_i\hat k_i\xvec^l_i=\sum_{ij}R^l_{ji}k_i \xvec^{l'}_j$
%and
%$|\kvec\rangle=\sum_i\hat k_i|L_i\rangle=\sum_{ij}R^l_{ji}k_i|L'_j\rangle$.
Rotating the vector $\kvec$ and the $l$ state $|\kvec\rangle$ yield
$\dyadic{R}\!{}^l\cdot\kvec=\sum_j\hat k_j\dyadic{R}\!{}^l\cdot\xvec^l_j=$
$\sum_{i,j}R_{ij}\hat k_j\xvec_j$ and
$\op U^l|\kvec\rangle=\sum_j\hat k_j\op U^l|L_j\rangle=$
$\sum_{i,j}R_{ij}\hat k_j|L_i\rangle$.
Thus, under the $l$-rotations $\op U^l$, the components of $|\kvec\rangle$
transform just like the components of the vector $\kvec$. We may express this as
$\op U^l|\kvec\rangle=|\!\dyadic{R}\!{}^l\!\cdot\kvec\rangle$ and
thus identify rotations of an $l$ state with the real-space rotations
of $\kvec$. %Note that this is so just because we happen to have $L=1$.

%$\kvec\cdot\op\vecb L=\hat k_1 \op L_1+\hat k_2 \op L_2+\hat k_3 \op L_3$

\levelthree{Spin angular momentum, $s=1$}

For spin, the procedure is completely analogous. We choose
another set of quantization axes $\{\xvec^s_\mu, \mu=1,2,3\}$ and define
the pair spin operator
$\op\vecb S=\xvec_1^s\op S_1+\xvec_2^s \op S_2+\xvec_3^s \op S_3$,
which decomposes into two parts, one for each spin:
$\op\vecb S=\frac{1}{2}(\op{\sigmavec}_1+\op{\sigmavec}_2)$.
For $\op S_3$, we have the triplet eigenstates 
$|\!\!\uar\uar\rangle$, $|\!\!\dar\dar\rangle$ and
$|\!\!\uar\dar\rangle+|\!\!\dar\uar\rangle$, with the eigenvalues $1$, $-1$
and $0$, respectively. Using these, we define
$|S_1\rangle\equiv-|\!\!\uar\uar\rangle+|\!\!\dar\dar\rangle$,
$|S_2\rangle\equiv\iu(|\!\!\uar\uar\rangle+|\!\!\dar\dar\rangle)$,
$|S_3\rangle\equiv|\!\!\uar\dar\rangle+|\!\!\dar\uar\rangle$,
for which $\op S_\mu|S_\mu\rangle=0,\mu=1,2,3$. A general
normalised spin triplet state may be expressed as
$|\dvec\rangle=d_1|S_1\rangle+d_2|S_2\rangle+d_3|S_3\rangle$, where 
$\dvec=d_1\xvec_1^s+d_2\xvec_2^s+d_3\xvec_3^s$ is again 
a vector.\footnote{Here $\dvec$ is not necessarily normalised to unity, but
may be proportional to an overall (complex) temperature-dependent ``energy gap'' $\Delta$.
Often, as below in the context of quasiclassical theory,
$\dvec$ is written as $\boldsymbol{\Delta}$.
Its name varies in the literature: gap vector, 
order parameter vector, spin vector, etc.}
Just as above, rotations of the axes $\dyadic{R}\!{}^s(\nvec,\theta)$
and spin rotations 
$\op U^s(\nvec,\theta)=$
%$\op U^s(-\iu\theta\nvec\cdot(\op{\sigmavec}_1+\op{\sigmavec}_2)/2)$ 
$\exp(-\iu\theta\nvec\cdot\op{\vecb S})$ 
are related through $\op U^s|\dvec\rangle=|\dyadic{R}\!{}^s\!\cdot\dvec\rangle$
and can be identified.

%$\dvec\cdot\op\vecb S=d_1 \op S_1+d_2 \op S_2+d_3 \op S_3$

\levelthree{Pair state and different forms of the order parameter}

Let $\rvec$ be the center-of-mass coordinate of a pair.
A general $l=s=1$ pair state $|P(\rvec)\rangle$ may now be expressed in
the direct-product space spanned by the above spin and orbital basis states:
%the space spanned by the direct products of the above spin and orbital
%basis states
\begin{equation} \label{e.pair}
|P(\rvec)\rangle=\sum_{\mu,i} A_{\mu i}(\rvec)|S_\mu\rangle|L_i\rangle.
\end{equation}
Here the amplitude $A_{\mu i}(\rvec)$, or pair wavefunction, is usually called the
\emph{order parameter} of $^3$He.
Note that we always write
spin indices with Greek letters and orbital indices with
Latin letters. Often, Eq. (\ref{e.pair}) is expressed in a spin-state form by defining
$|\dvec(\rvec,\kvec)\rangle=\langle\kvec|P(\rvec)\rangle$
$=\sum_{\mu}d_\mu(\rvec,\kvec)|S_\mu\rangle$, where 
$d_\mu=\sum_iA_{\mu i}\hat k_i$ and $\hat
k_i=\langle\kvec|L_i\rangle$. If we further introduce
$\Delta_{\alpha\beta}(\rvec,\kvec)=\langle\alpha\beta|\dvec(\rvec,\kvec)\rangle$,
where $\alpha,\beta=\uar,\dar$, and write
$\ul\Delta=[\Delta_{\alpha\beta}]$, we get the most common form of
presenting the order parameter \cite{ketterson,vollhardt}
($\ul\sigma_\mu$ are the Pauli matrices):
%\begin{equation}
%\Delta_{\alpha\beta}(\rvec,\kvec)=A_{\mu i}(\rvec)\hat k_i S_\mu(\alpha\beta)
%\end{equation}
\begin{equation} \label{e.pairamp}
\ul{\Delta}(\rvec,\kvec)=\sum_\mu d_\mu(\rvec,\kvec)(\ul{\vecb\sigma}_\mu
\iu \ul{\vecb\sigma}_2)=
\left(\begin{array}{cc}
-d_1+\iu d_2 & d_3 \\
d_3 & d_1+\iu d_2
\end{array}
\right).
\end{equation}
This ``anomalous pair ampilitude'', ``off-diagonal mean field'' or
``gap matrix'' 
%$\Delta_{\alpha\beta}(\kvec)$ 
should be defined
more rigorously from a second-quantized
viewpoint; it is a thermodynamic expectation value, and 
$A_{\mu i}$ need not be interpreted as a wavefunction at all.  
The quantity $\ul\Delta$ can also be written for a more general BCS
type superfluid \cite{vollhardt}. 
It is a second-rank spinor, and transforms under spin
rotations as $\ul{\vecb\Delta}'=\ul U\ul{\vecb\Delta}\ul U^T$,
where $\ul U$ is the $2\times 2$ Wigner rotation matrix. 
%a matrix representation of $\op U^s=\exp(-\iu\theta\nvec\cdot\op{\sigmavec}/2)$.
Finally, I note that we can also give a
``dyadic'' representation for the order parameter:
$\dyadic{A}=\sum_{\mu i}A_{\mu i} \xvec_\mu^s\xvec_i^l$,
such that $\dvec=\dyadic{A}\!\cdot\kvec$. This rests completely on the 
possibility of identifying the spin and orbital bases with the
real space vectors defining their quantization axes, and therefore works only 
because we have $l=s=1$.
Since $A_{\mu i}$ transforms
as a vector with respect to both indices, it is sometimes called a 
``bivector''. Under simultaneous rotations of both the spin and orbital spaces, it
transforms as a second-rank \emph{tensor}. We denote the $3\times3$ matrix
representation of this tensor with $\dol A=[A_{\mu i}]$.

\leveltwo{Broken symmetries}

We chose above the spin and orbital bases such that the corresponding
amplitudes would transform conveniently under rotations. It can be stated that they
transform according to three-dimensional representations of the 
rotation group, namely the rotation matrices. Apart from the rotations, 
another type of a (global) symmetry transformation
is also attributed to the order parameter. This is the global gauge
transformation, i.e., an overall shift in the complex phase of Eq. (\ref{e.pair}).
The two rotations plus global gauge transfomations form the 
group $G=SO(3)^s\times SO(3)^l\times U(1)$, and, neglecting any
spin-orbit coupling, all operations of $G$ leave the free energy
invariant.
But operations of $SO(3)^s$, $SO(3)^l$ or $U(1)$ separately
\emph{do} change $A_{\mu i}$, that is, the physical state:
these symmetries are therefore said to be \emph{broken} in
the superfluid phases of $^3$He. Combined operations of $G$ may or may
not be symmetries of $A_{\mu i}$, depending on the situation. For
example, in the bulk B phase (see below), simultaneous $l$ and $s$
rotations remain a symmetry and form the group $SO(3)^{l+s}$.
A further example of a broken symmetry is given in the next section.

\leveltwo{Order parameter in the B phase}

The spin and orbital states $|\dvec\rangle$ and $|\kvec\rangle$
have the following properties $\dvec\cdot\op\vecb S|\dvec\rangle=0$ and
$\kvec\cdot\op\vecb L|\kvec\rangle=0$, as one can easily check.
The vectors $\dvec$ and $\kvec$ thus point in directions of zero
spin and angular momentum projections, respectively, and these directions
are related by $\dvec=\dyadic{A}\!\cdot\kvec$. The simplest possible
case one can now consider is that where the quantization axes are
equal, $\xvec^l_i=\xvec^s_i, i=1,2,3$, and 
$\dyadic{A}=\Delta\!\dyadic{I}$, or $\dvec=\Delta\kvec$, where 
$\Delta$ is (for now) a real valued constant of proportionality. 
In this case $A_{\mu i}=\Delta\delta_{\mu i}$ and one finds that if
$\op\vecb J=\op\vecb L+\op\vecb S$ is the total angular-momentum
operator, then $\op\vecb J^2|P(\rvec)\rangle=0$, i.e., the state 
with $\dvec=\Delta\kvec$ is an eigenstate of $\op\vecb J$ with the eigenvalue $j=0$.

%For the pair state this means 
%$\kvec\cdot(\op\vecb L+\op\vecb S)|P(\rvec)\rangle=0$, and since
%$\kvec$ is arbitrary, we see that in this case the \emph{total}
%angular momentum $J=L+S=0$.

In fact, this corresponds to an important stable bulk phase of superfluid
$^3$He: the Balian-Werthamer (BW) state, or the B phase. The amplitude 
$\Delta=\Delta(T)$ is a
temperature-dependent ``energy gap'', which is related to the energy
needed for breaking a Cooper pair. It is independent of $\kvec$,
which makes the B phase \emph{isotropic} and in many ways similar 
to superfluid $^4$He, or $s$ wave superconductors. This is in contrast
to the other stable phase (the A phase), which is axially anisotropic,
but we are not concerned with that here. 

The fact that $j=0$ is specific to the arbitrary
choice that $\dvec$ is parallel to $\kvec$. %when $\xvec^l_i=\xvec^s_i, i=1,2,3$.
If we do a rotation
$\dvec=\Delta\dyadic{R}(\nvec,\theta)\cdot\kvec$,
then the resulting $|P(\rvec)\rangle$ is no longer a state of definite $j$, but
rather a superposition of the $j=0,1,2$ states.
However, the states obtained in this way are degenerate in
energy with the unrotated one and should thus be equally probable to occur.
Taking also into account an arbitrary phase factor,
the most general form for the order parameter in the bulk B phase
is $\dyadic{A}=\Delta\dyadic{R}(\nvec,\theta)e^{\iu\phi}$, or 
\begin{equation} \label{e.bphase}
A_{\mu i}(\rvec) = \Delta R_{\mu i}(\nvec,\theta) e^{\iu \phi}.
\end{equation}
A dipole-dipole interaction between the nuclear magnetic moments fixes 
the rotation angle to the value 
$\theta=\theta_0\approx 104^\circ$, but this still leaves the
direction of the rotation axis $\nvec$ arbitrary.
Near a wall, the bulk form, Eq. (\ref{e.bphase}), is modified so that $\Delta$
should be replaced by a more general tensor quantity, as will be
discussed below.

%Rotation can also include rotations+inversions!

\newpage

\levelone{Ginzburg-Landau calculation}

\begin{figure}[!bt]
\begin{center}
\includegraphics[width=0.4\linewidth]{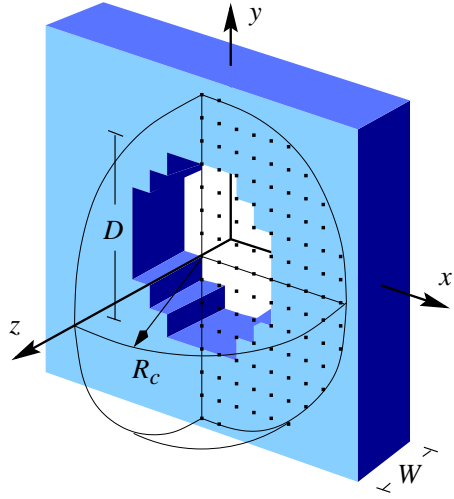}
\caption{A hole of diameter $D$ in a wall of thickness $W$.}
\label{f.hole3d}
\end{center}
\end{figure}

As a first case we considered a single cylindrical aperture (see Fig. \ref{f.hole3d}). 
The major task here is to calculate the order parameter 
self-consistently in and around the aperture. Doing this 
in the general quasiclassical formalism for all temperatures would
be a tremendous task. It would be even more so, if the full circular 
symmetry of a cylindrical aperture could not be assumed, which 
will turn out to be the case. However, near $T_c$ we may apply 
the Ginzburg-Landau (GL) expansion of the free energy and find the stationary
order-parameter field by minimising it. This procedure has
alredy been presented and described in more detail in Ref. 
\cite{etyo2}, but I include the results here for completeness.\footnote{The 
results presented in the special assignment were also slightly erroneous.}

\leveltwo{GL theory}
On small length scales (on the order of the coherence length,
$\xi_{\textrm{GL}}$) one may neglect the 
weak dipole-dipole interactions between the nuclear moments. 
Then, in the absence 
of a magnetic field, the GL free-energy expansion includes two 
terms \cite{vollhardt}
\begin{equation} \label{e.Fff}
F=F_b+F_k=\int\!d^3r [f_b+f_k],
\end{equation}
which are called the bulk and gradient energies, respectively. 
The bulk, or condensation free-energy density is given by
\begin{equation} \label{e.fbulk}
\begin{split}
f_{b} & = -\alpha~\Tr(\dol{A}\dol{A}\!\TC)
+ \beta_{1} 
        \vert\Tr(\dol{A}\dol{A}\!\T)\vert^2 
+ \beta_{2} 
        [\Tr(\dol{A}\dol{A}\!\TC)]^2 \\
& + \beta_{3}
        \Tr(\dol{A}\dol{A}\!\T\!\dol{A}\!\C\!\dol{A}\!\TC)
+ \beta_{4}
        \Tr(\dol{A}\dol{A}\!\TC\!\dol{A}\dol{A}\!\TC)
+ \beta_{5}
        \Tr(\dol{A}\dol{A}\!\TC\!\dol{A}\!\C\!\dol{A}\!\T).
\end{split}
\end{equation}
It includes all terms up to fourth order in $A_{\mu i}$ which are invariant
under separate rotations of basis vectors in the spin and orbital spaces,
and are linearly independent. Strictly speaking, in addition to the
dipole energy, we are also neglecting the weak spin-orbit
coupling of the Cooper pairs, both of which would violate this symmetry. 
The gradient, or kinetic energy density arises from a spatial bending of the 
order parameter field on the coherence-lenght scale:
\begin{equation} \label{e.fgrad}
%f_{k}= K \left[(\gamma -2)
%        \partial_{i}A_{\mu i}\partial_{j}A_{\mu j}^*
%        +\partial_{i}A_{\mu j}\partial_{i}A_{\mu j}^*
%        +\partial_{i}A_{\mu j}\partial_{j}A_{\mu i}^* 
f_{k}= K_1
        \partial_{i}A_{\mu i}^*\partial_{j}A_{\mu j}
        +K_2\partial_{i}A_{\mu j}^*\partial_{i}A_{\mu j}
        +K_3\partial_{i}A_{\mu j}^*\partial_{j}A_{\mu i} .
\end{equation}
These terms are also rotationally invariant with respect to both indices.
For a better justification of these forms for the energy densities, see
for example Refs. \cite{ketterson,vollhardt}. 

The supercurrent is related to the gradient energy by the de Gennes procedure for
ascribing to the Cooper pairs a fictitious ``gauge charge'' and
introducing a gauge field which couples to it \cite{degennes,cross}.
For a mass current, these are the Cooper-pair mass $2m_3$ and an external 
velocity field $\vecb{v_n}$ \cite{vollhardt}. To obtain the current
density $\vecb j$, one
must replace $\boldsymbol\nabla$ in $f_k$ by the ``gauge invariant'' 
form $\boldsymbol\nabla-\iu\frac{2m_3}{\hbar}\vecb{v_n}$ and  %\cite{schakel}
vary $F$ with respect to $\vecb{v_n}$
\begin{equation}
\vecb j(\rvec) =-\lim_{\vecb{v_n}\rightarrow\vecb 0}
\frac{\delta F}{\delta \vecb{v_n}(\rvec)},
\end{equation}
which means that $\vecb j$ and $\vecb v_n$ are, in some sense, 
conjugate variables.
To the first order in $\vecb{v_n}$ this yields for the $i$'th
component of the mass-current density the expression
%(in units of $\hbar\rho_s/2m_3\xi_{GL}$) 
\begin{equation} \label{e.curr}
%j_i=\frac{4m_3K}{\hbar}\mathrm{Im}\left[(\gamma-2)A_{\mu
%i}^*\partial_j A_{\mu j}+
%A_{\mu j}^*\partial_i A_{\mu j}+A_{\mu j}^*\partial_j A_{\mu i}
%\right],
j_i=\frac{4m_3}{\hbar}\mathrm{Im}\left[K_1A_{\mu
i}^*\partial_j A_{\mu j}+K_2
A_{\mu j}^*\partial_i A_{\mu j}+K_3A_{\mu j}^*\partial_j A_{\mu i}
\right],
\end{equation}
which is exactly conserved at a minimum of Eq. (\ref{e.Fff}).

\leveltwo{Implementation}
%\begin{equation} \label{glequred}
%\begin{split}
%\partial_j\partial_j A_{\mu i}
%+&(\gamma-1)\partial_i\partial_j A_{\mu j}\\
%=\big[-&\dol{A}
%+\beta_1\dol{A}\!\C\Tr(\dol{A}\dol{A}\!\T)
%+\beta_2\dol{A}\Tr(\dol{A}\dol{A}\!\TC)\\
%+&\beta_3\dol{A}\dol{A}\!\T\!\dol{A}\!\C
%+\beta_4\dol{A}\dol{A}\!\TC\!\dol{A}
%+\beta_5\dol{A}\!\C\!\dol{A}\!\T\!\dol{A} \big]_{\mu i}.
%\end{split} 
%\end{equation}
The minimum of the functional in Eq. (\ref{e.Fff}) was computed by deriving the
corresponding Euler-Lagrange equations, discretizing them in a 
3D lattice, and solving the resulting set of nonlinear equations by iteration.
The lattice is shown in Fig. \ref{f.hole3d}. Only a $m'm2'$ symmetry
($m$ denotes mirror reflection, $2$ a rotation by $180^\circ$, and prime a 
time inversion\footnote{These kinds of group operations conform to the 
International notation of crystallographic point groups; they should be
understood here as acting \emph{simultaneously} on both the spin and orbital spaces.}) was
used in the calculation.

At the wall we required the order parameter to vanish, which corresponds approximately
to a strongly diffusely scattering surface. Although this \emph{is} probably
the correct limit in most experiments, here the choice was
more a matter of convenience: any other boundary condition on the wall 
would have made the problem quite difficult. 
In the bulk, the boundary condition
\begin{equation}
\lim_{z\rightarrow\pm\infty}A_{\mu i}(x,y,z)=\Delta e^{\pm\iu \phi/2} \delta_{\mu i}
\end{equation}
was imposed, where $\phi$ is the phase difference between the two sides.
This requires the bulk superfluid to be in the B phase, with the
$\nvec$ vectors on the two sides sides being parallel. The general rotation
matrix need not be considered here, since we may neglect the dipole-dipole energy
on these length scales.

The current was calculated by integrating Eq. (\ref{e.curr}) over the aperture.
It is important here to keep track of current conservation, and the
self-consistency requirement that the total mass current through the
hole should be exactly given by
\begin{equation}
J(\phi)=\frac{2m_3}{\hbar}\frac{\partial F(\phi)}{\partial\phi} .
\end{equation}
This relation should be possible to confirm at least within GL theory, and is
really equivalent to current conservation in the stationary configuration
which minimises the free energy.

%At least within the GL theory, this can be shown to be a consequence
%of current conservation, $\nabla\cdot\vecb j=0$.

\leveltwo{Results: a trapped vortex state?}
\begin{figure}[!tb]
\begin{center}
\includegraphics[width=0.6\linewidth]{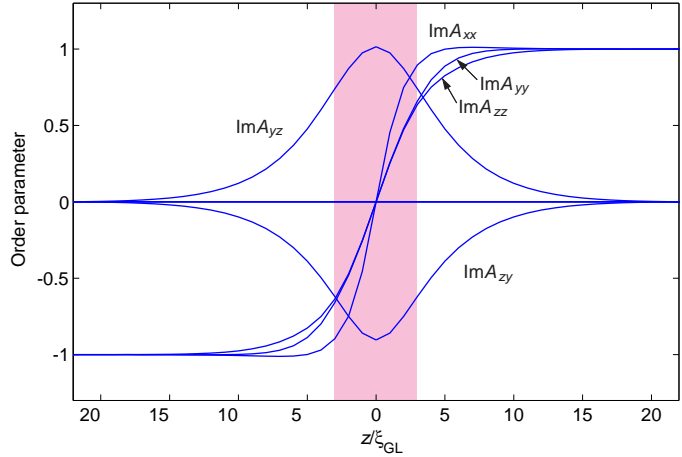}
\caption{The nonzero order-parameter components on the $\pi$ branch at
$\phi=\pi$. The large components in the middle are a product of
the broken symmetry: on the normal branch all components would vanish there
and the state would possess the full circular symmetry of the aperture.
The wall of width $W=6\xi_{GL}$ is shown shaded and the
diameter of the aperture was $D=10\xi_{GL}$. }
\label{f.profi}
\end{center}
\end{figure}
Figure \ref{f.profi} shows the nonvanishing components of the order
parameter in a particular new state which we found to be stabilised
for phase differences around $\phi=\pi$. The components $A_{zy}$ and
$A_{yz}$, which bulge in the middle, break the circular symmetry of
the aperture. The presence of the wall and the aperture has already
reduced the bulk symmetry of the B phase down to
$\frac{\infty}{m'}\frac{2'}{m}$, where
$\infty$ denotes continuous rotation (around $\zvec$), and this is
valid on any normal $J(\phi)$-branch. But now we have a case of 
\emph{broken symmetry}, where
even the simultaneous spin and orbital rotations around $\zvec$
result in new states 
$R_{\mu \nu}(\zvec)A_{\nu j}R^T_{ji}(\zvec)\neq A_{\mu i}$ 
with degenerate energies.
The remaining symmetry of the state is only $m'm2'$, which is just what 
we used in the calculation; assuming a symmetry somehow higher 
would have left the state undiscovered. 

An interesting fact is that the structure of the order parameter
in this state closely resembles
that of a double-core vortex \cite{thu87}. The state, which we
proposed to be related to the experimentally observed $\pi$ state,
would then correspond to a situation where a half-quantum vortex
has crossed the aperture and formed a trapped double-core vortex state
in it. In other words, a phase slip by $\pi$ has occurred. 
The analogy is perhaps more easily understood in the
case of an infinitely long narrow slit (a purely 2D situation), where a similar state was 
found. It is partly a mystery why that was not found already
in the 2D calculation reported in Ref. \cite{thu88}. 
%(There a similar
%structure was found for antiparallel spin-rotation axes.)

\begin{figure}[!btp]
\begin{center}
\includegraphics[width=0.65\linewidth]{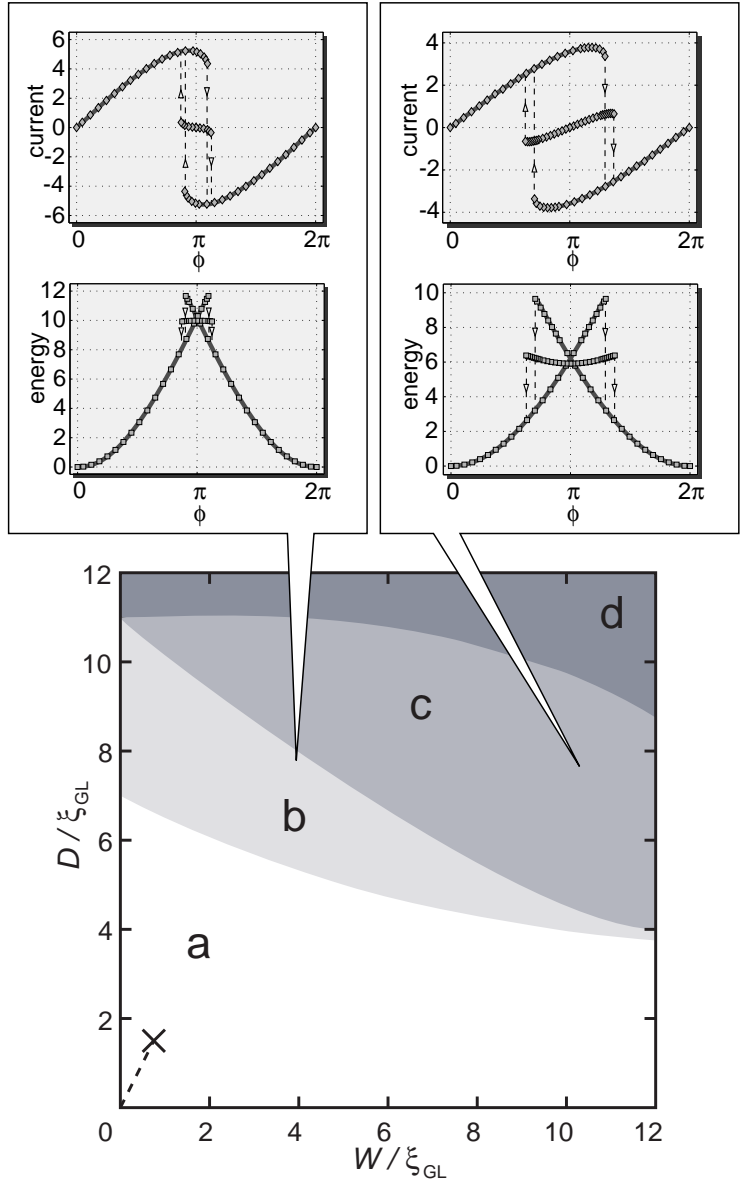}
\caption{Theoretical phase diagram for the $\pi$ state in a single
aperture. The current and the energy as a function of phase difference 
$\phi$ are also shown in two cases. Here $D$ is the diameter of the
aperture and $W$ the thickness of the wall. The $\pi$ branch is found
in regions (b)-(d), where it is locally stable at a fixed phase
difference of $\phi\approx\pi$. In regions (c) and (d), it is a local 
minimum of energy with respect to $\phi$ at $\phi=\pi$. In regions (b) 
and (c), it is also the absolute minimum of energy at $\phi=\pi$.
The parameters for one aperture of the experimental aperture array are 
indicated by the dashed line, and the observation of the $\pi$ state is
marked by the cross.}
\label{f.diagram}
\end{center}
\end{figure}
The lower part of Fig. \ref{f.diagram} shows a phase diagram describing the stability of 
the new state with varying junction parameters, namely the thickness
of the wall, $W$, and the diameter of the hole, $D$. In region (a), i.e., for
the smallest apertures tested, the state was not found. In region (b)
the ``$\pi$ branch'' appears, having a negative slope of the
current-phase relation at $\phi=\pi$  (insert at upper left). In regions
(c) and (d) the slope is positive and the $\pi$-branch is very strong
(insert at upper right). In (d) the state is only a local minimum
of energy at $\phi=\pi$, whereas in (b) and (c) it is a global minimum 
(although just barely). Marked with a line are the dimensions (in units of the
\emph{temperature-dependent} coherence length
$\xi_{\textrm{GL}}=\hbar\vF/\sqrt{10}\Delta(T)$) 
of one experimental aperture, and the approximate observation point of 
the $\pi$ state is denoted by a cross. These are deep in the region
where our $\pi$ branch is absent. Some of this discrepancy might be 
explained by the fact that the GL model we used works best only near
$T_c$. But this does not seem like the whole truth:
it appears that if this calculation is to have anything to do with 
the $\pi$ state of Ref. \cite{berkeley}, then the existence of a large
array of holes in this experiment should play a special role. 

On the other hand, our $\pi$ state could well be what is seen in the single 
narrow slit of Ref. \cite{singleaperture}. However, there are some
differences which require an explanation. 
The experimental $J(\phi)$'s would seem to have a $\pi$ branch even when 
the normal branches are apparently nonhysteretic. For our circular
apertures and the purely 2D slit, the normal branches are strongly hysteretic and
phase slips would appear to occur only between them. 
The $\pi$ branch in holes of this kind may thus not be experimentally 
attainable at all, unless one can somehow start directly from it at $\phi=\pi$. 
Smaller holes would get rid of the hystesesis, but in our case that
would also suppress the $\pi$ state.

Fortunately, there are also some other untested possibilities. 
(1) For a single slit of finite length, the calculation should probably be done
in three dimensions --- not just in the 2D approximation, which
assumes an infinite slit length (or a channel which restricts the 
flow to be strictly two-dimensional). This is suggested by the fact that
for such a slit the current density should be much larger at the ends of
the slit than in the middle. The results could differ from the 2D 
approximation and the 3D circular hole, which are the two cases we
have investigated so far.
(2) A more general quasiclassical calculation at low temperatures
could give a $\pi$ branch even for circular apertures with
nonhysteretic normal-branch behavior. 
(3) In the experimental cell of Ref. \cite{singleaperture},
there is no reason why the spin-rotation axes $\nvec$ on the different 
sides of the aperture should be parallel to each other, which is what
we have assumed. Instead, they may be free to point in just about 
any directions.

It would be interesting to see if some of these improvements would
allow direct transitions to a $\pi$ branch (like in those reported in Ref. \cite{thu88} for
antiparallel $\nvec$'s), or even a continuously evolving kink.

\newpage

\levelone{Tunneling model}

Next we deal with the details of the ``tunneling model'' of Ref. \cite{viljas}.
There we modeled the aperture array by the following phenomenological
expression for its coupling energy
\begin{equation} \label{e.fjg}
F_J=-\mathrm{Re}\sum_\mu[aA_{\mu z}^{L*}A_{\mu z}^{R}
+b(A_{\mu x}^{L*}A_{\mu x}^{R}+A_{\mu y}^{L*} A_{\mu y}^{R})],
\end{equation}
with $a,b$ constants having the same phase and assumed to be real.
This is obtained from the most general form of such coupling
\begin{equation} \label{e.general}
F_J=\sum_{\mu\nu i j} a_{\mu\nu i j} A^{L*}_{\mu i} A^{R}_{\nu j} +
{c.c.}, \qquad  (a_{\mu\nu i j}~\mathrm{real})
\end{equation}
by introducing the restrictions that (a) the barrier be ``achiral'' (only states 
with the same angular momentum projections on the two sides are
coupled), (b) no spin flips are produced (only states with same spin
projection on the two sides are coupled) and (c) the barrier is
isotropic in its plane \cite{degennes}.

In this simple model, which I call the ``18-state model'' (a
generalization of the Feynman two-state model \cite{barone,feynman}), the general
superposition state describing the two superfluids can be expressed as
\begin{equation} \label{e.state}
%|A\rangle=\sum_\mu\sum_i \td{A}_{\mu i}^{L}|\mu\rangle_L^l|i\rangle_L^s
%+\sum_\mu\sum_i \td{A}_{\mu i}^{R}|\mu\rangle_R^l|i\rangle_R^s,
|A\rangle=\sum_{\mu, i} \td{A}_{\mu i}^{L}|\mu,i\rangle_{\!L}
+\sum_{\mu,i} \td{A}_{\mu i}^{R}|\mu,i\rangle_{\!R},
\end{equation}
where the amplitudes are $\td{A}_{\mu i}=A_{\mu i}/\Delta=\langle\mu, i |A\rangle$.
The basis states are just the direct products of the spin and orbital angular
momentum states:
$|\mu, i\rangle=|\mu\rangle^s\otimes|i\rangle^l\equiv|\mu\rangle|i\rangle$.
We naturally assume them to be orthonormal, i.e.
$\langle\mu,i|\nu, j\rangle=\delta_{\mu \nu}\delta_{ij}$ 
on both sides.
In this basis, the tunneling Hamiltonian can be written
\begin{equation} \label{e.transfer}
\mathcal{H}_T=-\Delta^2
\sum_\mu [a(|\mu, z\rangle_{\!R} {{_L\!}\langle z,\mu|})
+b(|\mu, x\rangle_{\!R} {{_L\!}\langle x,\mu|}+|\mu, y\rangle_{\!R} {{_L\!}\langle y,\mu|})]
+{h.c}.
\end{equation}
Its expectation value in the state of Eq. (\ref{e.state}), namely
$\langle A|\mathcal{H}_T| A\rangle$, is now equal to Eq. (\ref{e.fjg}).
Note that the symmetry properties mentioned above are satisfied by
Eq. (\ref{e.transfer}): it only couples states with equal $l$ and $s$
projections on the two sides.
%\emph{i.e.} the barrier produces no ``spin-flips''.

\leveltwo{Interactions in the absence of a magnetic field}

For superfluid B phase on both sides of the barrier the coupling
energy, Eq. (\ref{e.fjg}), becomes
\begin{equation} \label{e.fj}
F_J=-[\alpha R_{\mu z}^LR_{\mu z}^R+\beta(R_{\mu x}^LR_{\mu
x}^R+R_{\mu y}^LR_{\mu y}^R)]\cos\phi.
\end{equation}
Here, and henceforth, we follow the usual summation convention for repeated 
indices, except for $x$, $y$, and $z$.
Formally this expression is obtained by substituting the order parameters
$A_{\mu i}^{L,R}=\Delta R_{\mu i}^{L,R} \exp{(i\phi^{L,R})}$ into
Eq. (\ref{e.fjg}), but actually there is no simple relation
between the coupling constants $a, b$ and $\alpha,\beta$. This is
because the order parameter of a $p$-wave superfluid is strongly
suppressed near walls and the exact meaning of Eq. (\ref{e.fjg}) is not
well defined. This will be considered in greater detail below,
when we present the self-consistent calculations of the order
parameter and a ``pinhole'' junction.
The result of the quasiclassical calculation is that,
near $T_c$, the coupling energy of a dense, coherent array of such pinholes is
of the form shown in Eq. (\ref{e.fj}) and the parameters $\alpha$ and $\beta$ can
be evaluated. More precisely, what comes out of the calculation is the
the total mass current $J$, but that should be related to $F_J$ through
\begin{equation} \label{e.currene}
J=\frac{2m_3}{\hbar}\frac{\partial F_J(R_{\mu i}^L,R_{\mu i}^R,\phi)}{\partial \phi},
\end{equation}
as already mentioned.
%\footnote{The approximation made here restricts
%us very close to $T_c$ and destroys some interesting effects, which 
%should only begin to appear at around $T=0.8 T_c$ or so.}

To analyse the coupling energy in Eq. (\ref{e.fj}) we have to parametrise the 
rotation matrices somehow. Typically one writes them in terms of the 
componenets of $\nvec$ and the rotation angle $\theta$
\begin{equation}
\begin{split}
R_{ij}(\nvec,\theta)&=\cos\theta\delta_{i j}
+(1-\cos\theta)\hat{n}_i\hat{n}_j-\sin\theta\epsilon_{i j k}\hat{n}_k \\
&=\frac{1}{4}[-\delta_{i j}+5\hat{n}_i\hat{n}_j-\sqrt{15}\epsilon_{i j k}\hat{n}_k].
\end{split}
\end{equation}
The latter form follows, because the long-range dipole-dipole
interaction \cite{mineev}
\begin{equation}
F_d = 8 g_d \Delta^2 \int\upd^3r \left(\frac{1}{4}+\cos\theta \right)^2
\end{equation}
fixes $\theta$ to its minimising value
$\theta=\theta_0=\arccos(-\frac{1}{4})\approx 104^\circ$.
Other interactions are here not strong enough to deflect the angle 
from this value considerably. Even at the wall it costs less energy
to deflect the direction of $\nvec$, instead. 
We may write $F_J$ in terms of
\begin{equation} \label{e.eka}
R_{\mu i}^L R_{\mu i}^R = \frac{1}{16}
[25(\nvec^L\cdot\nvec^R)^2+30(\nvec^L\cdot\nvec^R)-7]
\end{equation}
and
\begin{equation} \label{e.toka}
\begin{split}
R_{\mu z}^L R_{\mu z}^R = \frac{1}{16}[1&-5(\cos^2\eta^L+\cos^2\eta^R)
+25\cos^2\eta^L\cos^2\eta^R + \\
&+\sin\eta^L\sin\eta^R(25\cos\eta^L\cos\eta^R+15)\cos(\chi^L-\chi^R) \\
&-5\sqrt{15}\sin\eta^L\sin\eta^R(\cos\eta^L-\cos\eta^R)\sin(\chi^L-\chi^R)],
\end{split}
\end{equation}
where $\nvec^L$ and $\nvec^R$ have been represented in the polar ($\eta^{L,R}$) and
azimuthal ($\chi^{L,R}$) angles, the polar axis being the wall normal $\svec=\zvec$.
The energy only depends on the difference $\chi^L-\chi^R$, as required by symmetry.
%$R_{\mu i}^L R_{\mu i}^R$ varies between $-1$ and $+3$. It is maximised 
%by $\nvec^L=\nvec^R$ and minimised by $\nvec^L\cdot\nvec^R=-\frac{3}{5}$.
%$R_{\mu z}^L R_{\mu z}^R$ varies between $-1$ and $+1$.

In the presence of the wall we also have the surface dipole
interaction \cite{hydro}
\begin{equation} \label{eq.sd}
F_{\textrm{SD}}=\int_S \upd^2r[
b_4(\hat \vecb s\cdot \hat \vecb n)^4-
b_2(\hat \vecb s\cdot \hat \vecb n)^2],
\end{equation}
where $\svec$ is the surface normal.
This always tends to orient $\nvec$ perpendicular to the wall. 
In our case, however, the wall is locally
transparent at the junction, and 
the resulting coupling interaction $F_J$ outweighs $F_{\textrm{SD}}$ by
orders of magnitude. But away from the transparency, this is the
still dominant surface interaction (in the absence of a magnetic field) and 
should really leave only two choices for the orientation of the
$\nvec$ vectors there. It is now believed that this explains the experimentally
observed ``bistability'' of Ref. \cite{bistability}. In our model,
$F_{\textrm{SD}}$ 
is assumed to fix the left and right $\nvec$'s far away from the
junction perpendicular to the wall, either parallel or antiparellel,
but otherwise it is neglected --- it only appears in the form of a
phenomenological boundary condition as explained below.

As a result, there is a competition between the orienting effects of the
surrounding walls and the weak link. It is mediated by the gradual
bending of the left and right side $\nvec$ fields between their
orientations in the bulk and at the weak link. 
The bending energy of B phase is generally of the 
form \cite{hydro}
\begin{equation}
F_{\textrm{G}}=
\int\upd^3r\left [
\lambda_{\textrm{G1}}
\frac{\partial R_{\alpha i}}{\partial r_i}
\frac{\partial R_{\alpha j}}{\partial r_j}+
\lambda_{\textrm{G2}}
\frac{\partial R_{\alpha j}}{\partial r_i}
\frac{\partial R_{\alpha j}}{\partial r_i} \right],
\end{equation}
with a surface contribution
\begin{equation}
F_{\textrm{SG}}=\int_S\upd^2r \hat{s}_j R_{\alpha j}
\frac{\partial R_{\alpha j}}{\partial r_i}.
\end{equation}
The $\nvec(\rvec)$ field could be calculated exactly by 
a minimisation of these along with the coupling term, but that would
be overly complicated for our purposes. Instead, the following 
simplifications were made.\footnote{For the meaning of the parameters you
should consult Refs. \cite{thu87,hydro,serenerainer}, for example.}

\begin{figure}[!bt]
\begin{center}
\includegraphics[width=0.70\linewidth]{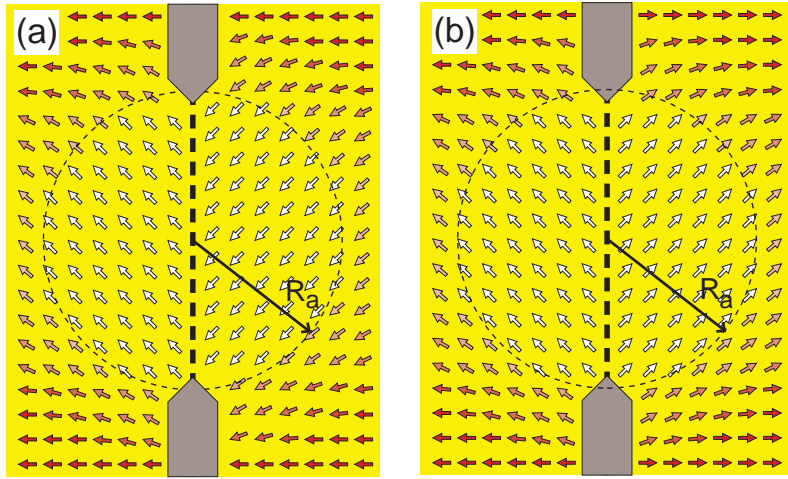}
\caption{Schematic view of $\nvec(\rvec)$ near the weak link in the $\pi$ state. 
Surrounding walls align $\nvec$ perpendicular to themselves, which 
gives rise to two possible relative orientations of $\nvec$'s far away from
the weak link: parallel (a) and antiparallel (b).}
\label{f.array}
\end{center}
\end{figure}

For the surface-gradient term, we assume the GL region, $\gamma=3$, and 
$\Delta_{\perp}\!=$ constant, which gives 
$\lambda_{\textrm{SG}}=4K\Delta^2=4\lambda_{\textrm{G2}}$. For the bulk 
terms we assume $\lambda_{\textrm{G1}}=2\lambda_{\textrm{G2}}$, and
expressing the rotation matrices in terms of $\nvec$ yields
\begin{equation}
F_{\textrm{Gtot}}=\frac{5}{8}\lambda_{\textrm{G2}}
\int\upd^3r \left[ 16(\partial_i\hat n_j)^2-
(\sqrt{3}\nabla\cdot\nvec+\sqrt{5}\nvec\cdot\nabla\times\nvec)^2
\right].
\end{equation}
This result was also obtained in Ref. \cite{smith}. Next we assume
that $\nvec$ varies only in one plane:
$\nvec=\sin\eta\xvec+\cos\eta\zvec$. Then we average over coefficients
in front of derivatives, for example 
$\sin^2\eta(\partial\eta)^2\rightarrow\frac{1}{2}(\partial\eta)^2$. In 
addition, we take the average
$\sum_ia_i(\partial_i\eta)^2\rightarrow\frac{1}{3}(\sum_ia_i|\nabla\eta|^2)$.
Doing all this we get
\begin{equation}
F_{\textrm{Gtot}}=\frac{25}{3}\lambda_{\textrm{G2}}
\int \upd^3r|\nabla \eta|^2.
\end{equation}
This can be minimised by a solution of the form
$\eta(\rvec)=\eta^L_\infty+c/r$ on the left side, and similarly on the
right. The divergence of the integral at
$r=0$ is cut off at the radius $R_a=a\sqrt{N/\pi}$, where $N=4225$ is
the number of apertures and $a=3~ \mu$m is the lattice constant of the
2D aperture array. Doing the integrals we find the quadratic forms
\begin{equation} \label{e.quadforms}
F_{\textrm{Gtot}}^L=\gamma(\eta^L-\eta^L_{\infty})^2 \qquad
\textrm{and} \qquad
F_{\textrm{Gtot}}^R=\gamma(\eta^R-\eta^R_{\infty})^2,
\end{equation}
where 
%\begin{equation} \label{e.gamma}
$\gamma=\frac{50}{3}\sqrt{\pi N}\lambda_{\textrm{G2}}a$.
%\end{equation}
They contain only the polar angles of the $\nvec$'s at the
junction and are thus convenient to handle. The
temperature-dependent $\gamma$ can be estimated from a detailed 
calculation but in practise it was ``fitted'' to experimental  data,
as were also $\alpha$ and $\beta$, the only other remaining free
parameters in our simple model. The infinity angles
$\eta_\infty^{L,R}$ are always assumed to be either $0$ or $\pi$ to
choose either the parallel or the antiparallel configuration.
Figure \ref{f.array} shows a schematic view of the expected spatial variation
of the $\nvec$ field around the weak link.
The dashed circles represent the cutoff radius $R_a$ used in the definition of the 
$\gamma$ parameter.

%\newpage

\leveltwo{Analysis of the tunneling model}

The goal was to find the $\nvec^{L,R}$ configurations which minimise the 
energy $F_J+F_{\textrm{Gtot}}$ for each fixed phase difference $\phi$.
The following analysis is done in two steps, first forgetting the gradient
energy altogether. It is immediately seen that the angle
$\phi=\frac{\pi}{2}$ is important, since $\cos\phi$ changes sign
there and the minimisation of the bracketed term in $F_J$,
namely $E_J$ in Eq. (\ref{e.brack}) below, changes to maximisation, or vice versa.

\levelthree{Coupling term only}

Near $T_c$, the coupling energy for the array was claimed to have the form
$F_J=-E_J\cos\phi$, where the Josephson energy $E_J$ is not just a
constant, but
\begin{equation} \label{e.brack}
\begin{split}
E_J
%&=[\alpha R_{\mu z}^LR_{\mu z}^R+\beta(R_{\mu x}^LR_{\mu
%x}^R+R_{\mu y}^LR_{\mu y}^R)] \\ %\cos\phi \\
&
=[(\alpha-\beta) R_{\mu z}^LR_{\mu z}^R+\beta(R_{\mu i}^LR_{\mu
i})]. %\cos\phi
\end{split}
\end{equation}
The current through this barrier is $J=J_c\sin\phi$ where $J_c=(2m_3/\hbar)E_J$.
We first summarize the stability criteria of the Josephson normal and
$\pi$ ''branches'' by taking into account this energy term alone.
Both the parallel and the antiparallel cases are considered.

Consider first $\cos\phi>0$.
The parallel normal state with
$\hat{\vecb{n}}^L=\hat{\vecb{n}}^R=\hat{\vecb{z}}$ 
has $J_c=(2m_3/\hbar)(\alpha+2\beta)$ and
is locally stable for all
$\alpha$ and $\beta$;
the antiparallel normal state with 
$\hat{\vecb{n}}^L=-\hat{\vecb{n}}^R=-\hat{\vecb{z}}$
has $J_c=(2m_3/\hbar)(\alpha-\frac{7}{4}\beta)$
and is locally stable only for $\alpha>\beta$. Note that at
$\alpha=\frac{7}{4}\beta$ this $J_c$ changes sign.
At $\cos\phi=0$, both normal branches become unstable towards a
(discontinuous) jump to the
$\pi$ branch. On the $\pi$ branch, which is now stable and has lower
energy for all $\alpha,\beta>0$ and $\cos\phi<0$, we have again two cases
\begin{equation}
J_c=\left\{\begin{array}{ll}
-(2m_3/\hbar)\alpha & \textrm{for} ~\alpha>\beta\\
-(2m_3/\hbar)(2\beta-\alpha) & \textrm{for} ~\alpha<\beta\\
\end{array} \right. .
\end{equation}
The first of these is achieved by
\begin{equation}
\nvec^{L,R}=\frac{1}{\sqrt{5}}(\mp\sqrt{3}\xvec\mp\yvec+\zvec)
\end{equation}
corresponding to $R_{\mu x}^LR_{\mu x}^R=+1$, 
 $R_{\mu y}^LR_{\mu y}^R=-1$, and  $R_{\mu z}^LR_{\mu z}^R=-1$.
The other case is 
\begin{equation}
\nvec^{L,R}=\frac{1}{\sqrt{5}}(\mp\xvec+\yvec\mp\sqrt{3}\zvec)
\end{equation}
with $R_{\mu x}^LR_{\mu x}^R=-1$, 
$R_{\mu y}^LR_{\mu y}^R=-1$, and  $R_{\mu z}^LR_{\mu z}^R=+1$.
That is more or less all there is to be said about this case.

\levelthree{Gradient energy included}
The only other significant energy term arises from the bending or
gradient energy, which we consider in the simplified form
\begin{equation}
F_{\textrm{Gtot}}=\gamma(\eta^{L}-\eta_\infty^{L})^2
+\gamma(\eta^{R}-\eta_\infty^{R})^2,
\end{equation}
as explained above.
%The meaning of the angles and the $\gamma$-parameter are.
When this is taken into account, certain changes to the stability
considerations arise. 
The normal branch with
$\hat{\vecb{n}}^L=\hat{\vecb{n}}^R=\hat{\vecb{z}}$ 
($\eta_\infty^L=\eta_\infty^R=0$)
remains unchanged for all $\alpha$ and $\beta$. 
A linear stability analysis shows that the branch is now locally stable for
\begin{equation}
%\phi_c^{\uparrow\uparrow}=\arccos(-\frac{2\gamma}{5(\alpha+\beta)})
5(\alpha+\beta)\cos\phi+2\gamma \geq 0,
\end{equation}
regardless of $\alpha$ and $\beta$. Thus the critical phase difference 
($\phi_c\in[0,\pi]$) is given by
\begin{equation}
%\phi_c^{\uparrow\uparrow}=\arccos(-\frac{2\gamma}{5(\alpha+\beta)})
\cos\phi_c^{\uparrow\uparrow}=-\frac{2\gamma}{5(\alpha+\beta)},
\end{equation}
If $\gamma=0$ this clearly reduces to the
case considered previously, $\phi_c=\pi/2$. Otherwise $\phi_c$ is 
moved closer to $\pi$. The transition from the normal to $\pi$ branch
can be shown to be continuous (second order).

The antiparallel normal branch with
$\hat{\vecb{n}}^L=-\hat{\vecb{n}}^R=-\hat{\vecb{z}}$ 
($\eta_\infty^L=\pi, \eta_\infty^R=0$)
also remains the same for all $\alpha>\beta$. 
For $\alpha<\frac{7}{4}\beta$, the critical current of
the branch still becomes negative. However, the situation where
$\alpha<\beta$ is more complicated, because then the bulk direction 
of the $\hat{\vecb{n}}$ vector is the only thing trying to keep the
branch stable. For the antiparallel case, a linear stability analysis
shows  that the branch is locally stable if
\begin{equation} \label{conds2}
\begin{split}
(25\alpha-5\beta)\cos\phi+16\gamma &\geq 0 \quad \textrm{and} \\
15(\alpha-\beta)\cos\phi+16\gamma &\geq 0,
\end{split}
\end{equation}
again regardless of the values of the parameters. 
Here comes an
important point: there are now \emph{two} critical phases whose positions depend 
on the parameters. For $\alpha>\beta$, the second condition is
automatically satisfied and only the (upper) critical phase difference
\begin{equation}
%\phi_c^{\uparrow\uparrow}=\arccos(-\frac{2\gamma}{5(\alpha+\beta)})
\cos\phi_{c1}^{\uparrow\downarrow}=-\frac{16\gamma}{(25\alpha-5\beta)}
\end{equation}
is relevant. The transition from the normal to 
the $\pi$ branch is discontinuous (first order).\footnote{The 
first one of the conditions (\ref{conds2}) is suppressed
if one requires that $\hat n^L_z=\pm \hat n^R_z$ always, and 
one obtains only the lower critical phase $\phi_{c2}$. On the other
hand, it can be seen that this requirement is usually statisfied on both
the normal and the $\pi$ branch.
This leads to the conclusion that the transition between the branches
at $\phi_{c1}$ proceeds via a route where it is \emph{not} satisfied 
and is therefore discontinuous.}
For $\alpha<\beta$ there is also a second (lower) critical phase
\begin{equation}
%\phi_c^{\uparrow\uparrow}=\arccos(-\frac{2\gamma}{5(\alpha+\beta)})
\cos\phi_{c2}^{\uparrow\downarrow}=-\frac{16\gamma}{15(\alpha-\beta)}.
\end{equation}
When $\frac{1}{5}\beta<\alpha<\beta$ the normal branch is stable \emph{between} 
the two critical phases. 
Finally, when $\alpha<\frac{1}{5}\beta$ the critical phase
$\phi_{c1}$ becomes irrelevant and the normal branch is
stable for $\phi>\phi_{c2}$ all the way to $\phi=\pi$.
% the condition for $\phi_{c1}$ becomes unbinding

On the $\pi$ branch itself, the changes introduced by a finite $\gamma$ 
are significant, but not much can be said about them with analytical
considerations. Whenever we are off the stable normal branch, the form of the
current-phase relation can only be obtained with a numerical
minimisation of $F_J+F_{\textrm{Gtot}}$. 
However, usually (not always) the
$\hat{\vecb{n}}$ vectors seem to be directed such that the
gradient energies on both sides of the junction are equal,
i.e., $\hat n^L_z=\pm \hat n^R_z$.
There is some hysteresis in $J(\phi)$ related to the 
discontinuous transition in the antiparallel case, but this is small and 
can also be studied only numerically. 

The cases where $\hat{\vecb{n}}$ is not perpendicular to the 
wall around $\phi=0$ can be quite similar to the usual $\pi$ branches
around $\phi=\pi$. There is essentially only a phase shift of $\pi$
separating the appearance of their current-phase relations. This should 
be taken into account in interpreting the experimental data: the
``$\pi$ state'' in the antiparallel case could as well be located
at $\phi=0$. We might even see a nonzero $1-|\hat n_z|$ as some kind of an ``order
parameter'' and the \emph{definition} of the $\pi$ state, regardless of the
values which $\phi$ might have there.
The Berkeley experiment can, indeed, only determine the
phase difference modulo $\pi$. This is in contrast to the toroidal cell
geometry of Ref. \cite{singleaperture} which can resolve absolute
phase differences.

%\newpage

\begin{figure}[!bt]
\begin{center}
\includegraphics[width=0.65\linewidth]{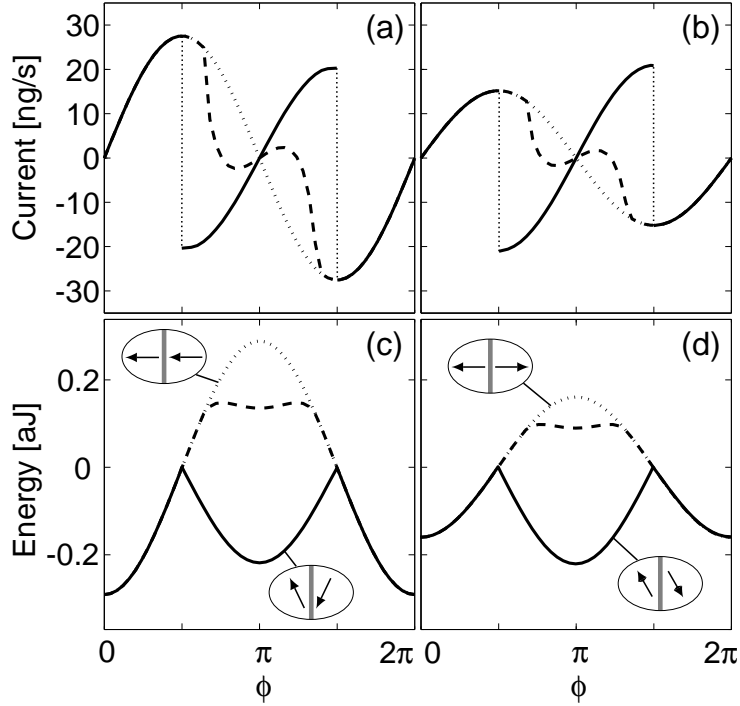}
\caption{Current-phase relationships and energies for the tunneling 
model. The left and right panels correspond to parallel and
antiparallel $\nvec$-vectors far away from the junction, respectively.
The directions near the junction are depicted schematically by arrows.
The curves correspond to different values of the gradient-energy
parameter $\gamma$: ideal $\pi$ state ($\gamma=0$, solid line), no
$\pi$ state ($\gamma=\infty$, dotted line), and an intermediate case 
($\gamma=0.245$ aJ, dashed line). The parameters $\alpha=0.2207$ aJ
and $\beta=0.0347$ aJ are chosen to imitate the the experiment 
\cite{bistability} at $T=0.55 T_c$.}
\label{f.gamma}
\end{center}
\end{figure}

\leveltwo{Results and experimental implications}
The essential features of the current-phase 
and energy-phase relations, with and without the $\pi$ state,
are shown in Fig. \ref{f.gamma}.
The ratio of the gradient-energy parameter $\gamma$ to the coupling
parameters $\alpha$ and $\beta$ determines their general form. If, due to a
large $\gamma$, the $\nvec$ vectors are fixed exactly parallel or antiparallel and
perpendicular to the wall at the junction, the resulting $J(\phi)$ is
sinusoidal. This situation is similar to what is found in $s$-wave
supeconductors where the extra degrees of freedom related to the
spin-orbit rotation are not present. With small enough $\gamma$,
there is some critical phase on the interval $[0,\pi]$ above which the
sinusoidal branch gives way to a lower-energy state, where
the $\nvec$-vectors near the junction have been deflected from their bulk
directions. 
The fact that the $\pi$ state only appears at low temperatures
is explained in this model with the different temperature dependencies 
of the parameters: $\alpha,\beta\propto(1-T/T_c)^2$, and
$\gamma\propto(1-T/T_c)$, such that $\gamma$ dominates close to $T_c$.

\begin{figure}[!btp]
\begin{center}
\includegraphics[width=0.75\linewidth]{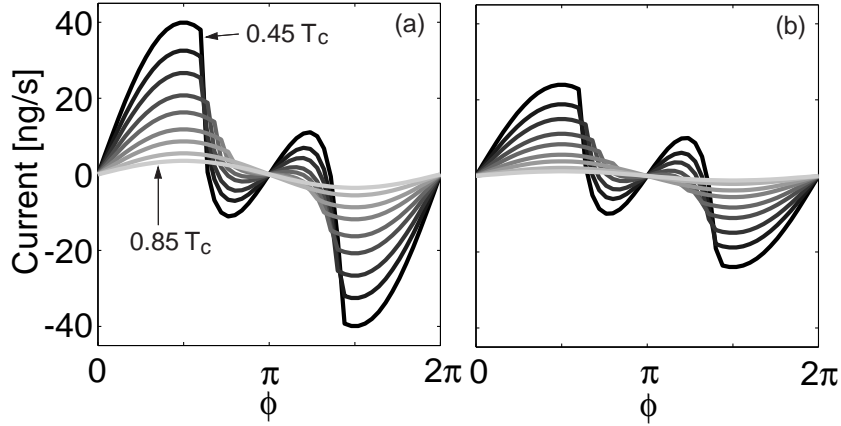}
\caption{Current-phase relationships for parallel (a) and
antiparallel (b) $\nvec$-vectors far away from the junction. The different
curves correspond roughly to temperatures from $0.45~ T_c$ to 
$0.85~ T_c$ with intervals of $0.05 ~T_c$. The parameters $\alpha$ and 
$\beta$ are calculated with the pinhole model and $\gamma$ is
estimated as explained in the text (see also Fig. \ref{f.array}). 
However, to get better correspondence with
experiments, we have multiplied the estimated values by factors 
of $7.5$, $1.3$ and $0.15$, respectively.}
\label{f.temps}
\end{center}
\end{figure}

The parameters $\alpha$ and $\beta$ were estimated using a
quasiclassical pinhole model. Their expressions are presented
below in the context of a general quasiclassical calculation.
Estimation of $\gamma$ involves evaluation of the gradient-energy 
parameter $\lambda_{\textrm{G2}}$ whose calculation
is discussed in Ref. \cite{hydro} and in Ref. \cite{serenerainer},
where
$\lambda_{\textrm{G2}}\equiv-(\hbar^2/4m_3)\rho_2^{\textrm{spin}}$; it
is not be presented here. 
\begin{figure}[!bt]
\begin{center}
\includegraphics[width=0.60\linewidth]{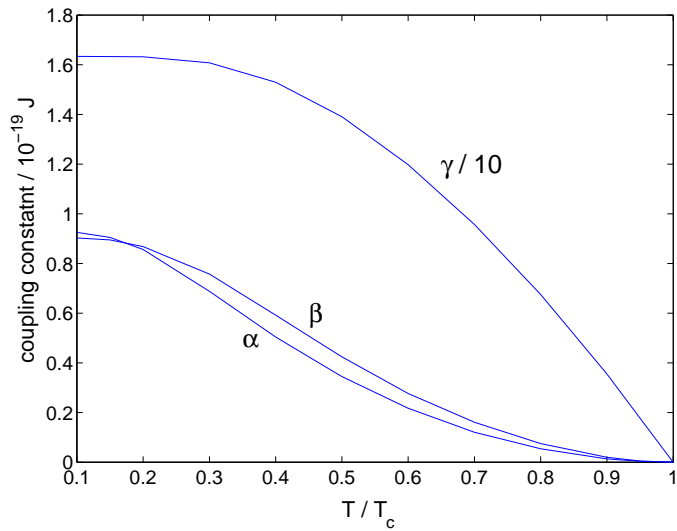}
\caption{Theoretical values for the parameters $\alpha$, $\beta$ and
$\gamma$ in the tunneling model. 
These were obtained from Eqs. (\ref{e.alpha}), (\ref{e.beta})
and (\ref{e.quadforms}), respectively. The following
experimental parameters were assumed: wall thickness $W=50$ nm, 
circular hole diameter $D=100$ nm, number of holes in the lattice
$N=4225$, lattice constant $a=3~\mu$m, and total lattice area
$S=3.8\cdot10^{-8}$ m$^2$. The values plotted here were modified in
subsequent calculations to give a better fit to the experiments.}
\label{f.tconst}
\end{center}
\end{figure}
Figure \ref{f.tconst} shows the temperature dependence
of the bare $\alpha$, $\beta$ and $\gamma$, as they come out 
from their respective equations using the experimental array parameters.
%The size of the experimental aperture 
%array and the true aspect ratio of its apertures were used in the estimations. 
As can be seen, there is more than an order-of-magnitude difference between 
$\gamma$ and the coupling parameters $\alpha$ and $\beta$. This
discrepancy can probably be decreased with a more realistic way of
estimating $\gamma$. However, at least the temperature dependencies should
be correct, and we dealt with the magnitude differences in a simple,
if not completely satisfying, way. To provide a more reasonable fit to the experimental
findings, the bare theoretical estimates were just scaled with some
constants. Figure \ref{f.temps}
shows the current-phase relationships for the best-fit scaling constants,
which are given in the caption. 

In the range of parameters which appeared to reproduce the
experimental results best, more complicated behavior than that depicted in
Figs. \ref{f.gamma} and \ref{f.temps} did not occur.
Unfortunately, the more unstable behavior described in the preceding
subsection is a real possibility within the model. It is realised for
the bare pinhole estimates, for which we usually have $\alpha<\beta$.
Here we avoided this problem by multiplying $\alpha$ with a larger
constant than $\beta$, which is an arbitrary choice and is not easily
justified. In the more
general pinhole calculation to be presented below, we do not have 
$\alpha$ and $\beta$ which could be separately adjusted; there we 
have to make do with the theory as it is.\footnote{We will see that 
restricting the angles of the transmitted quasiparticles will favor $\alpha$ over
$\beta$. This corresponds to increasing the aspect ratio $W/D$ of the
apertures. However, this not very useful, since it also leads to a
drop in the critical currents, which is not desirable.} 

Not only the relative sizes of $\alpha$ and $\beta$ pose a problem.
The pinhole estimates would appear to give slightly too low critical 
currents as well. 
%Considering apertures of finite size does not help,
%because they should always result in \emph{smaller} critical
%currents than a pinhole \cite{tks}. 
This can be explained in basically two ways: either the
experimental holes are slightly larger than those reported, or there
is some extra leakage present at the weak link, which the
experimenters are unaware of. Since the time these calculations were carried out
and published in Ref. \cite{viljas}, we have learned that one of these 
cases is indeed very possible (see Sect. 7). 

Furthermore, a slow variation of the
$\nvec$ field at the edges of the array and a slightly incoherent
behavior of the different apertures could easily smoothen the
features of $J(\phi)$ around the critical phases. Here the transition 
points are exactly defined, and appear very sharp and abrupt.
In addition, the finite aperture sizes probably cause some extra
slanting of $J(\phi)$ in the experiment. 
%The gradient-energy parameter $\gamma$ is also too large, but that is
%easier to believe, because of the crude approximation made in evaluating it. 
To put it more generally, the specific properties of the actual experimental cell
affect the measurements in ways that cannot be fully taken 
into account within our simple model. 
All of these things together could perhaps explain why the
details of the measured $J(\phi)$'s cannot be exactly reproduced 
with it.

In spite of these problems, the tunneling model makes at least two
clear predictions which should 
yield to experimental testing. The first one concerns the linear
dimension $L$ of the aperture array: the coupling and gradient
energies depend differently on $L$, namely 
$\alpha/\gamma, \beta/\gamma \propto L$. This means that the larger
the array is made, the more pronounced a $\pi$ state should be observed.
The other experimental prediction concerns magnetic fields: a strong enough
magnetic field should lock the directions of $\nvec^L$ and $\nvec^R$,
thereby making $J(\phi)$ strictly sine-like, although perhaps
slanted or hysteretic, as is usual for finite aperture sizes and low temperatures.
In the next subsection, the effects of an external magnetic field are given a
bit more complete and quantitative analysis.

\leveltwo{Effect of magnetic field}
In the bulk superfluid, a magnetic field will try to orient the
$\nvec$ vector parallel or antiparallel to itself. 
This gives rise to an energy term of the form
\begin{equation}
F_{\textrm{DH}}=-a\int \upd^3r(\hat\vecb n\cdot \vecb H)^2.
\end{equation}
This results from a combination of the depairing effects of the magnetic
field and the dipole energy \cite{smith}. 
A surface, on the other hand, will tend to orient $\nvec$ parallel
or antiparallel to its normal, because of the surface dipole
interaction in Eq. (\ref{eq.sd}). But in a magnetic field there emerges another 
surface energy which has the form
\begin{equation} \label{eq.sh}
F_{\textrm{SH}}=-d\int_S \upd^2r
(H_\alpha R_{\alpha i} \hat s_i)^2.
\end{equation}
For fields greater than about $5$ mT ($50$ G), it is usually the stronger
surface interaction \cite{yippi}.
It arises from depairing effects and can be understood as follows.
A magnetic field tends to break pairs with $m_s=0$ along the field
direction, whereas the surface breaks pairs with $m_l=0$ along the
surface normal. Both of these cost energy, and a minimum cost
can be achieved by choosing the direction of $\nvec$ such that these are 
exactly the same pairs. In the minimum of Eq. (\ref{eq.sh}) this is
satisfied as $R_{\alpha i}(\nvec)$ then rotates the surface normal to the 
magnetic field direction \cite{smith,cross}.

Let us first consider a homogeneous magnetic field $\vecb H_\perp$ 
perpendicular to the wall. In that case, the surface field energy is
minimized with $\hat \vecb n = \pm\hat\vecb z$. This energy is given by
\begin{equation} \label{eq.min1}
%F_{\textrm{SH}}^{\perp} \equiv
F_{\textrm{SH}}^{0\perp} = 
F_{\textrm{SH}}^{\perp}(\hat \vecb n = \pm\hat\vecb z)=
%- x g_Z \Delta^2 \xi_{\textrm{GL}}
-d|\vecb H^{\perp}|^2S,
\end{equation}
where $S$ denotes the surface area.
In the normal branch this will not affect the current-phase relation,
since there $\nvec$ is aligned perpendicular to the wall, anyway. 
Going to the $\pi$ state, however, this no longer
is the case: if $\alpha>\beta$, say, the Josephson coupling will tend
to orient $\nvec$ such that
$\hat n_z=\nvec\cdot\svec=\pm 1/\sqrt{5}$ (on both sides). In this case
$H_zR_{zz}(\hat n_z=1/\sqrt{5})\hat s_z = 0$
and hence $F_{\textrm{SH}}^{\perp}(\hat n_z=1/\sqrt{5})=0$, which is
higher than the value in Eq. (\ref{eq.min1}).
This discrepancy in the energies may act to suppress the $\pi$ state for
high enough perpendicular fields.
In the case of a field $\hvec^{\parallel}$ which is parallel to the
wall (no perperdicular component present), $F_{\textrm{SH}}$ is minimised with 
$(\nvec\cdot\svec)^2=(\nvec\cdot\hat\hvec)^2=1/5$ and again the minimum 
is
\begin{equation} \label{eq.min2}
F_{\textrm{SH}}^{0\parallel} = 
F_{\textrm{SH}}^{\parallel}(\hat n_z=\hat n_H=\pm 1/\sqrt{5})=
-d|\vecb H^{\parallel}|^2S.
\end{equation}
A field parallel to the wall can thus obviously affect the
normal state ($\nvec=\pm \hat\vecb z$) as well as the $\pi$ state
since $F_{\textrm{SH}}^{\parallel}(\nvec=\pm\hat\vecb z)=0$, which is 
again higher than the minimum in Eq. (\ref{eq.min2}). How exactly this
takes place is considered below  along with a general orientation of
$\hvec$.
%A general orientation of $\hvec$ will be considered later.

\levelthree{Estimation of field sizes}

To get a feeling for the magnitude of the fields which can 
affect the current-phase behavior significantly, we have to insert
some numerical values. The effect of a gradient energy or any other interaction but
$F_{\textrm{SD}}$ will be neglected.
Table \ref{tb.glpars} gives some of the required parameters in the
Ginzburg-Landau regime.
\begin{table}[!b] 
\caption{Ginzburg-Landau parameters for the calculation of
the surface magnetic interaction at zero pressure; 
$k_B T_c\approx1.38\cdot 10^{-26}$ J. }
\begin{center}
\begin{tabular}{|c|c|c|c|}
\hline 
$\alpha/(1-T/T_c)$ & $K$ & $g_z$ & $\Delta^2/(1-T/T_c)$ \\
$[10^{50}$J$^{-1}$m$^{-3}]$ & 
$[10^{34}$J$^{-1}$m$^{-1}]$ &
$[10^{50}$T$^-2$J$^{-1}$m$^{-3}]$ & 
$[(k_B T_c)^2]$\\
\hline
1.67 & 41.9 & 2.33 & 8.75 \\
\hline
\end{tabular}
\end{center}
\label{tb.glpars}
\end{table}
We also know that $d=d_0 (g_z\Delta^2\xi_{\textrm{GL}})$ where
$d_0\approx2.2$ for a diffusely scattering and $d_0\approx4.4$ for a
specularly scattering wall at zero pressure. The GL coherence length
$\xi_{\textrm{GL}}$ is defined as $\xi_{\textrm{GL}}=\sqrt{K/\alpha}$.
It can be extrapolated to lower temperatures with $\xi_{\textrm{GL}}(T)=\hbar
v_{\textrm{F}}/\sqrt{10}\Delta(T)$, if the behavior of the temperature-dependent
gap $\Delta(T)$ is known. We use the values 
$\Delta(0.45 T_c)\approx \sqrt{3}k_BT_c$ and 
$\Delta(0.8 T_c)\approx \sqrt{1.75}k_BT_c$. 
Above $T\approx 0.8 T_c$, the linear approximation
of Table \ref{tb.glpars} is sufficient.

We give the field estimates assuming the conditions of the
experiment \cite{bistability}. The letters $H$ and $L$ refer to
the high and low critical current cases in that experiment.
We denote the Josephson energy (in absence of the kink at
$\phi=\pi$) with $E_J\approx F_J(\pi/2)-F_J(0)$. 
Furthermore, we denote 
$E_\pi\equiv F_J(\pi)-F_J(0)$ such that 
$E_{\textrm{gain}}\approx 2E_J-E_\pi$ is the energy gained by dropping from the 
(imaginary) ``normal branch'' to the ``$\pi$ branch'' at $\phi=\pi$. 
This is the energy which a (possibly additional) perpendicular
magnetic field would have to win in order to kill the $\pi$ state and
make $J(\phi)$ sinusoidal. At low temperatures 
($T \approx 0.45 T_c$), these energies are 
$E_J^H=2.4\cdot10^{-19}$ J$=1.5$ eV, 
$E_{\textrm{gain}}^H=0.8\cdot10^{-19}$ J$=0.5$ eV and 
$E_J^L=1.0\cdot10^{-19}$ J$=0.6$ eV, 
$E_{\textrm{gain}}^L=1.1\cdot10^{-19}$ J$=0.7$ eV.
At high themperatures ($T\approx 0.8T_c$), the $J(\phi)$ is sinusoidal in
the experiments, but we may still consider it:
$E_J^H=0.6\cdot10^{-19}$ J$=0.35$ eV and
$E_J^L=0.2\cdot10^{-19}$ J$=0.12$ eV. In fact, if the gradient energy
and other interactions but the coupling are neglected, our model can allow for
nonsinusoidal behavior at any temperature.

From the theoretical pinhole calculation, one obtains an order-of-magnitude
estimate for the coupling parameters:
\begin{equation}
\alpha/\textrm{m}^2,\beta/\textrm{m}^2 \propto \hbar \vF 
\kB T_c \NF\frac{\pi D^2}{4a^2}=3.48\cdot10^{-11} \frac{\textrm{J}}{\textrm{m}^2}.
\end{equation}
Here the values $a=3~\mu$m, $D=100$ nm for the hole spacing
and diameter have been used. For an array of 65x65 holes of
this kind, the area of the array is about $3.8\cdot10^{-8}$ m$^2$ and
the order of magnitude of the coupling energies is $1.3\cdot10^{-11}$
J $\approx 8$ eV. Performing the actual calculation we have, for $T=0.45 T_c$,
$\alpha\approx 0.44\cdot10^{-19}$ J and $\beta\approx0.34\cdot10^{-19}$ J.
Generally speaking, these values give too small coupling energies and 
to get the model working properly, we need to multiply them with
factors like 7.5 and 1.3, respectively (see the discussion above). 
These give a Josephson energy
$E_J^H$ on the order of $2\cdot10^{-19}$ J $\approx1.3$ eV. The order of
magnitude is then about the same as in the experiment.
Based on the GL values in Table \ref{tb.glpars} we cannot really make accurate
estimates at low temperatures. The temperature dependence of $\alpha$
and $\beta$ in our model is also accurate only neat $T_c$ where
the coupling energy then drops as $(1-T/T_c)^2$. 
On the other hand, the experimental $\pi$ state is only visible at low temperature,
so that is the temperature region we are interested in.
Using the extrapolation of the coherence length, we assume that we can 
at least get the correct order of magnitude out.

If we know the total energy of the junction in its different 
configurations, we can estimate the fields which give surface energies 
of the same order. For example, if we want to know the perpendicular
field which will suppress the $\pi$ state by orienting $\nvec$ always
perpendicular to the wall, we can do the following: %for which the $|F_{\textrm{SH}}^{0\perp}|$ 
For the magnetic surface energy to win a coupling of the order $|E|$,
we should have $2|F_{\textrm{SH}}^{\perp}| \gtrsim |E|$.
(The factor of two is present because the wall has two sides here.) 
Then Eq. (\ref{eq.min1}) will yield a ``critical field''
\begin{equation}
|\hvec^\perp_{\textrm{c}}|\approx
\sqrt{\frac{|E|/2}
{d_0(g_z\Delta^2\xi_{\textrm{GL}}) S}}\approx
\sqrt{\frac{|E|/2\textrm{J}}{d_0(\Delta/k_BT_c) (S/\textrm{m}^2)}}
\cdot1.25\cdot10^{4} \textrm{T},
%\sqrt{\frac{|F_{\textrm{J}}|/\textrm{J}}{d_0(1-T/T_c)^{1/2}}}\cdot
%1.1738\cdot10^{11} \textrm{G},
\end{equation}
which goes as $\sim (1-T/T_c)^{3/4}$ near $T_c$.
Now $E\approx E_{\textrm{gain}}$, as explained before. 
We know this formula will give only a very rough estimate, so we just assume 
$E\approx 0.5$ eV$\approx0.8\cdot10^{-19}$ J.
Using the area $S=3.8\cdot10^{-8}$ m$^2$, and the gap values
mentioned above, this corresponds to
about $H_c\approx 6.5$ mT at $T=0.45T_c$ and $H_c\approx 7.7$ mT at 
$T=0.8 T_c$ for a diffuse wall.
Varying the parameters and using the estimated energies for the H and L
cases at these two temperatures, values of $H_c$ in the range of about
$3$ mT to some $15$ mT ($30$ G to $150$ G) are
obtained. In comparison, the Earth's magnetic field is
about $50~\mu$T ($0.5$ G). Our critical field values are at least much
larger this, and it thus seems that the Earth's field alone should not play a
decisive role.

\levelthree{General orientations}

At a very high field, the coupling energy becomes negligible and it
will be mostly the magnetic field which orients the $\nvec$ vectors.
This case has been analysed in Ref. \cite{yippi}. For a given angle between 
the magnetig field and the surface normal, there exist several
different possibilities for their orientation. If the coupling energy
is negligible, the orientations will be chosen at random
between the possibilities. Each of these cases may lead to a different
critical current. If the coupling
energy is also significant, jumps between these different
orientations may occur when the phase is varied. 

It has already been shown that a large enough perpendicular field will 
orient $\nvec$'s perpendicular to the wall and make the
current-phase relationship sinusoidal. This is not necessarily so when 
there is a parallel field component present. Assuming now \emph{only} a parallel
field, oriented in the direction of $\pm \hat \vecb y$, it can be shown
that the components of the allowed $\nvec$ vectors are given by
\begin{equation} \label{eq.allowed}
\left(+\sqrt{\frac{3}{5}},\pm\frac{1}{\sqrt{5}},\pm\frac{1}{\sqrt{5}}
\right),\quad
\left(-\sqrt{\frac{3}{5}},\mp\frac{1}{\sqrt{5}},\pm\frac{1}{\sqrt{5}} \right),
\end{equation}
where the upper and lower signs correspond to each other. 
On the other hand, as has been mentioned before, the orientations
preferred by the Josephson coupling when $\cos\phi<0$ are
\begin{equation} \label{eq.nvecs}
\left(\mp\sqrt{\frac{3}{5}},\mp\frac{1}{\sqrt{5}},+\frac{1}{\sqrt{5}} \right),
\end{equation}
where the upper and lower signs refer to the $\nvec$ vectors on the two 
sides of the junction. These orientations are, however, not unique: 
%The symmetries of the junction allows for some variation in the signs. 
the energy is degenerate with respect to 
any rotation around the $\hat\vecb z$ axis. 
If a small parallel magnetic field is switched on, the
degeneracy is lifted. But, as
can be seen from Eqs. (\ref{eq.allowed}) and (\ref{eq.nvecs}), some of the 
states allowed by the two interactions agree exactly. Thus,
a purely parallel magnetic field can in fact \emph{enhance} the
``$\pi$ type state'', where the $\nvec$-vectors are not perpendicular to
the wall. Unfortunately, although this coincidence is quite amusing
it is not very interesting, since (for large fields) it just results in another 
sinusoidal $J(\phi)$ and we have seen plenty of them. 

This concludes our study of the
phenomenological tunneling model and the
effects of an external magnetic field in the depth we have felt it necessary to
consider so far. Subsequent sections are devoted to a fully
quasiclassical study of a single pinhole, plus an enhancement of the
aperture array calculation on this basis. Effects of external magnetic fields
will be neglected.

%possibly with a negative critical current, i.e. a phase shift of $\pi$ . 
%With a large enough parallel field, the one of the above
%configurations remains for all $\phi$, and the $J(\phi)$ is sinusoidal 
%but possibly with a negative critical current, or a ``phase shift'' of $\pi$.

\newpage

%\section{Effect of walls}

\levelone{Quasiclassical theory}

The microscopic description of normal metals and superconductors has been
formulated completely in terms of Green's
functions \cite{abrikosov1}. The problem with the full interacting Green's function
is that is contains a lot of detailed microscopic information which is
often not needed for practical calculations. It was first shown by 
Eilenberger \cite{eilenberger} and Larkin and Ovchinnikov
\cite{larkin} that Gorkov's equations for the (stationary) full Green's function
can be transformed to ``transport-like'' equations for a
\emph{quasiclassical} Green's function, or propagator, where all the
unnecessary information concerning atomic length scales has been
integrated away at the outset. These are 
the so-called Eilenberger-Larkin-Ovchinnikov equations.
%These can be generalised to cover nonstationary cases 

The same approach can be used equally well for $^3$He which is also 
a degenerate fermion system, or a Fermi liquid \cite{serenerainer}. A condition for
the applicability of the quasiclassical approach is that the
characteristic length scales $q^{-1}$ are much 
larger than the Fermi wavelength
$\lambda_{\textrm{F}}=2\pi/\kF$. Similarly,  energies must be
much lower than the Fermi energy $E_{\textrm{F}}=\kB T_{\textrm{F}}$. 
If time dependence were to be considered, we would also require the
time scales to be much longer than the inverse Fermi 
frequency $\hbar/E_{\textrm{F}}$. We are, however, only concerned
with stationary effects. 
The length scale in superfluid $^3$He is
set by the coherence length $\xi_0=\hbar\vF/2\pi\kB T_c$, and energies 
by the transition temperature $T_c$ or the gap $\Delta\approx\kB T_c$.
At low pressures, for example, we have
$\xi_0\approx70$ nm $\gg\lambda_{\textrm{F}}\approx 0.8$ nm and 
$T_c\approx3$ mK $\ll T_{\textrm{F}}\approx 1$ K, so the requirements
are well satisfied. In what follows, we use a weak-coupling form of
the quasiclassical approach, where
quasiparticle-quasiparticle scattering is neclected.
We also restrict to the vapor pressure in all the 
numerical estimates (see Appendix B).

%This varies only on a length scale $q^{-1}\gg\kF^{-1}$ 
%\cite{degennes}

\leveltwo{Eilenberger equation} % Quasiclassical equations

The Green's functions being treated here are single-particle 
temperature (or Matsubara) Green's functions written in the 
``Nambu matrix'' representation. They are $4\times 4$ 
matrices in a direct-product space of particle-hole and spin spaces. 
We denote such Nambu matrices with a ``breve'' accent. 
The general stationary temperature Green's function in $\vecb k$-space 
is of the form $\nam G(\vecb k_1,\vecb k_2,\epsilon_m)$.
Here $\epsilon_m=\pi\kB T_c(2m+1)$ are the discrete Matsubara
energies, which are the Fourier variables of the imaginary time
parameter $\tau$. The fact that $(2m+1)$ assumes only odd integer
values is a consequence of the underlying Fermi statistics 
\cite{fetterwalecka}.

The most straightforward way to derive the Eilenberger equations is
to start with the general equation of motion for $\nam G$, the
Dyson equation, and do the ``left-right trick''. This begins by
writing down the left and right
side Dyson equations, which are, symbolically, 
$\nam G^{-1}\nam G=\nam 1$ and $\nam G\nam G^{-1}=\nam 1$.\footnote{The 
products are really so-called folding products, which include integration over
the ``internal vertices'', and a corresponding matrix summation
over the Nambu indices. Also, in the
$\rvec$-representation, $\nam G^{-1}$ is a differential operator, acting
either to the right or to the left \cite{serenerainer}.}
Here $\nam G^{-1}=\nam G_{0}^{-1}-\nam \Sigma$, where $\nam \Sigma$ is
the self-energy and
$\nam G_{0}^{-1}(\vecb k_1,\vecb k_2,\epsilon_m)=$
$[\iu\epsilon_m-\nam\tau_3\xi_{\vecb k_1}]\delta_{\vecb k_1,\vecb k_2}$
is the inverse Green's function for noninteracting fermions, with
$\xi_{\vecb k_1}=\epsilon_{\vecb k_1}-E_{\textrm{F}}$ and 
$\nam\tau_3=\textrm{diag}(1,1,-1,-1)$. Then one proceeds by rewriting 
these equations in the variables 
$\vecb k=(\vecb k_1+\vecb k_2)/2$ and $\vecb q=\vecb k_1-\vecb k_2$,
assuming that $q\ll \kF$. After this one multiplies them with $\nam\tau_3$ from
left and right, respectively, and subtracts them to make terms containing
$\xi_{\vecb k}$ cancel. Finally, one takes the Fermi-surface limit for the 
self-energy, defining
$\nam\sigma(\kvec,\vecb q,\epsilon_m)=a\nam\Sigma(\kF\kvec,\vecb q,\epsilon_m)\nam\tau_3$,
and ``$\xi$ integrates'' the whole equation, so that propagators will
only appear in the quasiclassical form
\begin{equation} \label{e.cg}
\nam g(\kvec,\vecb q,\epsilon_m)=\frac{\nam\tau_3}{a}\int_{-E_c}^{+E_c}
\upd\xi_{\vecb k} \nam
G(\vecb k,\vecb q,\epsilon_m).
\end{equation}
The ``quasiparticle renormalisation factor'' $a$ appearing here may be chosen
arbitrarily. %We choose it to get simple normalisations for $\nam g$. 
The cutoff energy $E_c\ll E_{\textrm{F}}$ is also arbitrary, but the results should not depend on 
how it is chosen: $\nam G$ is nonzero only near the Fermi surface.
A further Fourier transformation from $\vecb q$ to $\rvec$ plus
a small rearrangement of terms puts the Eilenberger equation into the
convenient final form
\begin{equation} \label{e.eilen}
\begin{split}
[i\epsilon_m \nam\tau_3-\nam\sigma,\nam g]+
\iu\hbar \vF \kvec\cdot\nabla_\rvec \nam g & = 0. %\\ 
%\nam g \nam g & = -1.
\end{split}
\end{equation}
This is a first-order differential equation for the Matsubara
propagator $\nam g(\kvec,\rvec,\epsilon_m)$ along classical
trajectories of quasiparticles. Some information concerning the
normalisation of $\nam g$ is lost
in the left-right subtraction process and, therefore, a separate
normalisation condition has to be introduced \cite{eilenberger}. 
With a suitable choice of $a$ this condition,
which is to be satisfied by all physical solutions of (\ref{e.eilen}), can be 
written
\begin{equation}\label{e.normal}
\nam g\nam g=-\nam1.
\end{equation}
To give a closed system of equations (\ref{e.eilen}) and (\ref{e.normal}) 
still need to be supplemented by some self-consistency equations
for the self-energy $\nam \sigma$. These are considered below.

\leveltwo{Impurity scattering}

The quasiclassical theory is perhaps most convenient for solving
static problems which involve nonuiformities, such as walls, interfaces or
impurities. But the above derivation is not really valid then, since for
these kinds of strong scatterers the condition $q\ll \kF$ is not satisfied. 
The proper way to proceed, then, is to utilise the formalism of
scattering $t$ matrices. This generally leads to a mess, and I cannot go very deep
into it here. The general idea is that the self-energy is divided into 
two parts $\nam\Sigma+\nam V$, where $\nam\Sigma$ contains the weak
interactions, like the pairing amplitude, and $\nam V$ is the strong
scattering potential. The perturbation series for $\nam G$ can be
written in the recursive form 
$\nam G=\nam G_0+\nam G_0(\nam\Sigma+\nam V)\nam G$, which is then 
decomposed into three parts \cite{ions,pinning}
\begin{equation}
\begin{align}
\nam G & = \nam G_1+\nam G_1\nam T\nam G_1 \label{e.eq1}\\
\nam T & = \nam V + \nam V\nam G_1\nam T \label{e.eq2} \\
\nam G_1 & = \nam G_0 + \nam G_0\nam\Sigma\nam G_1
~~\textrm{or}~~\nam G_1^{-1}=\nam G_0^{-1}-\nam\Sigma(\nam G).
\end{align}
\end{equation}
Here an unphysical intermediate Green's function $\nam G_1$ has been
introduced for convenience: the corresponding quasiclassical $\nam g_1$ can be solved 
with Eq. (\ref{e.eilen}), with the effect of $\nam V$ coming to
play only through $\nam G$ in a self-consistency equation 
$\nam\Sigma=\nam\Sigma(\nam G)$. If (an approximation for) the
$t$ matrix can be found, then $\nam g$, the quasiclassical counterpart of 
$\nam G$, can be solved iteratively.
The $t$ matrix Eq. (\ref{e.eq2}) is essentially the
Lippmann-Schwinger equation, presented in books 
of standard quantum mechanics \cite{landau}.
The quasiclassical form for this equation will depend on the type of the
scatterer; later on we are concerned with the $t$ matrix of a
specularly scattering wall.
The quasiclassical forms for $\nam T$ and $\nam V$ are obtained
by taking the Fermi-surface limits,
$\nam t(\kvec,\kvec',\epsilon_m)=a\nam T(\kF\kvec,\kF\kvec',\epsilon_m)\nam\tau_3$, 
and
$\nam v(\kvec,\kvec')=a\nam V(\kF\kvec,\kF\kvec')\nam\tau_3$, as
in the case of the self-energy $\nam\sigma$. All propagators are
transformed using Eq. (\ref{e.cg}).

%\subsection{What...}

\leveltwo{Self-consistency equations}

The propagator and the self-energy are usually decomposed into ``scalar''
and ``spin vector'' components, as can be done for an arbitrary Nambu
matrix. We write
\begin{equation}\label{e.gdecomp}
\begin{align}
\nam g & = \left [ \begin{array}{cc}
g+\gvec\cdot\ul\sigmavec & (f+\fvec\cdot\ul\sigmavec)\iu\ul\sigma_2 \\
\iu\ul\sigma_2(\td f + \td \fvec\cdot\ul\sigmavec) & 
\td g - \ul\sigma_2 \td \gvec\cdot\ul\sigmavec\ul\sigma_2
\end{array} \right] \\
%\end{equation}
%\begin{equation}
\nam \sigma & = \left [ \begin{array}{cc} \label{e.edecomp}
\nu+\nuvec\cdot\ul\sigmavec & \Deltavec\cdot\ul\sigmavec\iu\ul\sigma_2 \\
\iu\ul\sigma_2\Deltavec^*\cdot\ul\sigmavec & 
\td\nu - \ul\sigma_2 \td{\nuvec}\cdot\ul\sigmavec\ul\sigma_2
\end{array} \right]=\nam\Delta+\nam\nu,
\end{align}
\end{equation}
where
$\ul\sigmavec=\xvec\ul\sigma_1+\yvec\ul\sigma_2+\zvec\ul\sigma_3$, 
and $\ul\sigma_i, i=1,2,3$ are the Pauli matrices. There should also be
unit matrices $\ul 1$ multiplying the scalar components, but they are
omitted for brevity.
The upper right block in Eq. (\ref{e.edecomp}) now contains the pair amplitude
from Eq. (\ref{e.pairamp}), and this is where the information
about the order parameter comes in. The self-consistency equations
for the off-diagonal self-energy $\nam\Delta$ (the gap vector $\Deltavec$) and the 
diagonal self-energy $\nam\nu$ (the ``Landau molecular fields'' $\nu$, $\td\nu$ and 
$\nuvec$, $\td{\nuvec}$, which are all real valued) are \cite{serenerainer}
\begin{equation}
\begin{align}
\nu(\kvec,\rvec) & = \pi \kB T \sum_m \int \frac{\upd^2\kvec'}{4\pi}
A^s(\kvec\cdot\kvec')g(\kvec',\rvec,\epsilon_m) \label{e.nuse}\\
\nuvec(\kvec,\rvec) & = \pi \kB T \sum_m \int \frac{\upd^2\kvec'}{4\pi}
A^a(\kvec\cdot\kvec')\gvec(\kvec',\rvec,\epsilon_m) \label{e.nuvecse}\\
\Deltavec(\kvec,\rvec) & = \pi \kB T \sum_m \int \frac{\upd^2\kvec'}{4\pi}
V(\kvec\cdot\kvec')\fvec(\kvec',\rvec,\epsilon_m) \label{e.dse}
\end{align}
\end{equation}
with $\td \nu(\kvec)=\nu(-\kvec)$ and $\td{\nuvec}(\kvec)=\nuvec(-\kvec)$.
A scalar order-parameter part $\Delta(\kvec,\rvec)$ is missing because 
of triplet pairing. If we were to consider scattering from impurities, an
additional impurity self-energy $\nam\rho$ would have to be added to 
(\ref{e.edecomp}).
In the absence of mass currents $\nu$ vanishes, and
in the absence of spin currents $\nuvec$ vanishes. For simplicity, we
assume both of them to equal zero, although there are always spin
currents flowing along surfaces in the B phase, which are what we are
interested in \cite{zhang}.
The remaining self-consistency equation for the order parameter
$\Deltavec$ can be put into many different useful forms, such as 
\begin{equation} \label{e.selfconsist}
\Deltavec\ln\frac{T}{T_c}+\pi \kB T \sum_m \left [
\frac{\Deltavec}{|\epsilon_m|}-3\int\frac{\upd^2\kvec'}{4\pi}
\fvec(\kvec',\rvec,\epsilon_m)(\kvec'\cdot\kvec) \right] = \vecb 0.
\end{equation}

\leveltwo{General symmetries}

We denote $2\times 2$ matrices with an underline; for example the 
Pauli matrices in spin space are $\ul{\sigma}_i, i=1,2,3$. Similarly, 
the Pauli matrices in particle-hole space are denoted as
$\ul{\tau}_i$, and in the $4\times4$ Nambu space they take the form
$\nam \tau_i = \ul{\tau}_i\otimes\ul 1, i=1,2,3$, where $\otimes$
denotes a direct product.
With these notations, the propagator and the self-energy satisfy the
basic symmetry relations (and other forms can be easily derived from these)
\begin{equation}
\begin{align}
[\nam u(\kvec,\rvec,\epsilon_m)]^{T*} & = 
\nam\tau_3 \nam u(\kvec,\rvec,-\epsilon_m) \nam \tau_3  \label{e.sym1}\\
[\nam u(\kvec,\rvec,\epsilon_m)]^T & = 
\nam\tau_2 \nam u(-\kvec,\rvec,-\epsilon_m) \nam \tau_2 \label{e.sym2} 
\end{align}
\end{equation}
where $\nam u$ is either $\nam g$ or $\nam\sigma$.
They follow from the operator properties defining the Nambu matrices
$\nam G$ and $\nam\Sigma$, see Refs. \cite{buchholtz,zhang}.
In addition, the Eq. (\ref{e.eilen}) possesses the symmetry
\begin{equation} \label{e.equsymm}
[\nam u(\kvec,\rvec,\epsilon_m)]^T = 
-\nam\tau_2 \nam u(\kvec,\rvec,\epsilon_m) \nam \tau_2.
\end{equation}
More precisely, if one makes the transformations 
${\nam g}^T=\pm\nam\tau_2\nam g'\nam\tau_2$ and
${\nam\sigma}^T=-\nam\tau_2\nam\sigma'\nam\tau_2$ in Eq. (\ref{e.eilen}),
the equation is still satisfied for $\nam g'$ and $\nam\sigma'$. 
If the solutions are to be unique, we should have these satisfied as
symmetries: $\nam g'=\nam g$ and $\nam\sigma'=\nam\sigma$.
%This is a symmetry satisfied by \emph{physical} solutions.
%Replacing the $\nam g$-transformation by 
%$\nam g^T=+\nam\tau_2\nam g\nam\tau_2$
%also works, and that is a symmetry of \emph{unphysical} or diverging solutions.
The minus sign corresponds to \emph{physical} solutions and the plus
sign to \emph{unphysical} solutions, which also satisfy 
$\nam g\nam g=0$ instead of Eq. (\ref{e.normal}).
The off-diagonal self-energy (order parameter) is not affected by this 
requirement, but it generally causes some restrictions on $\nam\sigma$.
Note, for example, that any self-energy term proportional to $\nam 1$
does not satisfy the symmetry. This is not a real restriction either,
since such terms would cancel in the commutator of Eq. (\ref{e.eilen}) anyway.
If we redefine the propagator components by writing Eq. (\ref{e.gdecomp}) as
\begin{equation} \label{e.redef}
\nam g = \left [ \begin{array}{cc}
c+d+(\cvec+\dvec)\cdot\ul\sigmavec & 
(a+b+(\avec+\bvec)\cdot\ul\sigmavec)\iu\ul\sigma_2 \\
\iu\ul\sigma_2(a-b + (\avec-\bvec)\cdot\ul\sigmavec) & 
c-d -  \ul\sigma_2 (\cvec-\dvec)\cdot\ul\sigmavec\ul\sigma_2,
\end{array} \right]
\end{equation}
then the basic symmetries in Eqs. (\ref{e.sym1}) and (\ref{e.sym2}) can be written
componentwise:
\begin{equation} \label{e.basicsymm}
\begin{align}
a(-\kvec) & = +a(\kvec)^* & a(-\epsilon_m)&=+a(\epsilon_m)^* 
	\nonumber \\
b(-\kvec) & = -b(\kvec)^* & b(-\epsilon_m)&=-b(\epsilon_m)^* 
	\nonumber \\
c(-\kvec) & = +c(\kvec)^* & c(-\epsilon_m)&=+c(\epsilon_m)^* 
	\nonumber \\
d(-\kvec) & = -d(\kvec)^* & d(-\epsilon_m)&=+d(\epsilon_m)^* 
	\nonumber \\
\avec(-\kvec) & = -\avec(\kvec)^*   &
	\avec(-\epsilon_m)&=+\avec(\epsilon_m)^* \nonumber \\
\bvec(-\kvec) & = +\bvec(\kvec)^*  &
	\bvec(-\epsilon_m)&=-\bvec(\epsilon_m)^* \nonumber \\
\cvec(-\kvec) & = +\cvec(\kvec)^*  &
	\cvec(-\epsilon_m)&=+\cvec(\epsilon_m)^* \nonumber \\
\dvec(-\kvec) & = -\dvec(\kvec)^* & 
	\dvec(-\epsilon_m)&=+\dvec(\epsilon_m)^* .
\end{align}
\end{equation}
Further symmetries follow from the geometry of a specific problem 
through the order parameter and the Eilenberger equation.

\leveltwo{Physical and unphysical solutions --- the
multiplication trick}

As long as the symmetry in Eq. (\ref{e.equsymm}) is satisfied by the
self-energy, the
redefinition in Eq. (\ref{e.redef}) leads to a significant
simplification in the calculations.
This is because Eq. (\ref{e.eilen}) can then be decomposed into 
three independent blocks of equations
\begin{equation}
\begin{align}
\partial_u c & = 0 \\
\iu \me b+\iu\Deltavecim\cdot\cvec+
	\frac{\iu}{2}\hbar\vF\partial_ua&=0 \label{e.unphys1} \\
\iu \me a+\Deltavecre\cdot\cvec+
	\frac{\iu}{2}\hbar\vF\partial_ub&=0 \label{e.unphys2} \\
-\Deltavecre b + \iu \Deltavecim\cdot\cvec+
	\frac{\iu}{2}\hbar\vF\partial_u\cvec&=\boldsymbol{0} \label{e.unphys3} \\
\iu\me\bvec+\iu\Deltavecim d-\iu\Deltavecre\times\dvec+
	\frac{\iu}{2}\hbar\vF\partial_u\avec&=\boldsymbol{0} \label{e.phys1}\\
\iu\me\avec+\Deltavecre d+\Deltavecim\times\dvec+
	\frac{\iu}{2}\hbar\vF\partial_u\bvec&=\boldsymbol{0} \label{e.phys2}\\
-\Deltavecre\cdot\bvec+\iu\Deltavecim\cdot\avec+
	\frac{\iu}{2}\hbar\vF\partial_ud&=0 \label{e.phys3}\\
\iu\Deltavecre\times\avec+\Deltavecim\times\bvec+
	\frac{\iu}{2}\hbar\vF\partial_u\dvec&=\boldsymbol{0}, \label{e.phys4}
\end{align}
\end{equation}
where $\Deltavec=\Deltavecre+\iu\Deltavecim$ and $u$ parametrises
a quasiparticle trajectory: $\rvec=\rvec_0+u\kvec$ . The first equation 
forms its own block, and we may always set $c=0$. The
Eqs. (\ref{e.phys1})-(\ref{e.phys4}) form the \emph{physical} 
block whose solutions include the physical solution, which only has 
$\avec$, $\bvec$, $d$ and $\dvec$ nonzero. 
The normalisation $\nam g\nam g=-\nam 1$ for physical solutions
takes the component form
\begin{equation}
\begin{split}
-\iu d \dvec + \avec\times\bvec & = \vecb 0 \\
d^2+\dvec\cdot\dvec-\avec\cdot\avec+\bvec\cdot\bvec & = -1.
\end{split}
\end{equation}
In the bulk (constant $\Deltavec$ and $\nam g$), the normalised physical solutions are
\begin{equation}\label{e.bulkphys}
d=\frac{-\iu\me}{\sqrt{\me^2+|\Deltavec|^2}}, \quad
\avec=\frac{\Deltavecre}{\sqrt{\me^2+|\Deltavec|^2}}, \quad
\bvec=\frac{\iu\Deltavecim}{\sqrt{\me^2+|\Deltavec|^2}}, \quad 
\dvec=\vecb 0.
\end{equation}
To be exact, these forms are valid only for so-called 
\emph{unitary} states, for which $\nam\Delta\nam\Delta^+\propto\nam 1$,
or $\Deltavec\times\Deltavec^*=\vecb 0$. Fortunately, the unitarity requirement is
automatically satisfied for the B phase, as well as the A phase of $^3$He
\cite{vollhardt,serenerainer}. 
A more compact form for Eq. (\ref{e.bulkphys}) can be seen directly from
Eq. (\ref{e.eilen}) by setting the gradient term to zero and solving for
$\nam g$: 
%and by using
%$\nam\Delta\nam\Delta=-|\Deltavec|^2\nam 1$ with $\{\nam\Delta,\nam\tau_3\}=0$
%to satisfy the normalisation:
\begin{equation} \label{e.compact}
\nam g=\frac{\iu\me\nam\tau_3-\nam\Delta}{\sqrt{\me^2+|\Deltavec|^2}}.
\end{equation}
To satisfy $\nam g\nam g=-\nam 1$ for this, one must require 
$\nam\Delta\nam\Delta=-|\Deltavec|^2\nam 1$ and note that
$\{\nam\Delta,\nam\tau_3\}=0$.

However, the \emph{unphysical} block of Eqs. (\ref{e.unphys1})-(\ref{e.unphys3}) is
even more useful in practise than the physical one. This is because
the physical solutions of Eqs. (\ref{e.phys1})-(\ref{e.phys4}) can 
be obtained numerically more easily by finding the exploding and decaying
unphysical solutions $a$, $b$ and $\cvec$ and then using the so-called
\emph{multiplication trick}. %This is done as follows.
%Presence of a magnetic field would destroy the symmetry
%(\ref{e.equsymm}) and thus complicate calculations by removing this 
%possibility.

If we assume the order parameter to be real, $\Deltavec=\Deltavecre$,
we may choose 
all of the propagator components to be real, except for $\cvec$ and
$d$ which must then be purely imaginary. If we also
assume $\Deltavec$ to be constant, it is easy to see that
the unphysical block has the exploding ($<$, upper signs) and decaying 
($>$, lower signs) solutions 
\begin{equation} \label{e.exponential}
\left[ \begin{array}{c}
a^{(0)} \\ b^{(0)} \\ \cvec^{(0)}
\end{array}\right]_\gtrless= c_1
\left[ \begin{array}{c}
\me \\ \mp\sqrt{\me^2+|\Deltavec|^2} \\ \iu \Deltavec
\end{array}\right]
\exp{\left(\frac{\pm2\sqrt{\me^2+|\Deltavec|^2} u}{\hbar\vF} \right)},
\end{equation}
where $c_1$ is an undetermined constant.
(The unphysical block has also three constant solutions, but they
are of no interest here.) Similar ``exponential'' solutions exist for a general $\Deltavec$ as
well, although they no longer have this simple form. 
Such solutions always satisfy the normalisation condition 
$\nam g\nam g = 0$, or in component form $-a^2+b^2+\cvec\cdot\cvec=0$.
The multiplication trick now consists of the following property:
the physical solution along a given trajectory can be obtained
by taking the commutator of the exploding and decaying solutions on
that trajectory
\begin{equation} \label{e.getphysical}
\nam g(u)=\frac{\iu \nc}{2}[\nam g_<(u),\nam g_>(u)].
\end{equation}
Here the normalisation is given by
\begin{equation} \label{e.norm}
\begin{split}
\nc^{-1} & =\frac{1}{2}\{\nam g_<(u_0),\nam g_>(u_0)\} \\
	& = - a_<a_>+b_<b_>+\cvec_<\cdot\cvec_>,
\end{split}
\end{equation}
which is invariant along the trajectory, i.e. $u_0$ may be chosen freely.
The commutator gets conveniently rid of all terms proportional to the
unit matrix and the normalisation ensures that only the relative
proportions of the components of the unphysical solutions make a 
difference. In component form, the commutators become
\begin{equation} \label{e.multiply}
\begin{split}
\avec & =\iu \nc [\cvec_<b_>-b_<\cvec_>] \\
\bvec & =\iu \nc [\cvec_<a_>-a_<\cvec_>] \\
d & =\iu \nc [a_<b_>-b_<a_>] \\
\dvec & =-\nc\hspace{1mm}\cvec_<\times\cvec_>.
\end{split}
\end{equation}
One can readily check that this procedure works at least for 
Eqs. (\ref{e.bulkphys}) and (\ref{e.exponential}). For a more detailed 
justification of this trick see, for example, Ref.  \cite{kurkijarvi}.

\newpage

\levelone{The pinhole model}

\begin{figure}[!bt]
\begin{center}
%\psfrag{l}[][l]{$\Delta$}
\includegraphics[width=0.6\linewidth]{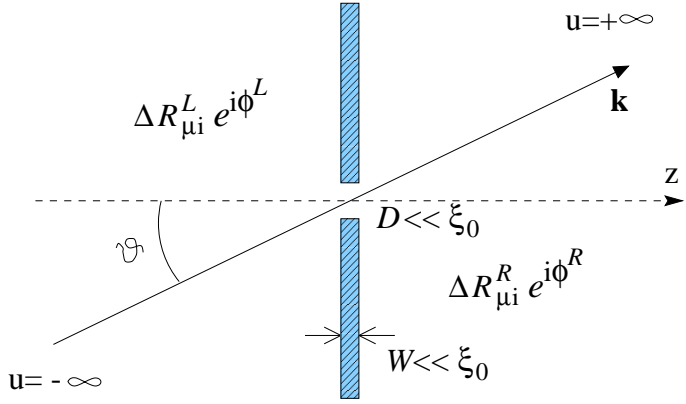}
\caption{
Quasiparticle trajectory through a pinhole aperture.}
\label{f.traj}
\end{center}
\end{figure}

In the pinhole model
one assumes a wall separating two volumes of $^3$He-B, whose 
thickness $W$ is much less than the coherence
length $\xi_0$. In this wall, the junction is formed by a hole, whose
diameter $D$ is also $\ll\xi_0$. This situation is depicted in Fig. \ref{f.traj}.
%\psfragscanon
%\psfragscanoff
The experimental holes of Ref. \cite{bistability} do not exactly count
as pinholes, but are not too far from them. The reason for considering
a pinhole here is in the relative simplicity of the resulting calculations. The
junction can be treated as a small perturbation whose effects on the
order parameter can be seen in the associated energies only
in the order $O([\textrm{open area}]^2)=O(D^4)$. Such effects can be
neglected to a first approximation. A pinhole 
was first considered as a model for superconducting microbridges
\cite{kulik}, and previous calculations for $^3$He also exist
\cite{kop86,kurkijarvi,tks}.\footnote{Recently, pinhole calculations for
$d$ wave superconductors have also been published 
\cite{yip,fogelstrom}.}
Here we generalise the previous calulations to find the current-phase
relations for both parallel and antiparallel $\nvec$ vectors at the
wall. A clear presentation on the critical currents of the pinhole model
has also been lacking for a long time, and this situation is corrected 
here. Finally, in the next section, we generalise the ``tunneling model''
by performing the corresponding calculations directly from the pinhole 
model for all temperatures.

\leveltwo{Symmetries of the problem}

The presence of a wall imposes symmetry restrictions on the order
parameter in the orbital space. If the spin and orbital axes are first
taken to be nonrotated relative to each other, the gap vector
${\Deltavec}(\kvec,\rvec)$ has the form ${\Deltavec}^{(0)}$
($\Delta_\perp$ and $\Delta_\parallel$ are real)
\begin{equation} \label{e.gap} 
\Deltavec^{(0)}(\kvec,z)=
\Delta_{\perp}(z)\zvec\zvec\cdot\kvec+
\Delta_{\parallel}(z)(\rhovec\rhovec+\chivec\chivec)\cdot\kvec,
\end{equation}
where $\{\zvec,\rhovec,\chivec\}$ is an orthonormal triad,
same for the spin and orbital spaces and with $\zvec$ 
perpendicular to the wall. This form is invariant under rotations of the
triad around $\zvec$, and in the bulk it should have the simple limit
$\Deltavec^{(0)}(\kvec)=\Delta\kvec$, where 
$\Delta=\Delta_{\perp}=\Delta_{\parallel}=\textrm{const.}$ is the
temperature-dependent B-phase bulk gap.
The general form of $\Deltavec$
is then obtained from this by taking into account general spin-orbit
rotations and overall phases. We assume these to be different on the
left ($L$) an right ($R$) sides of the thin wall, and write
\begin{equation} \label{e.genedelta}
\Deltavec(\kvec,z)=\left\{\begin{array}{ll}
\exp(\iu \phi^L)\dyadic{R}^L\!\cdot \Deltavec^{(0)}(\kvec,z) & \textrm{for}~  z < 0 \\
\exp(\iu \phi^R)\dyadic{R}^R\!\cdot \Deltavec^{(0)}(\kvec,z) & \textrm{for}~  z > 0 \\
\end{array} \right. .
\end{equation}
Note that all of these configurations are degenerate in energy only assuming
that spin-orbit coupling is negligible, but we \emph{do} assume that.
The actual $z$ dependencies of the order-parameter components 
$\Delta_\parallel(z)$ and $\Delta_\perp(z)$ have to be
calculated self-consistently, and they depend on the type of the wall.
We return to this shortly.

Assume now that a small pinhole is made to the wall, which
provides coupling between $L$ and $R$ by letting quasiparticles 
travel from one side to another. If the origin of the coordinates is
placed in the center of the hole, then any quasiparticle trajectory though the hole 
can be parametrised as $\rvec=u\kvec$, where $\kvec$ is the direction
of $\vecb v_{\textrm{F}}$. From the symmetry of the problem it follows that
\begin{equation}
\Deltavec^{(0)}(u)=\Deltavec^{(0)}(-u),
\end{equation}
and due to the symmetry of the unphysical block of Eqs. 
(\ref{e.unphys1})-(\ref{e.unphys3}) this implies
\begin{equation} \label{e.extrasymm}
\begin{split}
a_>^{(0)}(u,\me) & =+a_<^{(0)}(-u,\me) \\
b_>^{(0)}(u,\me) & =-b_<^{(0)}(-u,\me) \\
\cvec_>^{(0)}(u,\me) & =+\cvec_<^{(0)}(-u,\me).
\end{split}
\end{equation}
These can be used to find the decaying
solutions on one side from the diverging ones on the other, which 
is especially useful for $u=0$. Here (and henceforth) we denote by 
'$(0)$' the propagator solutions corresponding to $\Deltavec^{(0)}$, and
Eqs. (\ref{e.extrasymm}) \emph{do not} hold for the general order parameter.
But, in fact, we only need to solve the propagator in the simpler case.
Writing $\cvec=\iu\cvecim$, Eqs.
(\ref{e.unphys1})-(\ref{e.unphys3}) become (now for an arbitrary
$\rvec=\rvec_0+u\kvec$)
\begin{equation} \label{e.simple}
\begin{align}
\me b+\frac{1}{2}\hbar\vF\partial_ua&=0 \\
\me a+\Deltavec^{(0)}\cdot\cvecim+
	\frac{1}{2}\hbar\vF\partial_ub&=0 \\
\Deltavec^{(0)} b+
	\frac{1}{2}\hbar\vF\partial_u\cvecim&=\boldsymbol{0}. \label{e.cequ}
\end{align}
\end{equation}
From their solutions $a_\gtrless^{(0)}$, $b_\gtrless^{(0)}$
and $\cvec_\gtrless^{(0)}$, we get the solutions of Eqs.
(\ref{e.unphys1})-(\ref{e.unphys3}) for general $\Deltavec$, Eq.
(\ref{e.genedelta}), by forming the following linear combinations (on
either $L$ or $R$) %for $L$ and $R$ separately:
%\begin{equation}
%\begin{split}
%a_\gtrless^{L,R} & =
%a_\gtrless^{(0)}\cos\phi^{L,R}+\iu b_\gtrless^{(0)} \sin\phi^{L,R} \\
%a_\gtrless^{L,R} & =
%\iu a_\gtrless^{(0)}\sin\phi^{L,R}+b_\gtrless^{(0)} \cos\phi^{L,R} \\
%\cvec_\gtrless^{L,R} & = 
%\hspace{1mm}\dyadic{R}^{L,R}\!\!\cdot\cvec_\gtrless^{(0)}.
%\end{split}
%\end{equation}
\begin{equation} \label{e.generals}
\begin{split}
a_\gtrless^{} & =
a_\gtrless^{(0)}\cos\phi^{}+\iu b_\gtrless^{(0)} \sin\phi^{} \\
b_\gtrless^{} & =
\iu a_\gtrless^{(0)}\sin\phi^{}+b_\gtrless^{(0)} \cos\phi^{} \\
\cvec_\gtrless^{} & = 
\hspace{1mm}\dyadic{R}^{}\!\!\cdot\cvec_\gtrless^{(0)}.
\end{split}
\end{equation}
From Eq. (\ref{e.cequ}), we observe that $\cvec^{(0)}\parallel\Deltavec^{(0)}$, and
from Eqs. (\ref{e.genedelta}) and (\ref{e.generals}) we find that also $\cvec\parallel\Deltavec$, 
since both are obtained by the same rotation. However, note carefully that 
$\kvec\nparallel\Deltavec$, except in the bulk. If we write
 $\kvec=\cos\vartheta\zvec+\sin\vartheta\rhovec$,
we instead have the relations 
$\Deltavec^{(0)}=\Delta_\perp\cos\vartheta\zvec+\Delta_\parallel\sin\vartheta\rhovec$ 
and
$\cvec^{(0)}=c_z^{(0)}\zvec+c_\rho^{(0)}\rhovec$, i.e., the vectors 
$\kvec$, $\Delta^{(0)}$ and $\cvec^{(0)}$ are coplanar with
$\{\zvec,\rhovec\}$.
Furthermore, the norm of $\cvec$ is always restricted by 
$\cvec^2=a^2-b^2$, so that one of the unknowns ($a^{(0)}$, $b^{(0)}$, 
$\im~ c^{(0)}_z$ and $\im~ c^{(0)}_\rho$) can always be found in terms 
of the others, as long as one can be sure about the sign.
Eqs. (\ref{e.simple}) are thus effectively only a set of three linear
differential equations for three real functions ($a$, $b$ and 
$\im c_z$, say) %and $\im c_\rho$) 
and the solution must be integrated only for
all polar angles $\vartheta$ of $\kvec$ in one plane.\footnote{In % footnote
practise, it is much simpler to solve for all of the unknowns
separately and use the normalisation just to check the accuracy.
Note also that if (and only if) $\Delta_\perp=\Delta_\parallel$ we have
$c_z=c\cos\vartheta$, $c_\rho=c\sin\vartheta$, 
(i.e. $\kvec\parallel\Deltavec\parallel\cvec$)
such that one only needs to solve $a$, $b$ and $c$, which is trivial
and has been done analytically in Eq. (\ref{e.exponential}).}
For a $\kvec$ rotated from this plane around $\zvec$, the
components of $\cvec^{(0)}$ in the
fixed basis $\{\zvec,\rhovec,\chivec\}$ are obtained by the same rotation:
$c^{(0)}_i(\dyadic{R}\!(\zvec,\chi)\cdot\kvec)=R_{ij}(\zvec,\chi)c^{(0)}_j(\kvec)$.
The scalar components $a$ and $b$ are invariant under such rotations.
This, as well as Eq. (\ref{e.extrasymm}), is an additional symmetry which
followed from a specific form of the \emph{order parameter}, through
the Eilenberger equation; again, it does not hold for $\cvec$ in Eq.
(\ref{e.generals}), since rotations do not usually commute.
Similar rules apply to the physical vector and scalar components.
They are useful for doing some angular integrations over the directions
$\kvec$.

This could be stated in a different way. 
Under rotations of $\kvec$ around $\zvec$, the order parameter transforms
like $\Deltavec^{(0)}(\kvec')=\dyadic{R}^l_z\!\!{}\cdot\Deltavec^{(0)}(\kvec)$, if 
$\kvec'=\dyadic{R}^l_z\!\!{}\cdot\kvec$. In other words, the $\kvec$ rotation
around $\zvec$ is equivalent to a spin rotation.
Other kinds of $\kvec$ rotations do not have this symmetry. 
But now, \emph{all} spin rotations of $\Deltavec^{(0)}$ resulted in the same
rotation of $\cvec^{(0)}$ according to Eq. (\ref{e.generals}). The upper
left $2\times2$ spin block
$\cvec^{(0)}(\kvec)\cdot\ul\sigmavec$ of the unphysical \emph{propagator}
thus tranforms as follows:
\begin{equation} \label{e.transforms}
\cvec^{(0)}(\kvec)\cdot\ul\sigmavec~=~
\cvec^{(0)}(\dyadic{R}^l_z\!\!\cdot\kvec')\cdot\dyadic{R}^s\!\!\cdot\ul\sigmavec'~=~
\dyadic{R}^s\!\!{}^{-1}\!\cdot\dyadic{R}^l_z\!\!\cdot
\cvec^{(0)}(\kvec')\cdot\ul\sigmavec'~=~
\cvec^{(0)}(\kvec')\cdot\ul\sigmavec',
\end{equation}
%\begin{equation}
%\cvec^{(0)}(\kvec)\cdot\ul\sigmavec=
%\begin{cases}
%\dyadic{R}^s\!\!\cdot\cvec^{(0)}(\kvec)\!\cdot\ul\sigmavec', & 
%	\ul\sigmavec'=\dyadic{R}^s\cdot\ul\sigmavec  \\
%\dyadic{R}^l_z\!\!\cdot\cvec^{(0)}(\kvec')\!\cdot\ul\sigmavec, &
%	\kvec'=\dyadic{R}^l_z\!\!{}^{-1}\cdot\kvec.
%\end{cases}
%\end{equation}
where the final equality follows only if $\dyadic{R}^l_z=\dyadic{R}^s$.
Thus, only a spin rotation $\dyadic{R}^s_z$ around $\zvec$ can be ``undone''
with a simultaneous $\kvec$ rotation $\dyadic{R}^l_z=\dyadic{R}^s_z$.

\leveltwo{Propagator at the discontinuity}

At the pinhole, the order parameter ``jumps'' from one value to another 
over a vanishingly small distance. As long as there is only a
discontinuity of this kind and no delta-function potentials, the
physical propagator should nevertheless be \emph{continuous} along a 
trajectory crossing the pinhole, since the Eilenberger equation is of first order.
Let us consider trajectories $\rvec=u\kvec$, such that $u=0$ is at the
center of the pinhole. We further restrict to cases where $\kvec$ is
directed from \emph{left} to \emph{right}; for the opposite direction 
the roles of $L$ and $R$ should be interchanged.
Since an overall phase cannot affect any physical properties of a quantum
system,
we can also safely restrict to the symmetric case
$\phi^L=-\phi/2$, $\phi^R=+\phi/2$ from the start;
all results should be independent of this choice and only depend on
$\phi^R-\phi^L=\phi$.

We define new uppercase symbols
$A(\kvec)=a^{(0)}_<(\kvec,u=0)$, $B(\kvec)=b^{(0)}_<(\kvec,u=0)$ and
$\Cvec(\kvec)=\cvec^{(0)}_<(\kvec,u=0)$. All the physical and unphysical 
components for a general $\Deltavec$ can be expressed in terms
of these. Inserting the general diverging solutions
(\ref{e.generals}) to the multiplication formulas (\ref{e.multiply})
such that the diverging solutions ($<$) are taken from the left and decaying
ones ($>$) from the right, and applying the extra
symmetries in Eq. (\ref{e.extrasymm}) leads to 
%\begin{equation}
%\begin{split}
%\avec(0) & =\iu\td c[\dyadic{R}^L\!\!(\iu A\sin\phi_R-B\cos\phi_R)
%-(\iu A\sin\phi_L+B\cos\phi_L)\dyadic{R}^R ] \cdot\Cvec \\
%\bvec(0) & =\iu\td c[\dyadic{R}^L\!\!(A\cos\phi_R-\iu B\sin\phi_R)
%-(A\cos\phi_L+\iu B\sin\phi_L)\dyadic{R}^R ] \cdot\Cvec \\
%d(0) &=\iu \td c[\iu (A^2+B^2)\sin\phi-2AB\cos\phi] \\
%\dvec(0) &=-\td c (\dyadic{R}^L\!\!\!\cdot~\Cvec)\times
%(\dyadic{R}^R\!\!\!\cdot~\Cvec)
%\end{split}
%\end{equation}
\begin{equation}
\begin{align}
\avec(\kvec,0) & =\iu~\nc~(\Cvec^L+\Cvec^R)~
(\iu A\sin\frac{1}{2}\phi-B\cos\frac{1}{2}\phi) \label{e.physprop1} \\
\bvec(\kvec,0) & =\iu~\nc~(\Cvec^L-\Cvec^R)~
(A\cos\frac{1}{2}\phi-\iu B\sin\frac{1}{2}\phi) \label{e.physprop2} \\
d(\kvec,0) &=\iu~\nc~[\iu (A^2+B^2)\sin\phi-2AB\cos\phi] \label{e.physprop3}\\[1.1mm]
\dvec(\kvec,0) &=-\nc ~\Cvec^L\times\Cvec^R \label{e.physprop4}
\end{align}
\end{equation}
%\begin{equation} \label{e.physprop}
%\begin{split}
%\avec(0) & =\iu\nc~(\dyadic{R}^L\!\!+\dyadic{R}^R)\cdot\Cvec~
%(\iu A\sin\frac{1}{2}\phi-B\cos\frac{1}{2}\phi)  \\
%\bvec(0) & =\iu\nc~(\dyadic{R}^L\!\!-\dyadic{R}^R)\cdot\Cvec~
%(A\cos\frac{1}{2}\phi-\iu B\sin\frac{1}{2}\phi)  \\
%d(0) &=\iu \nc~[\iu (A^2+B^2)\sin\phi-2AB\cos\phi] \\[1mm]
%\dvec(0) &=-\nc ~(\dyadic{R}^L\!\!\!\cdot~\Cvec)\times
%(\dyadic{R}^R\!\!\!\cdot~\Cvec)
%\end{split}
%\end{equation}
where we also defined $\Cvec^{L,R}=\dyadic{R}^{L,R}\!\cdot\Cvec$.
The normalisation constant $\nc$ satisfies $\nc(-\kvec)=\nc^*(\kvec)$ and is given by
\begin{equation}
\nc(\kvec,0) =
[-(A^2+B^2)\cos\phi+2\iu AB\sin\phi+\Cvec^L\!\cdot\Cvec^R]^{-1}.
%\Cvec\cdot(\dyadic{R}^L)^T\!\cdot\dyadic{R}^R\!\!\cdot\Cvec]^{-1}
\end{equation}
Note that for $\nvec$ vectors satisfying $\nvec^L=\pm\nvec^R=\pm\zvec$ these
expressions still simplify considerably. 
To check that the $\kvec$ inversion symmetries in Eq. (\ref{e.basicsymm}) 
hold for Eqs. (\ref{e.physprop1})-(\ref{e.physprop4}), one must notice that the sign of $\phi$
must be reversed on interchanging $L$ and $R$, i.e., when
$\kvec\leftrightarrow-\kvec$.
These propagators can now be used to calculate physical quantities at
the junction. Most of all, we need the mass current.

%\begin{equation}
%d=\frac{(a^2+b^2)\sin\phi+2\iu a b\cos\phi}
%{(a^2+b^2)\cos\phi-2\iu a b\sin\phi-(a^2-b^2)\cos\theta}
%\end{equation}

\leveltwo{Mass current for the pinhole}

The general quasiclassical equation for mass-current density is \cite{serenerainer}
\begin{equation} \label{e.genecurrent}
\vecb{j}(\rvec)=2m_3 \vF \NF\pi \kB T \sum_m \int
\frac{\upd^2\kvec}{4\pi}\kvec g(\kvec,\rvec,\epsilon_m).
\end{equation}
This can be written in terms of $\re~ d(\kvec)$
alone, and for one pinhole with open area $A_o$ the current %($\hat k_z=\cos\vartheta$)
$J=A_o(\zvec\cdot\vecb j)$ becomes
\begin{equation}  \label{e.progcurrent}
J=A_o2m_3\vF\NF\pi\kB T \sum_{m}\int_0^1 \upd(\cos\vartheta)
(\cos\vartheta) p(\vartheta)\int_0^{2\pi}\frac{\upd\chi}{2\pi} %\sum_m.
~\re~ d(\vartheta,\chi,\phi).
%\frac{(b^2-a^2)}{a^2+b^2}
\end{equation}
Here we add an extra $p(\vartheta)$ to describe the possibly different
transmission probabilities for different trajectories; unless
otherwise stated, this should be set to unity in all the formulas where
it appears. Several forms for $d(\kvec)$ can be derived.
With a little trigonometric trickery, one finds
\begin{equation} \label{e.sophisticated}
d(\vartheta,\chi,\phi)=\frac{1}{4}
%\left[ 
\sum_{\delta=\pm1}
\frac{(B^2-A^2)\sin(\phi+\delta\theta)+2\iu A B}
{A^2\sin^2[\frac{1}{2}(\phi+\delta\theta)]+B^2\cos^2[\frac{1}{2}(\phi+\delta\theta)]},
%+\frac{(b^2-a^2)\sin(\phi-\theta)+2\iu a b}
%{a^2\sin^2[\frac{1}{2}(\phi-\theta)]+b^2\cos^2[\frac{1}{2}(\phi-\theta)]}
%\right] \nonumber
\end{equation}
where $\theta=\theta(\vartheta,\chi)$ is defined by 
$\Cvec^L\cdot\Cvec^R=\cos\theta\Cvec^2$. This is interesting for
analytical considerations, but not good for numerics since it is unnecessarily
complicated to calculate the $\theta$ angle in general.
A less sophisticated but more easily programmable form is
\begin{equation} \label{e.unsophisticated}
\re~ d(\vartheta,\chi,\phi)=-\frac{\Cvec^L\cdot\Cvec^R(A^2+B^2)\sin\phi
-\frac{1}{2}(A^2-B^2)^2\sin2\phi}{
[(A^2+B^2)\cos\phi-\Cvec^L\cdot\Cvec^R]^2+4A^2B^2\sin^2\phi}.
\end{equation}
For parallel and perpendicular $\nvec$'s we have
$\Cvec^L\cdot\Cvec^R=\Cvec^2=A^2-B^2$ and
for antiparallel
$\Cvec^L\cdot\Cvec^R=\Cvec^2-\frac{15}{8}\Cvec_\rho^2$,
which are the two cases of special interest. In the general case, 
the following is probably the simplest form one can get:
\begin{equation} \label{e.horrendous}
\begin{split}
\Cvec^L\cdot\Cvec^R &=(1+15(\nvec^L\cdot\nvec^R))(A^2-B^2)
-5[(\Cvec\cdot\nvec^L)^2+(\Cvec\cdot\nvec^R)^2] \\
&+(25(\nvec^L\cdot\nvec^R)-15)(\nvec^L\cdot\Cvec)(\nvec^R\cdot\Cvec) \\
&+5\sqrt{15}[(\nvec^L\cdot\Cvec)-(\nvec^R\cdot\Cvec)]
(\nvec^L\times\nvec^R)\cdot\Cvec.
\end{split}
\end{equation}
Note that the numerator is just half the $\phi$ derivative of the
denominator:
\begin{equation}
\re~ d(\kvec,\phi)=-\frac{1}{2}\frac{\partial}{\partial \phi}\ln|\nc(\kvec,\phi)|^2.
\end{equation}
It would seem that, apart from a constant, an analytical expression for
the \emph{energy} of the junction could be obtained by a simple integration 
over $\phi$. But unless we have equal rotation matrices on the
two sides of the junction, there should in general also be \emph{spin
currents} present, which could contribute to the energy in some unknown
way \cite{serenerainer}. For now we do not need the energy and will
return to the problem below. 
%[Add definition of spin current?]

\levelthree{Constant order parameter case}

%\begin{equation}
%J=\frac{1}{2}m_3\vF \NF \pi \Delta(T)\sin(\phi/2)
%\tanh\left(\frac{\Delta(T)\cos(\phi/2)}{2\kB T} \right)
%\end{equation}
If we assume that the order parameter is constant all the way to the 
wall, the current can be given a neat
analytical expression even in the general case.
Inserting Eqs. (\ref{e.exponential}) with $u=0$ into
Eqs. (\ref{e.sophisticated}) and (\ref{e.progcurrent}) and using
a tabulated formula for doing the Matsubara summation, one finds
\begin{equation} \label{e.siisti}
J=A_o\frac{1}{4}m_3\vF\NF\pi\Delta\sum_{\delta=\pm1}
\int\frac{\upd^2\kvec}{4\pi}\sin((\phi+\delta\theta_{\kvec})/2)
\tanh\left(\frac{\Delta\cos((\phi+\delta\theta_{\kvec})/2)}{2\kB T} \right),
\end{equation}
where $\Delta=|\Deltavec|$.
Setting $\theta_{\kvec}=0$ gives a formula which is essentially that
derived by Kulik and Omel'yanchuk for
superconducting microbridges \cite{kulik}. The more general form is 
exactly the result obtained by Yip \cite{yippi} as an explanation 
for the $\pi$ state. Unfortunately, while the assumption of a constant 
order parameter is valid for superconductors, it is poor for 
$^3$He, as we shall see.

% ************************************************
\levelthree{Parameters for the tunneling model}

We are finally ready to present how the $\alpha$ and 
$\beta$ parameters of the tunneling model were obtained.
In the Ginzburg-Landau region, the amplitude of 
$\Deltavec$ should be small. But since 
$|\Deltavec|^2\sim|\Cvec|^2=|A^2-B^2|$, we should have 
$|A^2-B^2|\ll A^2+B^2$. In this limit, we can integrate the current 
in Eq. (\ref{e.progcurrent}) with respect to $\phi$ into the form of
Eq. (\ref{e.fj}), where $\alpha$ and $\beta$ are proportional to
\begin{equation} 
\begin{align}
\td\alpha & =\hbar\vF\NF\pi\kB T\int_0^1 \upd(\cos\vartheta)(\cos\vartheta)
p(\vartheta)\sum_m\frac{(\textrm{Im}~ C_z)^2}{A^2+B^2}, \label{e.alpha}\\
\td\beta & =\frac{1}{2}\hbar\vF\NF\pi\kB T\int_0^1 \upd(\cos\vartheta)(\cos\vartheta)
p(\vartheta)\sum_m\frac{(\textrm{Im}~ C_{\rho})^2}{A^2+B^2}. \label{e.beta}
\end{align}
\end{equation}
%and $S$ is the total area of the array.
Here we hace taken advantage of the transmission probability $p(\vartheta)$ to
calculate the parameters for a pinhole which has the same aspect
ratio as the experimental apertures. If we assume that any trajectory
hitting the wall inside the aperture gets scattered diffusely, i.e., into
a random direction, it does not contribute to the current. Then, for a 
circular aperture of diameter $D$ and wall thickness $W$, we find
\begin{equation} \label{e.probability}
p(\vartheta)=\left\{\begin{array}{ll}
%\frac{D^2}{2 a^2}
\frac{2}{\pi}
(\gamma-\cos\gamma\sin\gamma) & \textrm{for} 
	\quad  \vartheta<\arctan(D/W)\\
0 & \textrm{for} \quad  \vartheta>\arctan(D/W),\\
\end{array} \right.   
\end{equation}
where $\gamma=\arccos(W/D)\tan\vartheta$. This gives the probability
that a quasiparticle gets transmitted \emph{given that} it hit the ``open 
area'' of the hole. Thus, $\td\alpha$ and $\td\beta$ here are the
original coupling constants $\alpha$ and $\beta$ \emph{per open area}
$S_o$ of a junction: $\alpha=S_o\td\alpha$, $\beta=S_o\td\beta$. Here
$S_o=d_oS$, where $S$ is the
total area of the array and $d_o=\pi D^2/4a^2$ is the fraction of open
area per total area in one primitive lattice cell, $a$ being
the lattice constant. We used the values $S\approx3.8\cdot10^{-8}$ m$^2$,
$W=50$ nm, $D=100$ nm and $a=3 ~\mu$m. The propagators in Eqs.
(\ref{e.alpha}) and (\ref{e.beta}) were calculated only for order
parameters corresponding to a fully diffuse wall. All the
tunneling model parameters were shown in Fig. \ref{f.tconst}.

\leveltwo{Boundary conditions}

The order parameter, or the
$z$ dependence of the  functions $\Delta_\perp(z)$ and
$\Delta_\parallel(z)$ near a wall, is needed before we can proceed to
do any of the other things just described. These have to be calculated
self-consistently, assuming some general properties for the wall 
and devising a boundary condition for the quasiclassical propagator.
The ``correct'' approach here would be to think about scattering
$t$ matrices \cite{buchholtzrainer,buchholtz}, but this would
be quite complicated, and it has been shown that much simpler
models essentially reproduce the same results 
\cite{zhang,kurkirainersauls,kopnin}.
We used perhaps the simplest of all, the ``randomly oriented mirror'' 
(ROM) model \cite{rom}. 
There are severe misprints in the original publication, so it is better to 
express the whole algorithm here anew. The unit vector $\svec$
is perpendicular to the wall. 

\begin{quote}

\textbf{ROM algorithm for calculating the propagator}

%- consider elastic scattering.
(1) Calculate up to the wall the solutions growing exponentially
toward the wall on all trajectories, i.e. calculate $\nam g_{<}$ for
$\kvec\cdot\svec<0$ and $\nam g_{>}$ for $\kvec\cdot\svec>0$. 
(The latter can actually be obtained from the former by applying the
$\kvec$-inversion symmetries in Eq. (\ref{e.basicsymm}).)

(2) Calculate the solutions growing exponentially out of the wall
with the initial values
\begin{equation} \label{e.bcon}
\begin{split}
& \sum_{\kvec'\cdot\svec<0} w_{\kvec,\kvec'} 
\frac{\nam g_{<}(\kvec',\vecb 0, \me) }{
\{\nam g_{<}(\kvec',\vecb 0,\me), \nam g_{>}(\kvec,\vecb 0,\me) \}},
\qquad \textrm{for} ~\kvec\cdot\svec>0 \\
& \sum_{\kvec'\cdot\svec>0} w_{\kvec',\kvec} 
\frac{\nam g_{>}(\kvec',\vecb 0, \me) }{
\{\nam g_{<}(\kvec,\vecb 0,\me), \nam g_{>}(\kvec',\vecb 0,\me) \}},
\qquad \textrm{for} ~\kvec\cdot\svec<0 
\end{split} \nonumber
\end{equation}
where $\vecb 0$ denotes the point at the wall.
(Again, the $\kvec\cdot\svec<0$  initial condition is not 
actually needed because of symmetries.)

(3) Evaluate the normalised physical propagator 
for all directions and at all positions from the commutator of the 
converging and diverging solutions, as explained above. 

\end{quote}

\begin{figure}[!bt]
\begin{center}
\includegraphics[width=0.6\linewidth]{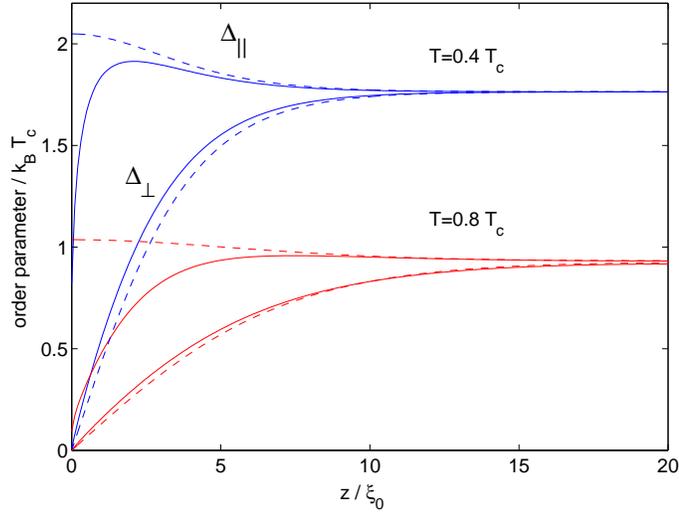}
\caption{Order parameter at the wall for two temperatures, $T=0.4T_c$
(upper bulk values) and $T=0.8T_c$ (lower bulk values). For each temperature,
the lower curve is $\Delta_{\perp}(z/\xi_0)$ and the upper
$\Delta_{\parallel}(z/\xi_0)$. The solid lines are for a diffuse wall and the
dashed lines for a specular one.}
\label{f.ops}
\end{center}
\end{figure}

\noindent 
The specular scattering limit is obtained from this by choosing
\begin{equation}
w_{\kvec,\kvec'}=\begin{cases}
1, & \textrm{for}~ \kvec-\kvec'=2\svec(\kvec\cdot\svec)\\
0, & \textrm{otherwise},
\end{cases}
\end{equation}
which just requires the propagators to be continuous on
mirror-reflected trajectories: 
$\nam g_\gtrless(\kvec,0)=\nam g_\gtrless(\ul\kvec,0)$, where 
$\ul\kvec=\kvec-2\svec(\kvec\cdot\svec)$. This is a simple and 
intuitive case, but not very realistic. The totally diffuse limit can be
modelled by replacing
\begin{equation}
\sum_{\kvec'\cdot\svec<0} w_{\kvec,\kvec'} \longrightarrow 
\frac{1}{\pi}\int_{\kvec'\cdot\svec<0}\upd^2\kvec' |\kvec'\cdot\svec|.
\end{equation}
It describes a rough wall where an incoming quasiparticle can be
scattered into any angle, irrespective of the original direction.
This limit is considered to be the most realistic one in most cases.
Fig. \ref{f.ops} shows examples of the resulting order parameters for these two 
limiting cases at two temperatures.
The component $\Delta_\perp$ is more strongly suppressed
than $\Delta_\parallel$, as is characteristic for any wall.

\leveltwo{Numerics}

Computation of the order parameter is based on an iterative process.
Here an initial guess is first taken for $\Deltavec^{(0)}$. Then,
using the ROM boundary condition, the 
Eqs. (\ref{e.simple}) are integrated to get
the diverging (and decaying) solutions and the multiplication trick of 
Eq. (\ref{e.multiply}) is applied to obtain the physical solutions.
Finally the self-consistency Eq. (\ref{e.selfconsist})
is used to get a new approximation for $\Deltavec^{(0)}$ and the
process is repeated until convergence %(as judged by the eye) 
is obtained.
Having a converged order parameter and the propagator components
$A$, $B$ and $\Cvec$, the currents, energies and other physical
properties can be evaluated by simple integrations over quasiparticle directions.
All calculations were performed in reduced units, where lengths
appear in units of the coherence length $\xi_0=\hbar\vF/2\pi\kB T_c$
and energies in units of $\kB T_c$.

In the self-consistency equation and elsewhere, numerical integrations over the polar 
angles were carried out using a 32-point (or less) Gaussian quadrature. 
%but
%using the basic symmetries (\ref{e.basicsymm}) integrations could be
%restricted to one of the half-spaces $\kvec\cdot\svec\gtrless 0$. 
Accordingly, the diverging solutions had to be calculated for
up to 16 trajectories toward ($\kvec\cdot\svec<0$)  and away 
($\kvec\cdot\svec>0$) from the wall.
For solutions diverging toward the wall, the bulk forms
of Eq. (\ref{e.exponential}) were used as the initial values. The initial values
for solutions diverging toward the bulk were obtained
from the ROM prescription. \emph{Only} the diverging solutions for all
$\kvec$ directions needed to be calculated, because after these were 
known, the decaying solutions could be found by using the
symmetries in Eq. (\ref{e.basicsymm}): $a_>^{(0)}(\kvec)=a_<^{(0)}(-\kvec)$,
$b_>^{(0)}(\kvec)=-b_<^{(0)}(-\kvec)$ and 
$\im~\cvec_>^{(0)}(\kvec)=-\im~ \cvec_<^{(0)}(-\kvec)$ at each point
in space. All of this had to be done for a range of
positive Matsubara energies, the negative ones being obtained through
symmetries. Ten energies was usually enough, although a hundred would
have been no problem either, since the calculation is not otherwise very demanding.
Summing up, all that needed to be caculated explicitly can be 
described with $\nam g_<^{(0)}(\vartheta,z,\epsilon_m)$. For general
azimuthal angles, the symmetries described in connection with
Eq. (\ref{e.transforms}) could be applied.

The fourth-order Runge-Kutta method was used to integrate the
Eilenberger equations. The diverging solution is easy to find because this explicit (or
``forward-type'') method should always be unstable towards finding 
the solution of Eq.  (\ref{e.simple}) corresponding to the largest
eigenvalue of its coefficient matrix. Because of this, the
diverging solution is in fact the only solution to be found with the method.
Proceeding along the trajectory the solutions will diverge
approximately as (\ref{e.exponential}). The step size $h_u$ should then
be small enough to satisfy $2\sqrt{\epsilon_m^2+|\Deltavec|^2}h_u/\hbar\vF\ll 1$
in order to have reasonable accuracy.
Nevertheless, the sensibility of computing such diverging solutions numerically is a bit
questionable. On long trajectories the solutions tend to overflow on 
any computer, and some intermediate rescaling
of the propagator has to be done. One way to circumvent the
divergencies completely is to rescale $a$, $b$ and $\im~\cvec$ at each step, 
such that $a=1$ everywhere: after all, only their relative magnitudes
make a difference when the normalised physical propagator is formed in
Eq. (\ref{e.multiply}). Another way would be to absorb the leading 
divergence into an exponential by defining $\nam g(u)=\nam
g'(u)\exp(Gu)$, where $G<2\sqrt{\epsilon_m^2+|\Deltavec|^2}/\hbar\vF$ 
is some positive constant. Then one would rewrite Eqs.
(\ref{e.simple}) for $\nam g'(u)$, and solve them, instead. The
exponential factors cancel in Eq. (\ref{e.multiply}) and the correct
physical propagator should result. For some reason, this approach
seemed to be prone to a numerical instability towards the
\emph{original} divergence and was not used. It is worth 
investigating further, perhaps with another integration algorithm.

\leveltwo{Results for a single pinhole}

\begin{figure}[!tbp]
  \begin{minipage}[t]{.49\linewidth}
    \centering \includegraphics[width=1.0\linewidth]{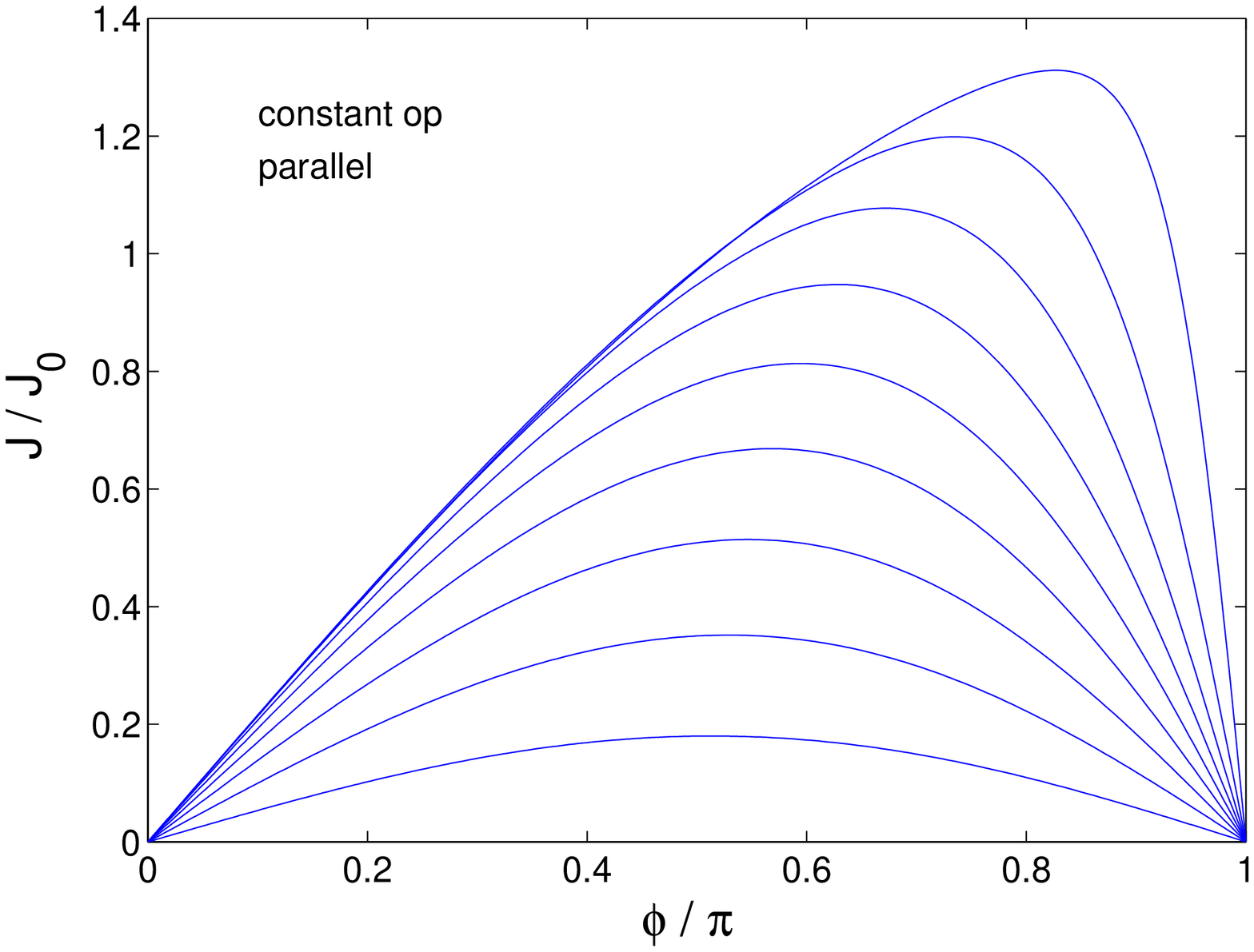}
  \end{minipage}
  \hfill
  \begin{minipage}[t]{.49\linewidth}
    \centering \includegraphics[width=1.0\linewidth]{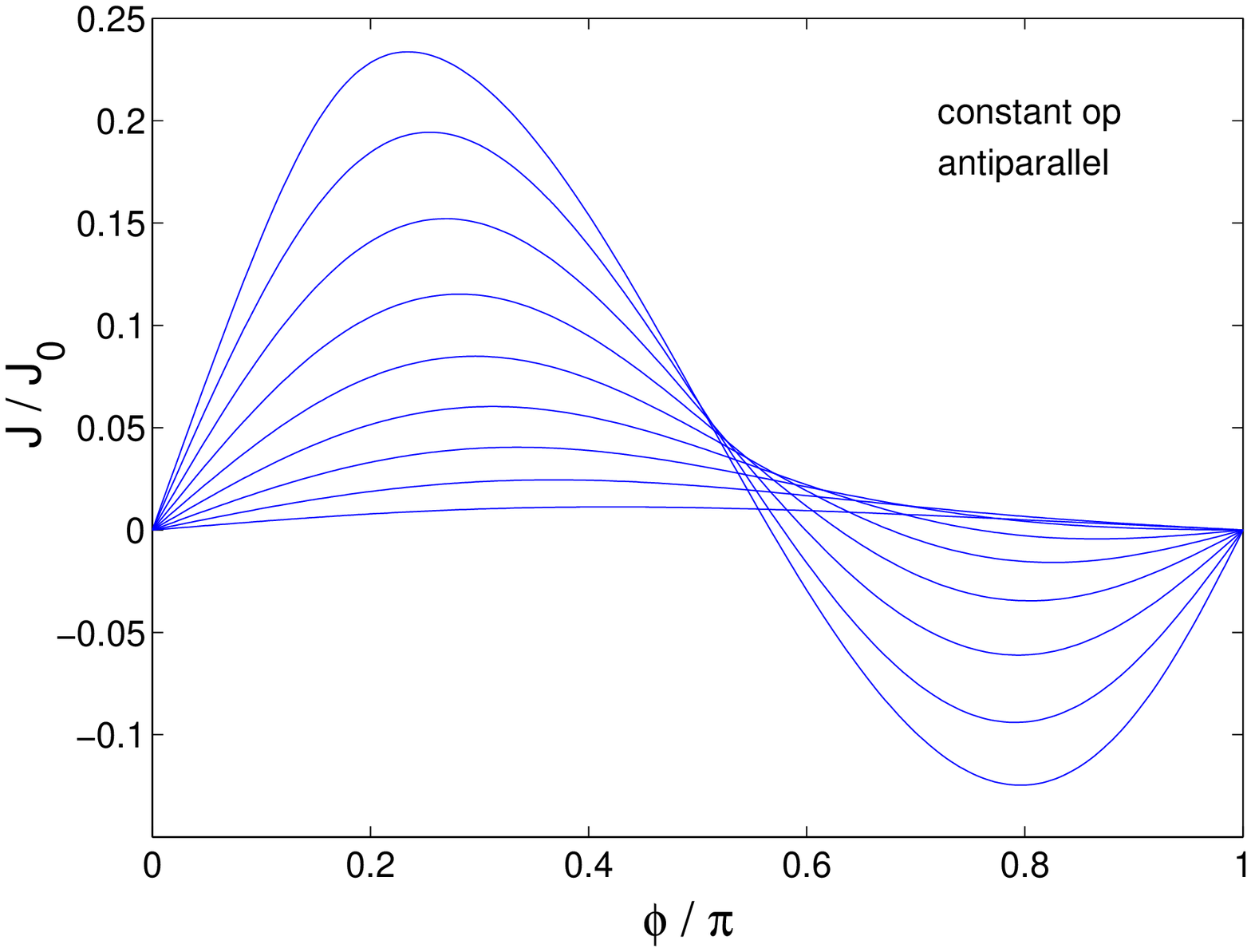}
  \end{minipage}
\caption{Current-phase relations for a constant order parameter.} \label{f.kiphi}
\end{figure}

\begin{figure}[!tbp]
  \begin{minipage}[t]{.49\linewidth}
    \centering \includegraphics[width=1.0\linewidth]{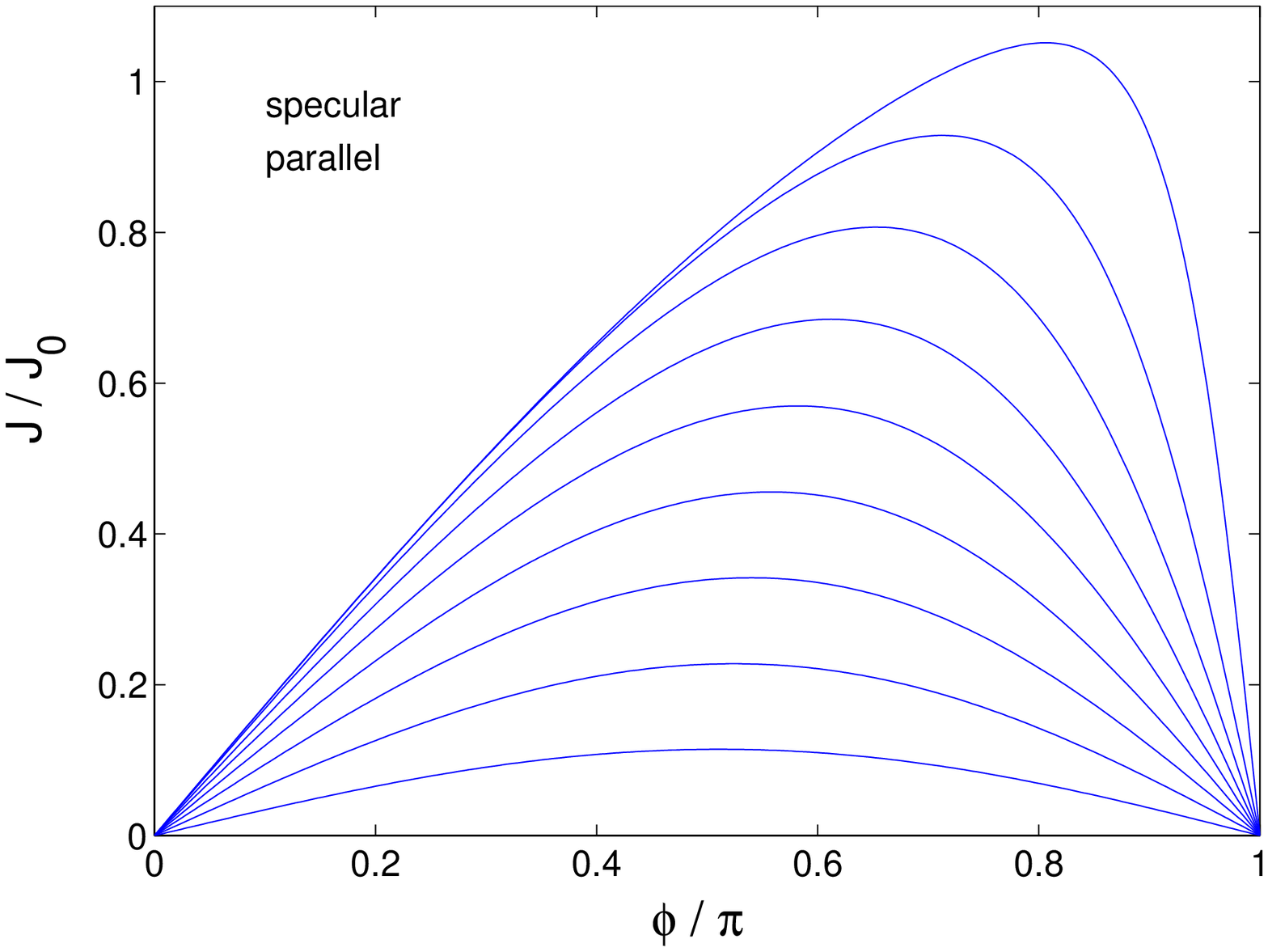}
  \end{minipage}
  \hfill
  \begin{minipage}[t]{.49\linewidth}
    \centering \includegraphics[width=1.0\linewidth]{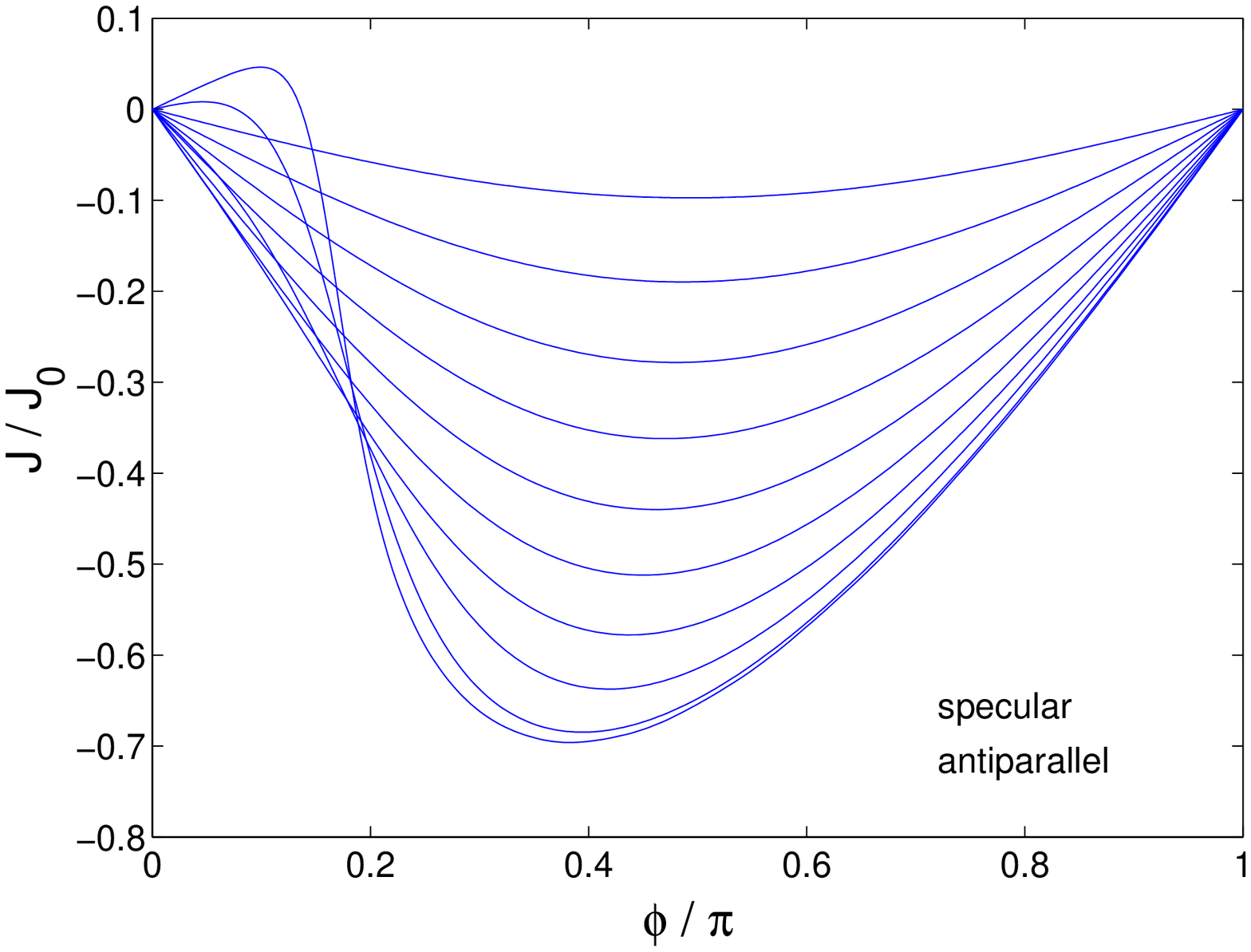}
  \end{minipage}
\caption{Current-phase relations for a specularly scattering wall.} \label{f.siphi}
\end{figure}

\begin{figure}[!tbp]
  \begin{minipage}[t]{.49\linewidth}
    \centering \includegraphics[width=1.0\linewidth]{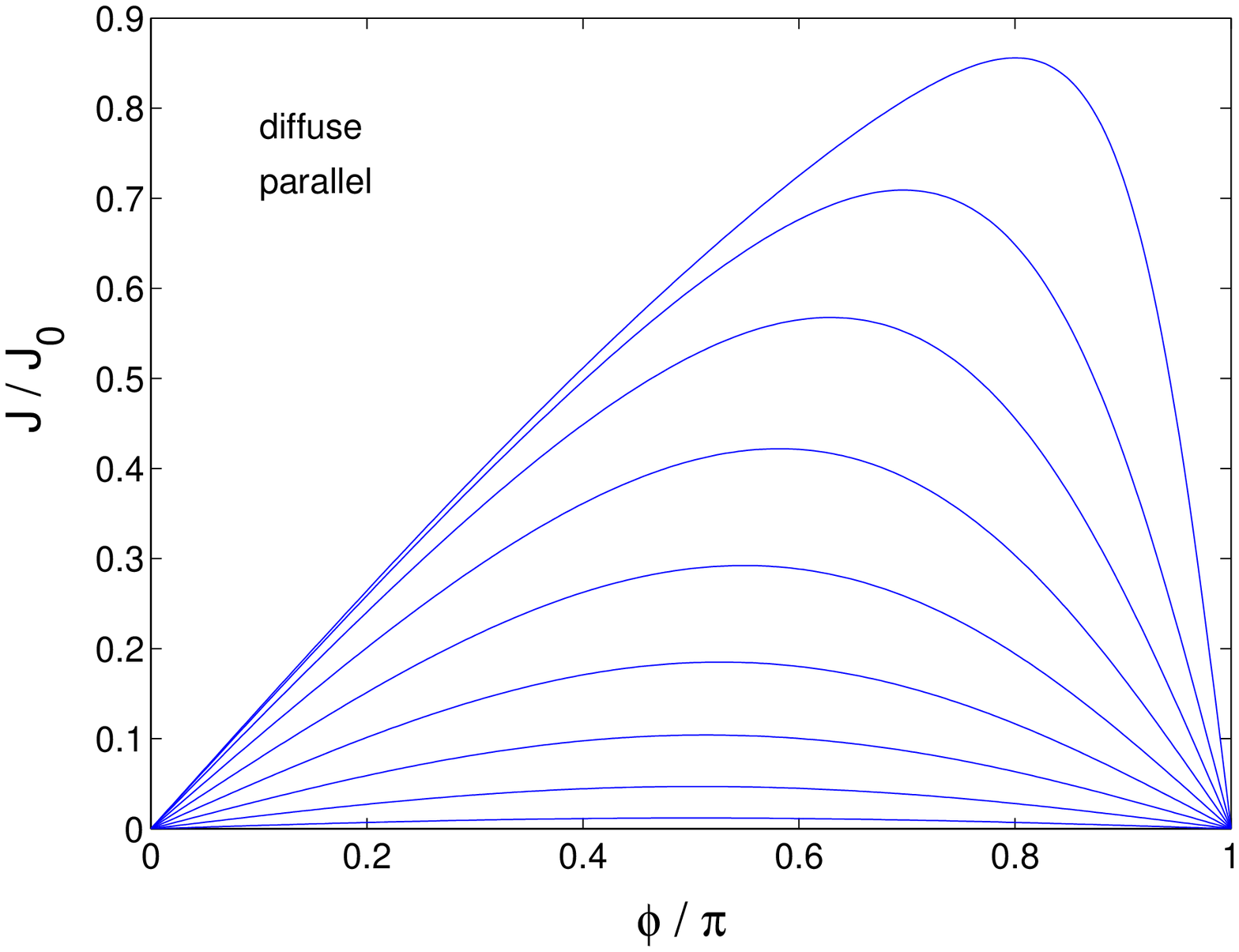}
  \end{minipage}
  \hfill
  \begin{minipage}[t]{.49\linewidth}
    \centering \includegraphics[width=1.0\linewidth]{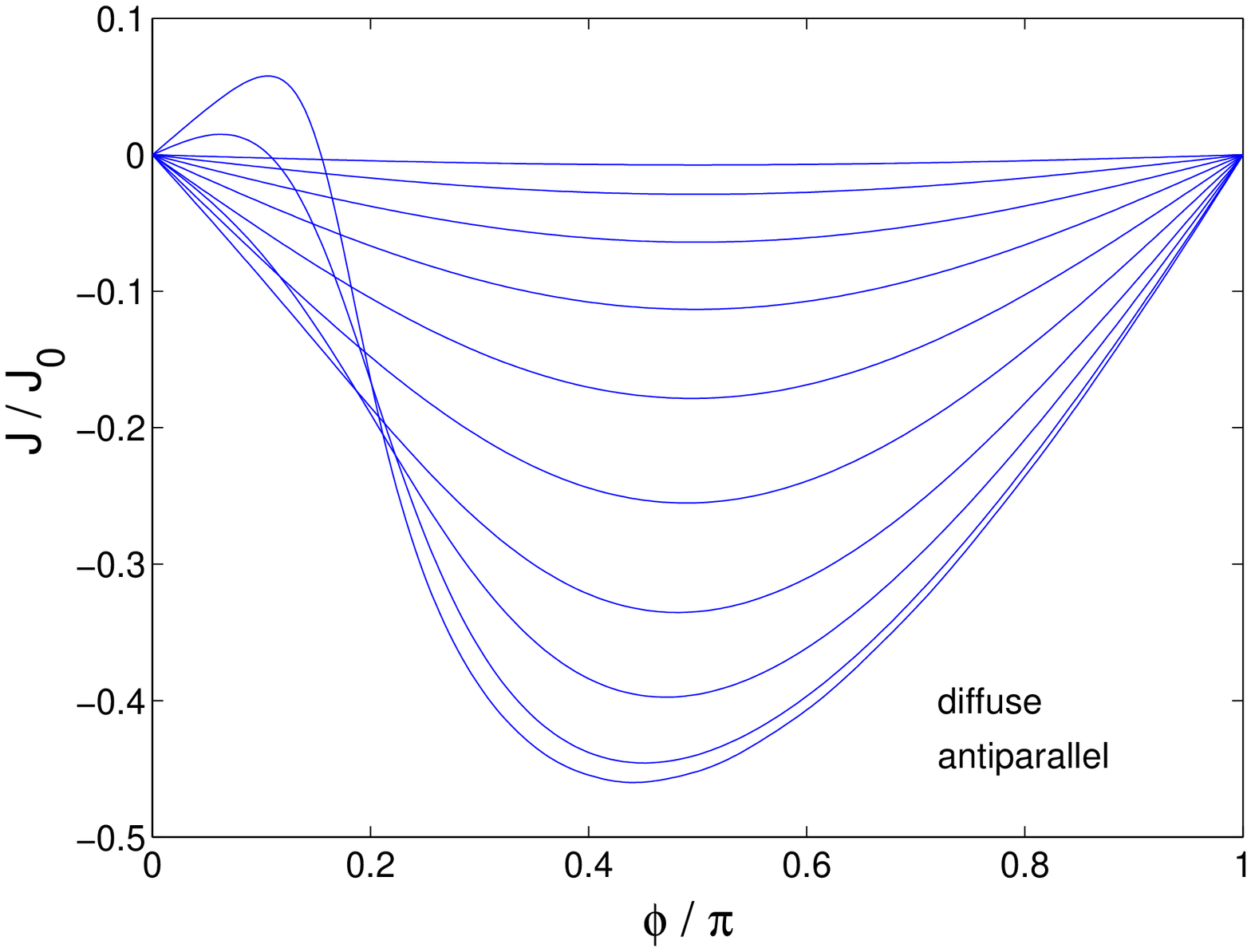}
  \end{minipage}
\caption{Current-phase relations for a diffusely scattering wall.} \label{f.diphi}
\end{figure}

Here we consider the results for calculations of the Josephson current
through a single pinhole aperture, assuming that the surrounding walls 
have fixed the spin-orbit rotation axes $\nvec^{L,R}$ perpendicular to the
wall. As discussed above, there are then two cases: $\nvec^L=\nvec^R$ 
(parallel) and $\nvec^L=-\nvec^R$ (antiparallel).
Three different boundary conditions on the wall were used for both
cases: (1) a case where the order parameter was assumed to be constant 
all the way to the wall, (2) a specular wall and (3) a diffuse wall.
The corresponding current-phase relations are shown in Figs.
\ref{f.kiphi}-\ref{f.diphi}, with the parallel case always on the left 
and the antiparallel on the right. The (mass) currents are in units of 
$J_0=2m_3\vF\NF\kB T_c\times\textrm{[open area]}$, and only phase
differences in the range $[0,\pi]$ are shown due to the symmetry
$J(2\pi-\phi)=-J(\phi)$.

\levelthree{Current-phase relations}

(1) \emph{Constant order parameter} --- This is the case discussed by
Yip \cite{yippi}, and the current-phase relations shown in Fig.
\ref{f.kiphi} are exactly the same as those obtained by him. They can be
simply plotted from Eq. (\ref{e.siisti}) provided that one knows the
temperature dependence of the bulk gap $\Delta$. These curves are for 
$T/T_c=0.9, 0.8,\ldots, 0.1$ in order of increasing critical current.
The parallel case is well known 
\cite{kurkijarvi}, but the new feature found by Yip is seen in the
antiparallel case on the right: very close to $T_c$ the $J(\phi)$ is 
sinusoidal, but at temperatures below about $0.5 T_c$ a new point 
on $[0,\pi]$ will appear, where $J=0$ and a very strong 
extra kink in $J(\phi)$ will form around $\phi=\pi$. This has a simple 
explanation in terms of Eq. (\ref{e.siisti}) where the
phase difference $\phi$ appears only in the combination
$\phi+\delta\theta_{\kvec}$. For the antiparallel $\nvec$ vectors,
$\theta_{\kvec}$ depends strongly on the polar angle $\vartheta$ of $\kvec$ 
($\cos\theta_{\kvec}=1-\frac{15}{8}\sin^2\vartheta$), and different
quasiparticle directions contribute to the current with a different
effective phase differences. The currents then cancel each other in
such a way that a kink will appear --- this strong calcellation is also why the critical
currents are so much smaller in the antiparallel than in the parallel case.
For a related effect in $d$ wave superconductors, see
Ref. \cite{yip}.

(2) \& (3) \emph{Self-consistent order parameter} ---
For the more realistic surface models, the results are essentially different.
Fig. \ref{f.siphi} shows $J(\phi)$'s for a specular surface and
Fig. \ref{f.diphi} for the diffuse surface.
The curves are again for $T/T_c=0.9, 0.8,\ldots, 0.1$ in order of
increasing critical current; for the antiparallel case also curves for 
$T/T_c=0.05$ are shown to emphasize the behavior at low temperature. 
The parallel current-phase relations look exactly as before, although
their critical currents are slightly reduced. But a striking difference is seen 
in the antiparallel cases: they remain sinusoidal down to very low
temperatures. As can be seen from Eq. (\ref{e.sophisticated}),
the same current-cancellation effect is still possible in principle,
but here it is strongly reduced: 
an additional kink appears only at around $0.2 T_c$.
Now it also occurs around $\phi=0$, instead of $\phi=\pi$.
These results have been obtained for a bare pinhole without
adjusting any effective parameters or restricting transmission angles by the 
probability $p(\vartheta)$.
%Temperatures above $T/T_c\approx 0.5$ in the diffuse case 
High temperatures in the diffuse case 
thus correspond to the tunneling model with
$\alpha<\frac{7}{4}\beta$. This is why the critical current in the
antiparallel case is ``negative''. From this point of view it seems
that the strong rescaling of $\alpha$ and $\beta$, which made it
positive in the original tunneling
model calculation, was not  well justified.

%\newpage   % THIS NEWPAGE MAY BE ESSENTIAL

 % used to be height=55mm

\levelthree{Critical currents}

\begin{figure}[!tbp]
  \begin{minipage}[t]{.49\linewidth}
    \centering \includegraphics[width=1.0\linewidth]{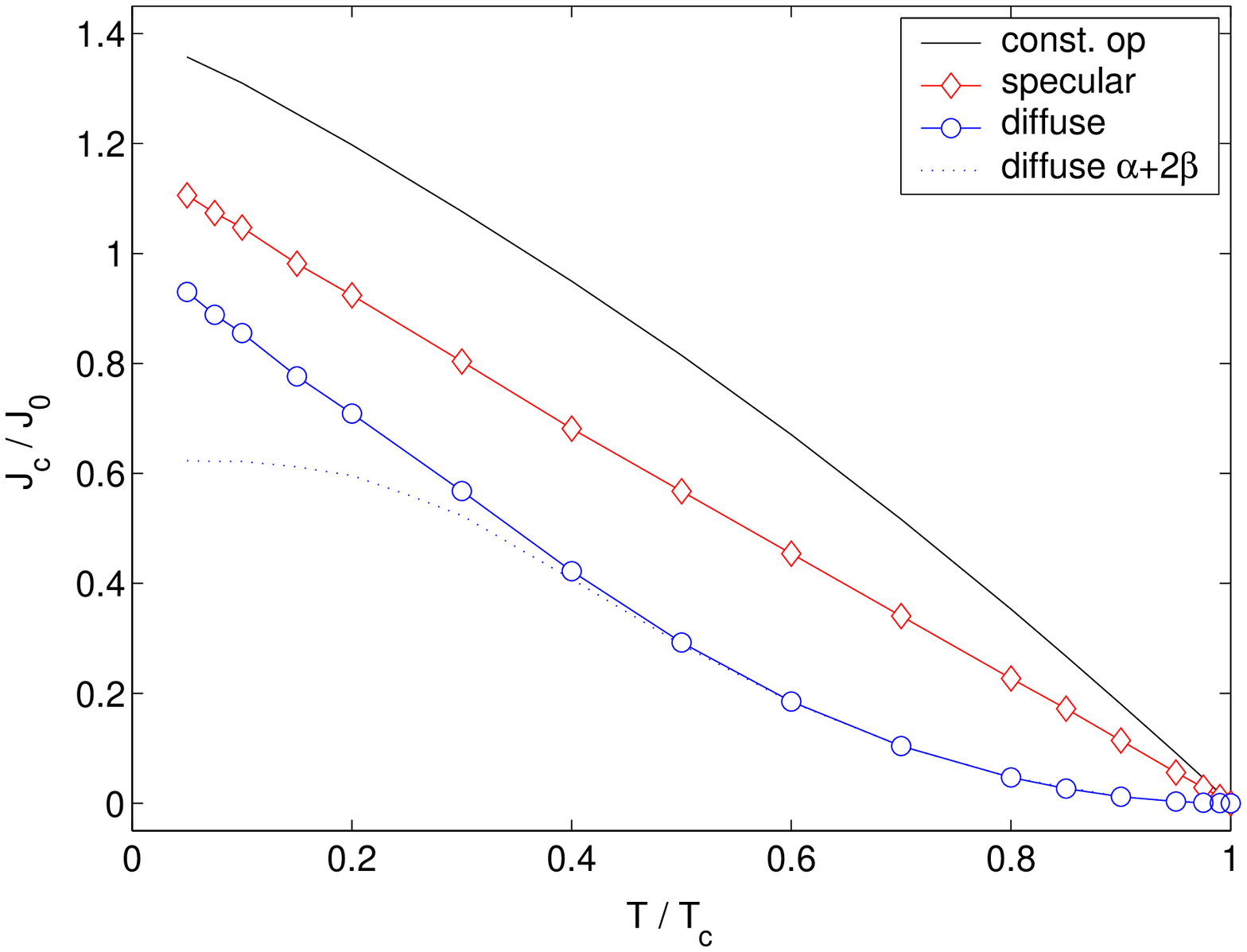}
  \end{minipage}
  \hfill
  \begin{minipage}[t]{.49\linewidth}
    \centering
      \includegraphics[width=1.0\linewidth]{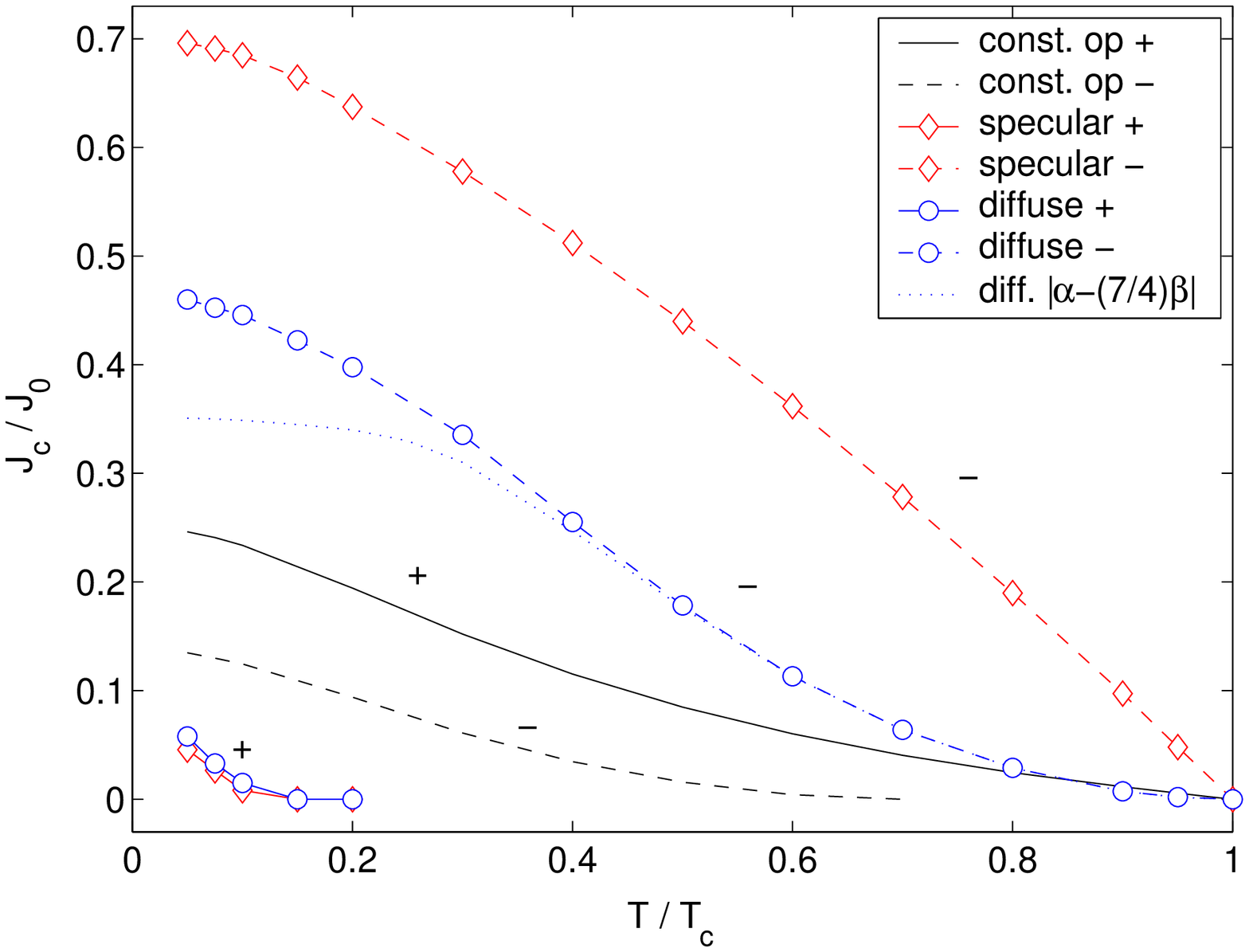} 
  \end{minipage}
\caption{Critical currents for different boundary
conditions. \textbf{Left} (Parallel $\nvec$'s): constant order
parameter (solid line), specular surface (diamonds) and diffuse surface
(circles). \textbf{Right} (Antiparallel $\nvec$'s): solid lines
correspond to positive extrema of $J(\phi)$ and the dashed lines to
negative extrema; plain curves are for the constant order parameter,
and again diamonds for a specular wall and circles for a diffuse wall.
In both figures, the dotted lines give the estimates
obtained from the ``tunneling model parameters'' $\td\alpha$ and
$\td\beta$.} \label{f.ictemps}
\end{figure}

Figure \ref{f.ictemps} further illustrates the critical currents $J_c$
and the possible additional extrema of $J(\phi)$ 
for a pinhole junction in all of the above cases and as a function of temperature.
For parallel $\nvec$ vectors such a plot has been published in Ref.
\cite{tks}, but those results were erroneous. In this case we see that, close
to $T_c$, $I_c(T)\propto(1-T/T_c)$
for the constant order parameter and the specular surface, and for a
diffuse surface $I_c(T)\propto(1-T/T_c)^2$, as expected; see Ref.
\cite{kop86}. The critical current for a
constant order parameter is always the highest and for a diffuse wall
it is the lowest. For antiparallel $\nvec$ vectors the roles change:
the constant order
parameter case has the \emph{lowest} $J_c$, due to the strong
cancelling between different quasiparticle directions, but the
negative extremum around $\phi=\pi$ is nearly as pronounced as
the positive one. For the diffuse and specular surfaces, it is seen
quite clearly that
the other extrema appear only at much lower temperatures.
The dotted lines correspond to the high-temperature approximations obtained from 
Eqs. (\ref{e.alpha}) and (\ref{e.beta}) for a single pinhole with
$p(\vartheta)=1$ in a diffuse wall, i.e. $(\td\alpha+2\td\beta)/E_0$ for parallel and
$(\td\alpha-\frac{7}{4}\td\beta)/E_0$ for antiparallel $\nvec$'s.
Here the energy unit is $E_0=\hbar\vF\NF\kB T_c\times[\textrm{open area}]$.
These lines follow the correct critical currents amazingly well down to temperatures less than 
$T=0.4 T_c$.

\begin{figure}[!btp]
\begin{center}
\includegraphics[width=0.6\linewidth]{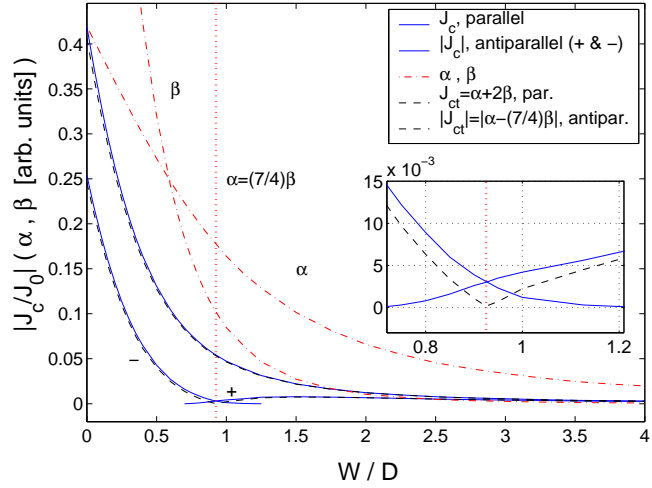}
\caption{
Absolute values of critical currents for increasing
$W/D$ at $T=0.4 T_c$ (see Fig. \ref{f.wdrat}).
Solid lines are exact results and the dashed
lines are obtained from the high-temperature approximation. The small
differences in details are
depicted in the inset, which corresponds to the region where
$\alpha\approx\frac{7}{4}\beta$. The dash-dotted lines 
represent the relative sizes of $\alpha$ and $\beta$.}
\label{f.wdeff}
\end{center}
\end{figure}

\begin{figure}[!btp]
\begin{center}
\includegraphics[width=0.6\linewidth]{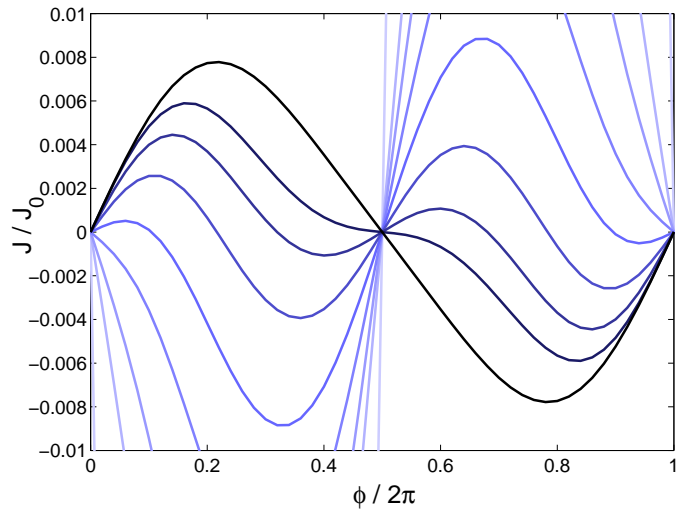}
\caption{Effect of restricting quasiparticle transmission in the antiparallel case at
$T=0.4~T_c$. The curves correspond to
$W/D=0.0,0.5,0.6,0.7,0.8,0.9,1.0,1.1,1.5$
from lighter shades to darker. The $0.0$-curve (only partly visible) is the one 
shown in Fig. \ref{f.diphi}; its critical current is some 30
times larger than that of the $1.5$-curve, and has a different
sign. The crossing regime corresponds roughly to the tunneling
parameters $\alpha\approx\frac{7}{4}\beta$, but instead of going through
zero, $J(\phi)$ exhibits the transition by forming a ``$\pi$ kink''.}
\label{f.wdrat}
\end{center}
\end{figure}

In Figs. \ref{f.wdeff} and \ref{f.wdrat}, we illustrate the effect of
restricting angles with a nonzero $W/D$ in
Eq. (\ref{e.probability}). 
Most importantly, increasing its value results in a drastic reduction
of the critical currents. But
for the antiparallel case, there is also some subtle fine structure
involved, and it is interesting to compare the details of the exact results and the
high-temperature approximation. 
In this approximation, the tunneling model, the critical current
changes its sign at
$\alpha=\frac{7}{4}\beta$ where the current-phase relation
$J(\phi)\equiv 0$ for all $\phi$. In the exact case, however, this
never takes place. Instead, around values of $W/D$ where
$\alpha\approx\frac{7}{4}\beta$ the current-phase relation develops a
kink, and for a short range of $W/D$ there exist \emph{two} extrema of $J(\phi)$ on
the interval $\phi\in[0,\pi]$ (see Fig. \ref{f.wdrat}). This is really just Yip's
``$\pi$ state'', which thus occurs at high temperatures also, but 
only at considerable cost in the critical currents.

For large $W/D$ in Fig. \ref{f.wdeff}, we see that both critical
currents are positive and very close to each other. In terms of
the tunneling model, this is easy to understand.
In this limit $\alpha\gg\beta$, and the coupling energy
is $F_J=-\alpha R_{\mu z}^LR_{\mu z}^R\cos\phi$. 
Now, for fixed $\nvec^L=\pm\nvec^R=\pm\zvec$, we have 
$R_{\mu z}^LR_{\mu z}^R=1$ and $F_J$ reduces into the same simple
Josephson relation for both the parallel and antiparallel $\nvec$ vectors.

\levelthree{Preliminary conclusions}

Based on the above results, we now see that the simple tunneling
model which we initially considered can reproduce most of the
characteristics of a pinhole junction, or a coherent array of such pinholes,
down to very low temperatures. We may now also safely state that the
current-cancellation mechanism of Yip is \emph{not} the mechanism underlying the
$\pi$ state, although it is in principle quite interesting.
The diffusely scattering
wall is likely to be the closest model to a real surface, and for that 
a purely sinusoidal behavior is found to much lower temperatures
than $0.6 T_c$, where the Berkeley $\pi$ state is already observed. Of course,
there is also no way Yip's model alone could explain the fact that the
weak link can be found in \emph{two} different states, \emph{both} of which show an
additional kink in their current-phase relations. Yip had to employ a
magnetic field to explain this, although the fields which were present 
in the Berkeley experiment should not have been large enough to affect 
the $J(\phi)$ significantly (see the previous discussion on magnetic
field strengths). 

Although we already stated that the tunneling model should be a good
description of the pinhole array, it is still worth while to do the
array calculation more accurately. Not least because the antiparallel $\pi$ state in the 
results presented above is probably wrong due to the arbitrary
resacaling of the tunneling parameters.

\newpage
%%%%%%%%%%%%%%%%%%%%%%%%%%%%%%%%%%%%% ARRAY

\levelone{Pinhole array }

The final challenge was to construct a quasiclassical free-energy functional for the
pinhole junction. Once a suitable form was found, the goal was
to repeat all the calculations of the tunneling
model as well as possible using that. Although the new functional
is more accurate, it is also, in some sense, more restricted, which leads 
to some changes in the results. 
The idea underlying the derivation is that the small perturbations 
caused by insertion or removal of a pinhole in a nontransparent wall is analogous
to, say, inserting ionic impurities in a bulk superfluid or superconductor. These kinds of
problems have already been succesfully solved \cite{ions,pinning}, 
and a very similar approach was taken here.

\leveltwo{The pinhole free energy}

We start from the well-known expression for the energy difference
between states with one impurity and no impurity, $\nam V$ being the
impurity potential \cite{free,serenerainer}
\begin{equation} \label{e.enefu}
\begin{split} 
\delta\Omega^{\textrm{tot}}&=-\frac{1}{2} \Tr [\ln(-\nam G_0^{-1}+\nam \Sigma+\nam
V)-\ln(-\nam G_0^{-1}+\nam \Sigma)]. %\\
\end{split}
\end{equation}
To eliminate the logarithm, we may apply some form of the ``$\lambda$-trick''
\cite{fetterwalecka}. We choose to integrate over the strength of
$\nam V$ by making the substitution $\nam V\rightarrow\lambda\nam V$
and writing
\begin{equation} \label{e.lambdatrick}
\begin{split}
\delta\Omega^{\textrm{tot}} = -\frac{1}{2}\Tr\int_0^1\frac{\upd\lambda}{\lambda}
(\nam G_0^{-1}-\nam\Sigma-\lambda\nam V)^{-1}\lambda \nam V
=\frac{1}{2}\Tr\int_0^1\frac{\upd\lambda}{\lambda}\nam G_{1}\nam
T_{\lambda}.
\end{split}
\end{equation}
Here the latter equality follows from a formal application of the
$t$ matrix equation $\nam T_{\lambda}=\lambda \nam V+\nam T_{\lambda}\nam
G_{1}\lambda\nam V$ and the relation $\nam G=\hat G_{1}+\nam G_{1}\nam 
T_{\lambda}\nam G_{1}$. Here $\nam G=(\nam G_0^{-1}-\nam\Sigma-\lambda\nam
V)^{-1}$  gives the full propagator in the
presence of an impurity scattering potential and $\hat G_{1}$ is an
``intermediate'' Green's function which does not include the effect of 
the impurity.\footnote{See Sec. 5. For some further discussion 
on this type of operator formalism, see for example
Refs. \cite{fetterwalecka,landau} and some of the papers
listed in the references.} The trace operation $\Tr$ %in (\ref{e.enefu})
is defined as \cite{ions}
\begin{equation} \label{e.trace}
\begin{split}
\Tr \nam F(\vecb k,\vecb k',\epsilon_m) & =
\kB T \sum_{m} \int\frac{\upd^3k}{(2\pi)^3} \Tr_4
\nam F(\vecb k,\vecb k,\epsilon_m) \\
& =\kB T\sum_{m} \int\frac{\upd^2\kvec}{4\pi} \int \upd\xi_{|\vecb k|} N(\xi)
\Tr_4 \nam F(\vecb k,\vecb k,\epsilon_m),
\end{split}
\end{equation}
where $\nam F$ is a $4\times 4$ Nambu matrix and $\Tr_4$ is its trace.
Equation (\ref{e.lambdatrick}) is now in a form where the propagator
can be $\xi$-integrated directly. However, to avoid a divergence in
the Matsubara summation, we have to subtract from Eq. (\ref{e.lambdatrick})
the normal-state contribution $\delta\Omega^N$, which is obtained by
setting $\nam\Sigma=0$ in Eq. (\ref{e.enefu}). We define 
$\delta\Omega=\delta\Omega^{\textrm{tot}}-\delta\Omega^N$
and transform this to the quasiclassical form
\begin{equation} \label{e.quasifree}
\begin{split}
\delta\Omega & =\frac{1}{2}\hbar\NF\pi \kB T \sum_{m} \int
\frac{\upd^2\kvec}{4\pi} \int_0^1 \frac{\upd\lambda}{\lambda} \times\\
&\times \Tr_4[\nam g_1(\kvec,\vecb r_{\textrm{imp}},\epsilon_m)
\nam t_{\lambda}(\kvec,\kvec,\epsilon_m)-
\nam g^N_1(\kvec,\vecb r_{\textrm{imp}},\epsilon_m)
\nam t^N_{\lambda}(\kvec,\kvec,\epsilon_m)],
\end{split}
\end{equation}
where $\vecb r_{\textrm{imp}}$ is the location of the impurity and
$\nam t(\kvec,\kvec,\me)$ is the forward-scattering part of the
$t$ matrix. This formula is simpler to use than those in Refs. \cite{ions,pinning},
since it does not involve an integration over $\rvec$. More
importantly, $\nam g_1$ is constant in the $\lambda$ integration.

Now we specialise the above approach to our particular problem.
The coupling energy we wish to know is, by definition, the difference
in energies between an open pinhole and a blocked pinhole. It should be
irrelevant how the hole is blocked as long as the transmission of
quasiparticles is prevented. Changing the type of blockage should only
change some  constant terms in the energy, which do not
depend on the phase difference or the rotation matrices.
We might, for example, block it with a piece of specularly scattering
surface, which corresponds to a delta-function scattering potential 
$\mathcal{V}\delta(z)$ in the limit
$\mathcal{V}\rightarrow\infty$. 
The $t$ matrix of this type of ``impurity'' is of the particularly simple form
\cite{buchholtzrainer,serenerainer} 
\begin{equation} \label{e.tmatrix}
\nam t_{\lambda}(\kvec,\kvec')=\frac{2\vF|\kvec\cdot\zvec|\lambda
\mathcal{V}A_o\delta^2_{\vecb k_{||},\vecb k'_{||}}}
{2\vF|\kvec\cdot\zvec|-\lambda\mathcal{V}[
\nam g_1(\kvec,z=0,\epsilon_m)+
\nam g_1(\ul\kvec,z=0,\epsilon_m)]},
\end{equation}
where $A_o$ is the area of the blocking piece wall with normal $\zvec$ 
(equal to the open area of one open pinhole), 
$\vecb k_{\parallel}=\kvec-(\kvec\cdot\zvec)\zvec$ denotes the
parallel component of $\kvec$, and 
$\ul\kvec=\kvec-2(\kvec\cdot\zvec)\zvec$.
On inserting this into Eq. (\ref{e.quasifree}) and performing the
$\lambda$-integration, we find
\begin{equation}
\begin{split}
F(\mathcal{V})&=\frac{1}{2}A_o\hbar\vF\NF\pi\kB T \sum_{m} \int 
\frac{\upd^2\kvec}{4\pi}|\kvec\cdot\zvec|\Tr_4\times\\
&\times\ln
\frac{2\vF|\kvec\cdot\zvec|-\mathcal{V}[
\nam g_1(\kvec,z=0,\epsilon_m)+
\nam g_1(\ul\kvec,z=0,\epsilon_m)]}
{2\vF|\kvec\cdot\zvec|+2\iu\mathcal{V}\nam \tau_3 \Sgn(\epsilon_m)}
\end{split}
\end{equation}
where the normal-state propagator $\nam g^N=\iu\nam\tau_3\Sgn(\me)$ 
was obtained from the bulk form, Eq. (\ref{e.compact}), by setting
$\Deltavec=\vecb 0$, $\nam\Delta=0$. 
%\begin{equation}
%F(\mathcal{V})=\frac{1}{2}S\hbar\vF\pi \Tr'
%|\kvec\cdot\nvec|\ln
%\frac{2\vF|\kvec\cdot\nvec|-\lambda\mathcal{V}[
%\nam g_1(\kvec,R_{\perp}=0,\epsilon_m)+
%\nam g_1(\ul\kvec,R_{\perp}=0,\epsilon_m)}
%{2\vF|\kvec\cdot\nvec|+2\iu\mathcal{V}\nam \tau_3 \Sgn(\epsilon_m)}
%\end{equation}
Finally, in the limit $\mathcal{V}\rightarrow\infty$ we have
\begin{equation} \label{e.detf}
\begin{split}
F_{\textrm{pinhole}}
%&=\frac{1}{2}S\hbar\vF\NF\pi T\sum_{\epsilon_m} \int\frac{\upd^2\kvec}{4\pi}
%|\kvec\cdot\nvec|\Tr_4\ln\{\frac{\iu}{2}\Sgn(\me)\nam\tau_3
%[\nam g_1(\kvec,0,\epsilon_m)+
%\nam g_1(\ul\kvec,0,\epsilon_m)]\} \\
&=\frac{1}{2}A_o\hbar\vF\NF\pi \kB T\sum_{m} \int\frac{\upd^2\kvec}{4\pi}
|\kvec\cdot\zvec|\ln\{\mathrm{Det}_4 \frac{1}{2}
[\nam g_1(\kvec,0,\epsilon_m)+
\nam g_1(\ul\kvec,0,\epsilon_m)]\},
\end{split}
\end{equation}
where $\nam g_1$ is the physical propagator inside the open pinhole,
which has already been ``solved'' in Eqs. (\ref{e.physprop1})-(\ref{e.physprop4}).
Here we used the general properties 
%$\Tr\{\ln[A]\} = \ln\{\Det[A]\}$ and $\Det AB=\Det A ~\Det B$ and noted 
$\Tr \ln A = \ln \Det A$ and $\Det AB=\Det A ~\Det B$ and noted 
that $\Det[\iu\Sgn(\me)\nam\tau_3]=1$. 

Evaluation of the determinant in Eq. (\ref{e.detf}) is straightforward in
principle, but very complicated in practise. So, instead of actually
doing that, we considered a further simplification. Our choice to
block the hole with a piece of specularly scattering wall was already 
arbitrary, so there is no reason to stick with that in case an even
simpler surface is found. It is now possible to think of an imaginary 
type of surface, which \emph{retroreflects} all quasiparticle
directions instead of mirror-reflecting them. It would seem at least
\emph{intuitively plausible} that even this choice should
lead to the correct coupling energy terms. The t-matrix for such
a surface is obtained from
Eq. (\ref{e.tmatrix}) by replacing the mirror-reflected directions
$\ul\kvec$ by the reversed directions $-\kvec$.  
The same replacements should then be done in the expression for 
the energy in Eq. (\ref{e.detf}) as well, but that is all. 
Doing this will simplify things considerably, and the 
determinant becomes relatively easy to calculate;
the process of evaluation is described in more detail in Appendix A. 
The end result of the calculation is that the coupling energy takes 
the simple form
\begin{equation} \label{e.final}
F_{\textrm{pinhole}}=
\frac{1}{2}A_o\hbar\vF\NF\pi\kB T\sum_{m} \int\frac{\upd^2\kvec}{4\pi}
|\kvec\cdot\zvec|\left[\ln|\nc(\kvec,\epsilon_m)|^2+
\ln(4A^4)\right],
\end{equation}
where
\begin{equation} \label{e.again}
|\nc(\kvec)|^{-2}=[(A^2+B^2)\cos\phi-\Cvec^L\cdot\Cvec^R]^2 +4A^2B^2\sin^2\phi.
\end{equation}
As expected, $J=j_zA_o=(2m_3/\hbar)\partial F_{\textrm{pinhole}}/\partial\phi$ gives
exactly the mass current in Eqs. (\ref{e.genecurrent}) and
(\ref{e.unsophisticated}) for a pinhole of open area $A_o$.
The spin current contribution seems to have
been absorbed into Eq. (\ref{e.final}) completely through a proper choice
of the ``limits of $\phi$ integration''.

\leveltwo{Spin current}

A mass current is associated with broken gauge symmetry $U(1)$
and occurs when the phase of the order parameter varies in
space. Such a variation increases the energy and leads to
``phase rigidity'', since the phase field tends to be as uniform as
possible. Similarly, due to the 
broken $SO^l(3)\times SO^s(3)$-symmetry, a spatial
variation of the spin-orbit rotation $R_{\mu i}(\nvec,\theta)$ (in the
B phase) causes \emph{spin currents} \cite{cross}. This is because
there is then a position-dependent phase difference
between ``spin up'' and ``spin down'' Cooper pairs and, although the total mass
current due to this vanishes, there can be a net transfer of angular momentum.
In our case, there is a discontinuous jump 
between the rotation matrices $R_{\mu i}^{L,R}$
and spin currents should in general be present. 
%although they did not reveal themselves
%explicitly in $F_{\textrm{pinhole}}$.

\levelthree{Quasiclassical expression for a spin current}

In quasiclassical theory, the spin current density has the expression
\begin{equation} \label{e.spincurr}
\vecb j_{\textrm{spin}}^{\gamma} (\rvec)=
\hbar\vF\NF\pi\kB T\int\frac{\upd^2\kvec}{4\pi}\kvec g_\gamma(\kvec,\rvec,\epsilon_m),
\end{equation}
where $g_\gamma$ is the $\gamma$-component of the propagator vector
$\gvec$ \cite{serenerainer}. 
For a pinhole of open area $A_o$, the spin current
$J_{\textrm{spin}}^\gamma=A_o(\zvec\cdot\vecb j^{\textrm{spin}}_\gamma)$ 
may be expressed as
\begin{equation} \label{e.spinc}
J_{\textrm{spin}}^\gamma=A_o\hbar\vF\NF\pi\kB T\sum_{m}
\int\frac{\upd^2\kvec}{4\pi}\kvec\cdot\zvec \re~ d_\gamma(\kvec,0,\me)
\end{equation}
where $d_\gamma$ is the $\gamma$-component of Eq. (\ref{e.physprop4}):
\begin{equation}
\begin{split}
\re~d_\gamma&=\re\left\{-\nc~[\Cvec^L\times\Cvec^R]_\gamma\right\}
=-|\nc|^2(\re~\nc)~[\Cvec^L\times\Cvec^R]_\gamma \\
&=\frac{[-(A^2+B^2)\cos\phi+\Cvec^L\cdot\Cvec^R]
[\Cvec^L\times\Cvec^R]_\gamma }
{[(A^2+B^2)\cos\phi-\Cvec^L\cdot\Cvec^R]^2 +4A^2B^2\sin^2\phi}.
\end{split}
\end{equation}
We may express this in terms of the rotation matrices with
$\Cvec^L\cdot\Cvec^R=R^L_{\mu i}R^R_{\mu j}C_iC_j$ and 
$[\Cvec^L\times\Cvec^R]_\gamma=\epsilon_{\alpha\beta\gamma}R^L_{\alpha k}R^R_{\beta l}C_kC_l$.
There should now exist a general way of writing the spin current 
in terms of the energy $F=F_{\textrm{pinhole}}$ and these rotation
matrices, analogously to Eq. (\ref{e.currene}) for the phase
difference and mass current. 
With the help of such a relation, one would be able to confirm the correctness 
of Eq. (\ref{e.final}), although it is otherwise not needed here.

\levelthree{General spin current expression at a discontinuity}

Consider again the spin triplet state of a Cooper pair, now in the
form 
$|\dvec\rangle=(-d_x+\iu d_y)|\!\uar\uar\rangle$
$\!+(d_x+\iu d_y)|\!\dar\dar\rangle+d_z(|\!\uar\dar\rangle+|\!\dar\uar\rangle)$.
If we apply to this state a spin rotation $\exp(\iu\theta_z\zvec\cdot\op{\vecb S})$
around the quantization axis $\zvec$, we find 
$\exp(\iu\theta_z\zvec\cdot\op{\vecb S)}|\dvec(\rvec,\kvec)\rangle=$
$(-d_x+\iu d_y)\exp(\iu\theta_z)|\!\uar\uar\rangle$
$\!+(d_x+\iu d_y)\exp(-\iu\theta_z)|\!\dar\dar\rangle+d_z(|\!\uar\dar\rangle+|\!\dar\uar\rangle)$.
If the rotation angle $\theta_z$ possesses a nonvanishing gradient into some 
direction, then we can see that the up-spin states have a velocity
into that direction while the down-spin states have a velocity in the
oppsite direction: mass currents cancel, but angular momentum is being 
transported. This is the essence of the concept of spin currents.
Similar considerations apply for general spin rotations 
$\boldsymbol{\theta}=\theta\nvec=\theta_x\xvec+\theta_y\yvec+\theta_z\zvec$, and at
each point in space we define a \emph{spin velocity} 
$\vecb v_{\textrm{spin}}^\gamma=(\hbar/2m_3)\boldsymbol\nabla\theta_\gamma$,
analogously to the B-phase superfluid velocity $\vecb
v_s=(\hbar/2m_3)\boldsymbol\nabla\phi$, where $\boldsymbol\nabla\phi$ is the gradient of an
overall phase.
Now, just as the mass current density is related to $\vecb v_s$ by
$\vecb j(\rvec)=(2m_3/\hbar)(\delta F/\delta \boldsymbol\nabla\phi(\rvec))$, we
also get the current density of angular momentum by
$\vecb j_{\textrm{spin}}^\gamma(\rvec)=\delta
F/\delta\boldsymbol\nabla\theta_\gamma(\rvec)$ \cite{cross}.

In our case, however, we have discontinuous jumps of the phase 
and the spin rotation angles at the junction. They are of a
step-function type so that $[\nabla\phi]_z=\phi\delta(z)$ and 
$[\nabla\theta_\gamma]_z=\theta_\gamma\delta(z)$,
where we defined the differences
$\phi=\phi^R-\phi^L$ and
$\theta_\gamma=\theta_\gamma^R-\theta_\gamma^L$.
The problem is that the latter of these bears no meaning in this global 
sense: we cannot add rotation vectors, since three-dimensional 
rotations do not usually commute.
But for infinitesimally small additional relative rotations we can safely write
%$\boldsymbol{\delta\theta}=\boldsymbol{\delta\theta}^R-\boldsymbol{\delta\theta}^L$.
$\delta\theta_\gamma=\delta\theta^R_\gamma-\delta\theta^L_\gamma$,
and we expect to have the total mass and spin currents at the
junction to be given by
\begin{equation} \label{e.massspin}
J=\frac{2 m_3}{\hbar}\frac{\partial F(\phi+\delta\phi)}{\partial\delta\phi}\bigg|_{\delta\phi=0}
\quad\textrm{and}\quad J^\gamma_{\textrm{spin}}=
\frac{\partial F(\boldsymbol{\theta}^L,\boldsymbol{\theta}^R,
\boldsymbol{\delta\theta})}{\partial\delta\theta_\gamma}
\bigg|_{\boldsymbol{\delta\theta}=\vecb 0}.
\end{equation}
Assuming now that $F$ can be written in 
terms of the combinations $R^L_{\mu i}R^R_{\mu j}$ as above,
we can represent such additional relative infinitesimal rotations by the 
expression $R_{\alpha i}^LR_{\beta j}^RR_{\alpha \beta}$,
where $R_{\alpha\beta}(\boldsymbol{\delta\theta})$
$=\delta_{\alpha\beta}+\epsilon_{\alpha\beta\gamma}\delta\theta_\gamma$.
By using the chain rule of partial differentiation, we then find from
Eq. (\ref{e.massspin})
\begin{equation}
J^\gamma_{\textrm{spin}}=\frac{\partial F}{\partial (R_{\mu i}^LR_{\mu j}^R)}
\frac{\partial [R_{\alpha i}^LR_{\beta j}^R(\delta_{\alpha\beta}+
\epsilon_{\alpha\beta\gamma}\delta\theta_\gamma)]}{\partial\delta\theta_\gamma}
\bigg|_{\boldsymbol{\delta\theta}=\vecb 0},
\end{equation}
which is nothing but
\begin{equation} \label{e.plausible}
J_{\textrm{spin}}^\gamma=\epsilon_{\alpha\beta\gamma}
R_{\alpha i}^LR_{\beta j}^R
\frac{\partial F}{\partial(R_{\mu i}^LR_{\mu j}^R)}.
\end{equation}
This is certainly satisfied by the expressions in Eqs. (\ref{e.final}) and
(\ref{e.spinc}) and it was easy to guess by mere inspection
before any calculation. Note that this gives again a particularly
simple expression, if calculated for the high-temperature form,
Eq. (\ref{e.fj}), of the coupling energy $F$.

\leveltwo{Parameters and implementation}

In the case of a single pinhole we could simply assume the
$\nvec$-vectors to be always parallel or antiparallel and perpedicular to the 
wall, and the energy of Eq. (\ref{e.final}) was not needed. 
But as we saw in the tunneling model, if we take a large array 
of such pinholes, the total coupling energy $F_J$ can overcome 
the gradient energy $F_{\textrm{Gtot}}$ and there will then be a copmpetition
between these energies to find a stable equilibrium state for each given $\phi$.
Here we want to find the mass currents corresponding to
those equilibrium states. This is done by minimising the total coupling 
energy, namely the ``sum'' of $F_{\textrm{pinhole}}$'s for all holes in the 
array, plus the same model gradient energies (\ref{e.quadforms}) we
used in the tunneling model. As \emph{parameters} in our calculations
we have the total area of the junction $S$ and the open area density
$d_o$. The coupling energy $F_J$ is obtained from
$F_{\textrm{pinhole}}$ by replacing $A_o$ with the total open area 
$S_o=d_o S$, which means that we assume the pinholes in the array to operate
fully coherently. We may also study the effect of restricted angles,
so the aspect ratio $W/D$ in Eq. (\ref{e.probability}) is in principle third
possible adjustable parameter. The value of the gradient-energy
parameter $\gamma$ is set by the total area $S$ and it is given in
Fig. \ref{f.tconst}, but this can be scaled also. We are, nevertheless, in a
more restricted situation now, because the sizes of ``$\alpha$ and
$\beta$'' cannot be adjusted freely.

We again parametrise $F_J$ in terms of the polar ($\eta$) and
azimuthal ($\chi$) angles of $\nvec^{L,R}$ with respect to some fixed
axes. The polar axis
$\zvec$ is perpendicular to the wall.
Owing to the symmetry of the wall, the absolute azimuthal angles
$\chi^L$ and $\chi^R$ do not matter, only their difference does.
We thus choose $\chi^L=-\chi$ and $\chi^R=\chi$, i.e.
$2\chi=\chi^L-\chi^R$, and insert the expressions
\begin{equation}
\begin{split}
\nvec^L & =\sin\eta^L\cos\chi\xvec-
\sin\eta^L\sin\chi\yvec+\cos\eta^L\zvec\\
\nvec^R & =\sin\eta^R\cos\chi\xvec+
\sin\eta^R\sin\chi\yvec+\cos\eta^R\zvec\\
\end{split}
\end{equation}
into the energy, Eq. (\ref{e.final}), where $\Cvec^L\cdot\Cvec^R$ is given by
Eq. (\ref{e.horrendous}).
The task is then to minimise the total energy with respect to the three
real angular variables $\{\eta^L,\eta^R,\chi\}$. The vectors attain
all possible values on the intervals $\eta^L,\eta^R,\chi\in[0,\pi]$,
but it is easier not to restrict their values in the numerical algorithm.

The numerical minimisation of $F(\phi)=F_J(\phi)+F_{\textrm{Gtot}}$ was
carried out using the standard NAG Fortran
library routine E04JYF. The order parameters and propagators needed for 
the calculation of $F_J$ were obtained in the same way as in the
single pinhole case. Practically speaking, only the minimisation
routine had to be added to the program at this final stage. 

\begin{figure}[!btp]
\begin{center}
\includegraphics[height=0.9\linewidth,width=0.8\linewidth]{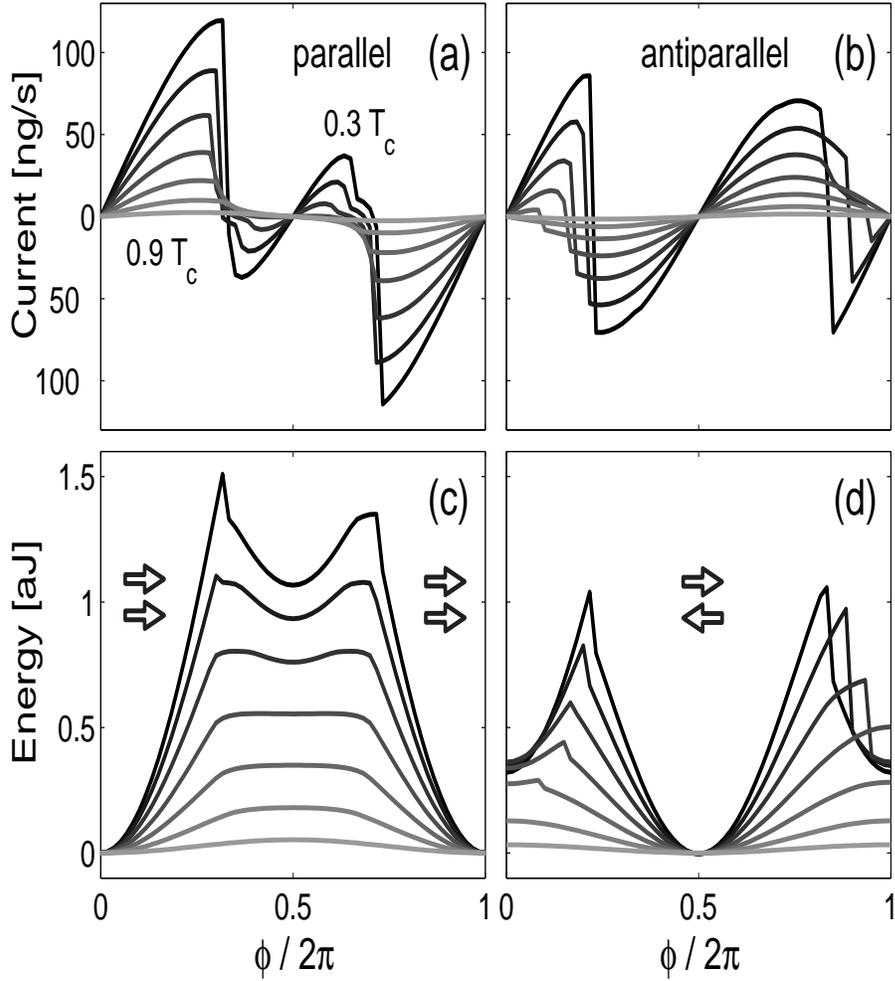}
\caption{Current-phase and energy-phase relations for the coherent
pinhole array at temperatures between $0.3 T_c$ and $0.9 T_c$ in intervals 
of $0.1 T_c$. The left
panels (a),(c) are for parallel $\nvec$ vectors at infinity and the
right panels (b),(d) for antiparallel ones. The parameters
used are $S=3.8\cdot 10^{-8}$ m$^2$, $d_0=14.69\cdot10^{-4}$,
$W/D=0.0$, and $\gamma$ was multiplied by a factor of
$0.1$. The $\phi$ sweep was done from left to right, and this is the
form of the curves as they came out of the minimisation routine.
No numerical noise was applied in the minimisation, and we find
strongly hysteretic behavior at the transitions, which makes the
antiparallel $J(\phi)$ look a bit ``unappealing''. In the parallel
case, the $\nvec$ vectors are perpendicular to the wall around
$\phi=0$, but in the antiparallel case they are perpendicular around
$\phi=\pi$. The two ``$\pi$ states'' are thus in different places.}
\label{f.parapar}
\end{center}
\end{figure}

\leveltwo{Results for the pinhole array}

Compared to the tunneling model calculation, 
here we chose to use slightly different parameters for the
experimental aperture array. This is because we have subsequently 
learned that, instead of cicular holes, the aperture array had
actually always consisted of approximately square holes
(at least that was the form aimed at in the etching)
\cite{private}. Also, instead of a diameter of $100$ nm, these
squares were roughly of size $115$ nm $\times~ 115$ nm. There is still some
doubt as to whether these are the real hole dimensions, because they
have been deduced \emph{after} the etching to be consistent with an experiment
measuring the normal-state flow resistance, which should depend on the
open area of the holes \cite{kulik}.

Nevertheless, due to this change, the total open area has now been
multiplied by a factor of about $1.68$, which increases the critical
currents quite a lot and at least solves one of the previous problems
right away. 
Fig. \ref{f.parapar} shows the results of minimisations at
temperatures $T/T_c=0.3, 0.4, \ldots, 0.9$ for 
$S=3.8\cdot 10^{-8}$ m$^2$, $d_0=1.469\cdot10^{-3}$ and
$W/D=0.0$, which is the new ``basic configuration''. 
In addition to these values, the gradient-energy parameter $\gamma$
obtained from Fig. \ref{f.tconst} has been multiplied by $0.1$ to
get more pronounced $\pi$ states. Note, however, that this is now
the \emph{only} unjustified scaling that is being done.

There are some clear differences between the results of Fig.
\ref{f.parapar} and those in Figs. \ref{f.gamma} and
\ref{f.ictemps}. First of all, the \emph{antiparallel} ``$\pi$ state'', where
the $\nvec$ vectors are not perpendicular to the wall, has
moved from $\phi=\pi$ to $\phi=0,2\pi$. Moreover, the form of the corresponding
$J(\phi)$ is now also quite different from the parallel $J(\phi)$.
We know that in the experiments the low and high
critical current $J(\phi)$'s actually did look quite different, and it
was the low critical current case which had the relatively more
pronounced $\pi$ state. This is just what is seen in Fig.
\ref{f.parapar} and, therefore, at least one more problem is now corrected,
although the new curves are not quite as ``pretty'' as before.
I should stress that above $T=0.4 T_c$, essentially the 
same results as here could have been obtained also by the high temperature
form of $F_J$ used in the tunneling model. 

\begin{figure}[!bt]
\begin{center}
\includegraphics[width=0.65\linewidth]{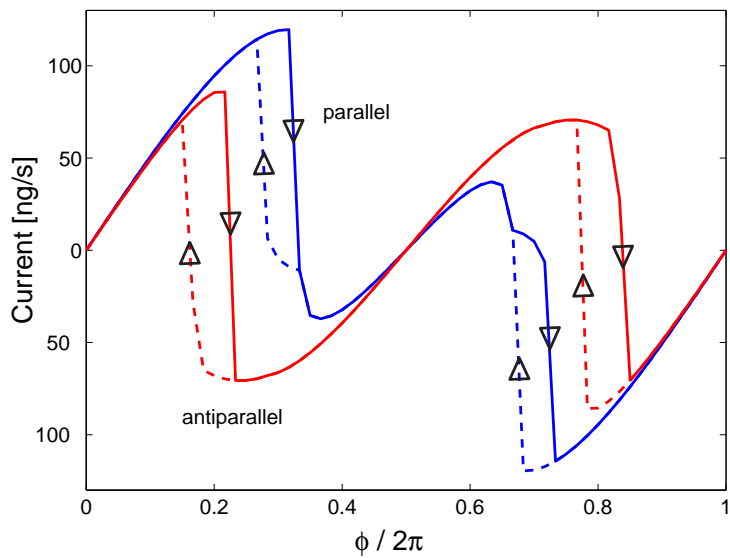}
\caption{Hysteresis associated with the jumps between the normal and
$\pi$ branches at low temperature ($T=0.3 T_c$). The arrows show the
routes taken when proceeding into different directions. How much of
this is real and how much just numerical is difficult to
study. All parameters were as in Fig. \ref{f.parapar}.}
\label{f.hyst}
\end{center}
\end{figure}

As is shown more clearly in Fig. \ref{f.hyst} for $T=0.3 T_c$, there
is again some hysteresis associated with the transitions. 
This is hard to study properly, because the analysis has to 
be done numerically. Some of the hysteresis can always be purely
numerical, such that the minimisation routine just gets stuck in an
unstable extremum of the energy. This could be prevented by
adding some random numerical noise in the initial $\nvec$ configurations for
each minimisation, instead of using the solution for the previous
$\phi$. Unfortunately, this might then induce some premature jumps 
between branches and wipe out some of the true hysteresis as well.
I postpone any further study of this until later (in the case it
should turn out especially interesting.)

\begin{figure}[!bt]
\begin{center}
\includegraphics[width=0.65\linewidth]{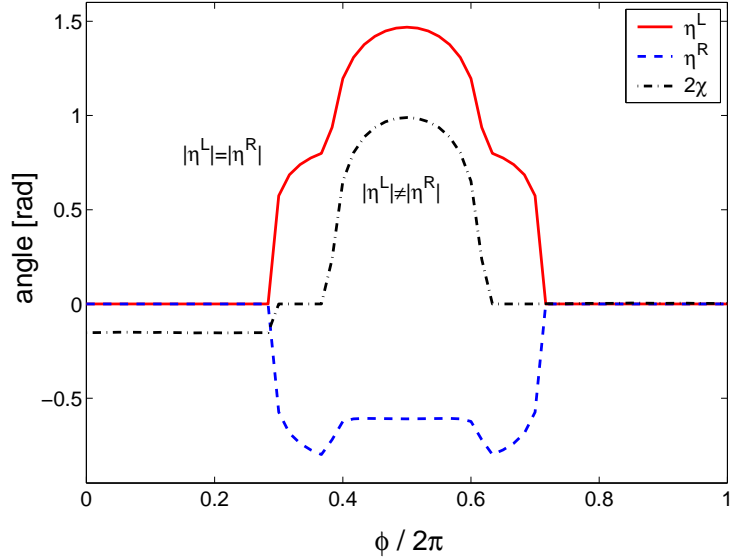}
\caption{Relative angles of the $\nvec^{L,R}$ vectors in one
minimisation sweep at $T=0.5 T_c$ for parallel $\nvec$'s at
infinity. All parameters were the same as in Fig. \ref{f.parapar}. The
interval with the strange bumps in the middle is a
region where the gradient energies arising from the left and right
sides of the junction are not equal. This is seen as an asymmetry in
the magnitudes of the angles $\eta^{L,R}$. 
%It also appears clearly inthe form of $J(\phi)$. 
The nonzero $2\chi$ angle in the beginning
follows directly from an initial guess, because for perpendicular
$\nvec$ vectors the angle is actually undetermined. 
%Negative polar angles are used here just for convenience.
}
\label{f.kulmat}
\end{center}
\end{figure}
Figure \ref{f.kulmat} shows the minimising angles $\{\eta^L,\eta^R,2\chi\}$
for the curve at $T=0.5 T_c$ for parallel $\nvec$ vectors in Fig. \ref{f.parapar}.
Previously, in connection with the tunneling model, we mentioned that in the
$\pi$ state the $\nvec$ vectors are \emph{usually} directed so that the
gradient energies on the left and right sides are
equal: $\hat n_z^L=\pm\hat n_z^R$. Here we have a case where this
is not obeyed. Up to about $\phi/2\pi=0.35$, the condition
$|\eta^L|=|\eta^R|$ is satisfied\footnote{No need to worry about the
negative polar angle: it is used here just for convenience.}, but then a smaller energy
configuration is found, where it fails.
This is seen also in the corresponding $J(\phi)$ as an extra bump.
Here is another ``inconvenient'' effect which was (unintentionally)
avoided in the tunneling model calculation, but which seems to be
present for these more realistic choices of parameters.

Many other combinations for the parameters than those used in Fig. \ref{f.parapar}
were also tested, but none resulted in any new kind behavior: Increasing
$W/D$ decreases the critical currents, which is not
very useful. Increasing the scaling factor of $\gamma$ over $0.2$ or so
begins to suppress the $\pi$ states too much and the discrepancy in the
relative sizes of the coupling and gradient energies thus remains.
A larger array than that in the experiments, or a better way of estimating
$\gamma$ would still be needed to correct this.

\newpage
\levelone{Conclusions and discussion}

We note that our new results for the aperture array (in Sec. 7) differ somewhat 
from those  published previously (Sec. 4). 
The main reason why the ``inconveniences'' related to them were originally 
avoided (by the rescaling of $\alpha$ and $\beta$)  
was that the resulting curves looked too suspicious at that time. And, 
admittedly, we were in a hurry to publish the results. However, the
essential physics is still the same,
irrespective of whether it is right or wrong --- and we 
believe it is right. I repeat the simple predictions of our model,
which should be easy to test experimentally: a smaller array size or 
a strong enough magnetic field should suppress the $\pi$ state. If 
even this does not happen, then we obviously need
something very different to explain the experiments.
Such tests have not yet been carried out, but we hope that the 
situation will soon change.

One rather trivial, and initially plausible explanation was already
given in Ref. \cite{spoil}. There the whole $\pi$ state was explained
by a simple argument where half of the holes in the array were assumed 
to be on one branch and the other half on another branch of a hysteretic $J(\phi)$.
According to this, the $\pi$ state would have resulted from the
currents of different holes cancelling each other --- much
like in Yip's explanation. But this explanation was soon deemed unlikely,
because it required the (incorrect) assumption of hysteretic ``normal branches''.
Nevertheless, it is not at all clear as to why the apertures should work
absolutely coherently. In this work we have consistently assumed it to be the
case, but mainly because the experimenters claim so (see Ref. \cite{bp_nat97}).
Even to themselves, it has always been a bit of a mystery
\emph{why} this holds. 
%, when there is no physical reason as to why it \emph{should}. 
But perhaps some minor incoherence is just one of the many possible
factors which make the experimental findings different from our 
theoretical ones.

In summary, our current-phase relationships would now seem to mimic
the experimental ones quite nicely, considering the crude
approximations made. They fall clearly into two classes according to
the size of the critical current, and the explanation for this is simple:
the parallel or antiparallel orientations of the $\nvec$ vectors.
Both cases show an additional kink at low temperatures, which is
stronger for the smaller $J_c$ case. The temperature dependence of the 
form of $J(\phi)$ is also explained in a simple fashion with the different temperature 
dependencies of $F_J$ and $F_{\textrm{Gtot}}$. The only remaining
problem is actually the too large magnitude of the latter (by a
factor of around $10$), as obtained from our estimate. We should try to
improve the estimate, hopefully still retaining the simple quadratic
form of $F_{\textrm{Gtot}}$. The problem with too small critical
currents was solved by having to increase the size of the apertures,
and now the $J_c$'s are perhaps even too large. But our knowledge of
the actual aperture sizes is too inaccurate to make any conclusions
based on this. On the other hand, the current densities in a pinhole
\emph{should} always be larger than for any finite-size aperture. 
The next step, hopefully after correcting also the problem with the
size of $\gamma$, could thus be the self-consistent quasiclassical
calculation for an aperture of finite size. Whether that is needed, however, still
remains to be seen.

Interest in the $\pi$ state of the Berkeley array experiment
\cite{berkeley} was the reason why we started working on this subject 
in the first place. The pinhole array model (Sects. 4 and 7) is now our primary
candidate for an explanation of the $\pi$ state (see Fig. \ref{f.parapar}). But
we should not forget the valuable
``byproducts'' of this project either. The $\pi$ branch found in the
single large aperture by the GL calculation (Sec. 3) is also very interesting
(see Fig. \ref{f.diagram}). And, as already mentioned, this
could in fact be the proper
interpretation of the results of Ref. \cite{singleaperture} measured for the
single narrow slit. We are eagerly waiting to hear more news on
the progress of these experiments. 

There is also a small chance that, instead of
having holes larger than they initially thought, the Berkeley
experimentalists have some extra leakage parallel to their aperture 
array. If this were the case, also their $\pi$ state could perhaps be
explained with the single aperture calculation.
Doing a new, more sophisticated GL calculation 
in the future is therefore also a possibility --- at least in the case that
the aperture array model turns out to fail the possible tests
concerning magnetic fields and array sizes.

Furthermore, the properties of a single
pinhole have now been investigated in some detail, taking into account
different surface models (Sec. 6). Although the pinhole problem is already an
old one, no very clear plots of the current-phase relations 
(Figs. \ref{f.kiphi}-\ref{f.diphi}) or the 
critical currents (Fig. \ref{f.ictemps}) have ever been published before.  
With the results of this calculation, we also consider Yip's model for
the $\pi$ state disproven.\\ \\

%Moreover, the results for parallel $\nvec$-vectors can be 
%applied for $s$-wave superconductors as well.

%- Although the tunneling model is a good description for a pinhole
%array, it may not be so for larger apertures.

\newpage

\appendix

\levelone{Evaluation of the determinant}

The coupling energy of the open pinhole, relative to the state blocked
with a specular wall was found to be given by
\begin{equation}
F_{\textrm{pinhole}}=
\frac{1}{2}A_o\hbar\vF\NF\pi\kB T\sum_{\epsilon_m} \int\frac{\upd^2\kvec}{4\pi}
|\kvec\cdot\nvec|\ln\{\mathrm{Det}_4 \frac{1}{2}
[\nam g_1(\kvec,0,\epsilon_m)+
\nam g_1(\ul\kvec,0,\epsilon_m)]\},
\end{equation}
where $\ul\kvec=\kvec-2(\nvec\cdot\kvec)\nvec$ is the reflected
quasiparticle direction.
The difficult part in evaluating this is to find the determinant of
the $4\times4$ matrix.
The expression to be calculated is of the form
\begin{equation} \label{e.detti}
%\ln\left\{
\textrm{Det}_4\frac{1}{2}[\nam g_1+\nam g_2],
\end{equation}
where in our special case $\nam g_1=\nam g(\kvec)$ and 
$\nam g_2=\nam g(\ul\kvec)$.
Writing the matrix sum directly in component form and evaluating the
determinant that way is difficult, but luckily there is a much simpler
detour. The determinant in Eq. (\ref{e.detti}) can be written 
\begin{equation} \label{e.detform}
\sqrt{\left(\text{Det}_4\frac{1}{2}[\nam g_1 +\nam g_2]\right)^2}
=\sqrt{\textrm{Det}_4\left(\frac{1}{2}[\nam g_1 +\nam g_2]\right)^2}
=\sqrt{\frac{1}{2^4}\textrm{Det}_4\left(-\nam 1+\frac{1}{2}\{\nam g_1,\nam g_2\}\right)},
\end{equation}
using the general property $\Det{AB}=\Det A~ \Det B$ and the physical normalisations
$\nam g_1 \nam g_1=\nam g_2 \nam g_2=-\nam 1$.
Note that if $\nam g_1=\nam g_2$, this equals unity:
indeed, since we require a physical propagator to have the
normalisation $\nam g\nam g=-\nam 1$, it follows that its determinant 
must be $\textrm{Det}_4\nam g=\pm 1$. This is consistent with the fact 
that the logarithm in the junction energy expression should be zero 
if $\ul\kvec$ is replaced by $\kvec$ (and thus $\nam g(\ul\kvec)=\nam
g(\kvec)$), i.e., if the energy of the transmitting junction is calculated 
relative to itself, and not to that of the blocked junction.

In any case, the commutator $\{\nam g_1, \nam g_2\}$ can be simplified
further without going into any particulal coordinate
representation. One finds that it is of the form
\begin{equation}
\frac{1}{2}\{\nam g_1,\nam g_2\} = \left [ \begin{array}{cc}
g+\gvec\cdot\ul\sigmavec & f\iu\ul\sigma_2 \\
\td f\iu\ul\sigma_2 & 
g + \iu\ul\sigma_2 \gvec\cdot\sigmavec\iu\ul\sigma_2
\end{array} \right],
\end{equation}
with
\begin{equation}
\begin{split}
g&=d_1d_2+\dvec_1\cdot\dvec_2-\avec_1\cdot\avec_2+\bvec_1\cdot\bvec_2\\
\gvec&=d_1\dvec_2+\dvec_1d_2+\iu\avec_1\times\bvec_2-\iu\bvec_1\times\avec_2\\
f&=\dvec_1\cdot\avec_2+\avec_1\cdot\dvec_2
+\dvec_1\cdot\bvec_2+\bvec_1\cdot\dvec_2 \\
\td f&=\dvec_1\cdot\avec_2+\avec_1\cdot\dvec_2
-\dvec_1\cdot\bvec_2-\bvec_1\cdot\dvec_2 
\end{split}
\end{equation}
Finding the determinant from this form is easy:
\begin{equation} \label{e.deter}
\textrm{Det}_4\left(-\nam 1+\frac{1}{2}\{\nam g_1,\nam g_2\}\right)
=\left(-f\td f-(g-1)^2+\gvec\cdot\gvec \right)^2.
\end{equation}
In principle, it is now straightforward to just insert the expressions
for $f$, $\td f$, $g$ and $\gvec$. In the present case we need the physical
propagators at the pinhole, Eqs. (\ref{e.physprop1})-(\ref{e.physprop4}),
namely %(assuming $\phi_R=\phi/2$ and $\phi_L=-\phi/2$)
\begin{equation}
\begin{split}
\avec(\kvec,0) & =\iu~\nc~(\Cvec^L+\Cvec^R)~
(\iu A\sin\frac{1}{2}\phi-B\cos\frac{1}{2}\phi)  \\
\bvec(\kvec,0) & =\iu~\nc~(\Cvec^L-\Cvec^R)~
(A\cos\frac{1}{2}\phi-\iu B\sin\frac{1}{2}\phi)  \\
d(\kvec,0) &=\iu ~\nc~[\iu (A^2+B^2)\sin\phi-2AB\cos\phi] \\[1mm]
\dvec(\kvec,0) &=-\nc ~\Cvec^L\times\Cvec^R
\end{split}
\end{equation}
where
\begin{equation}
\nc(\kvec,0) =
[-(A^2+B^2)\cos\phi+2\iu AB\sin\phi+
%\Cvec\cdot(\dyadic{R}^L)^T\!\cdot\dyadic{R}^R\!\!\cdot\Cvec]^{-1}
\Cvec^L\cdot\Cvec^R]^{-1}
\end{equation}
and the corresponding expressions for $\ul\kvec$. Unfortunately, the
calculation turns out to be very tedious in practice, because the
parallel and perpendicular components of $\Cvec$ behave differently
under the symmetry operations connecting $\kvec$ and $\ul\kvec$. 

However, intuition comes to rescue. As mentioned in the text, the
coupling energy should not really depend on how
the pinhole is blocked. The type of wall used to reflect the
quasiparticles should only be seen in an additional constant in the
free energy, which no longer depends on the phase difference or the
rotation matrices. Thus, if one imagines the hole being blocked by
some kind of material which \emph{retroreflects} all quasiparticle
directions, the interesting energy terms obtained should still be exactly
the same. Making this choice will simplify calculations considerably, since one can
now replace everywhere above $\ul\kvec$ with $-\kvec$ and apply
symmetries much more efficiently. With this
simplification one finds
\begin{equation}
\begin{split}
%f\td f&=0 \\
%|\nc|^{-2}g-|\nc|^{-2}&=-4A^2\left[(A^2+B^2)-
%\Cvec^L\cdot\Cvec^R\cos\phi\right] \\
%|\nc|^{-4}\gvec\cdot\gvec&=16A^4\sin^2\phi
%\left[(\Cvec^2)^2-(\Cvec^L\cdot\Cvec^R)^2 \right]
f\td f&=0 \\
g-1&=-4A^2\left[(A^2+B^2)-
\Cvec^L\cdot\Cvec^R\cos\phi\right]|\nc|^2 \\
\gvec\cdot\gvec&=16A^4\sin^2\phi
\left[(\Cvec^2)^2-(\Cvec^L\cdot\Cvec^R)^2 \right]|\nc|^4
\end{split}
\end{equation}
and after some algebra the determinant in Eq. (\ref{e.deter}) takes the
embarrassingly simple form $2^8A^8|\nc|^4$.
%\begin{equation}
%|\nc|^8\left[-(|\nc|^{-2}g-|\nc|^{-2})^2+
%|\nc|^{-4}\gvec\cdot\gvec\right]^2
%=2^8A^8|\nc|^4.
%$\left[-(g-1)^2+\gvec\cdot\gvec\right]^2=2^8A^8|\nc|^4$.
%\end{equation}
%The equality follows because the expression in the brackets
%turns out to be exactly $-2^4A^4|\nc|^{-2}$. 
The logarithm of Eq. (\ref{e.detform}) in the energy expression then becomes
\begin{equation}
\ln\sqrt{\frac{1}{2^4} 2^8A^8|\nc|^4} = 
\ln|\nc|^2+\ln 4A^4,
\end{equation}
where the second term is obviously
the extra constant term, whose form depends on the choice of the
blockage. This has to be retained in order to have convergence in the
Matsubara summation, and the final form for the coupling energy is
\begin{equation}
F_{\textrm{pinhole}}=
\frac{1}{2}A_o\hbar\vF\NF\pi\kB T\sum_{\epsilon_m} \int\frac{\upd^2\kvec}{4\pi}
|\kvec\cdot\nvec|\left[\ln|\nc(\kvec,\epsilon_m)|^2+
\ln(4A^4)\right].
\end{equation}

\newpage

\levelone{Important constants}

All calculations presented in the text were done at vapor pressure, 
which is the situation that was present in the experiment of
Ref. \cite{bistability}. Here I just tabulate the values
of some important constants in the SI units, and give some definitions.
First of all we have 
\begin{equation}
\begin{split}
m_3 & =5.00812\cdot10^{-27}  ~\textrm{kg}  \\
\hbar & =1.054573\cdot10^{-34} ~\textrm{Js} \\
\kB & =1.380658\cdot10^{-23} ~\textrm{J/K}
\nonumber
\end{split}
\end{equation}
where $m_3$ is the mass of a $^3$He atom, $\hbar$ is Planck's
constant divided by $2\pi$ and $\kB$ is Botzmann's constant.
Secondly, for the Fermi wavenumber and Fermi velocity we have the relation
$\kF=(m^*/\hbar)\vF$, where $m^*$ is the effective mass of a $^3$He
quasiparticle. At vapor pressure we have the values
\begin{equation}
\begin{split}
m^*&=2.8m_3 \\
\vF&=59.03 ~\textrm{m/s} \\ 
\kB T_c&=1.28\cdot10^{-26} ~\textrm{J},
\nonumber
\end{split}
\end{equation}
where $T_c$ is the critical temperature for pure $^3$He.
The density of states at the Fermi surface $N(\xi_{\vecb k}=0)$ has the expression
$\NF=\frac{m^*\kF}{2\pi^2\hbar^2}=\frac{{m^*}^2\vF}{2\pi^2\hbar^3}$,
which gives, at vapor pressure, the following energy and current units:
\begin{equation}
\begin{split}
2m_3\vF\NF\kB T_c &= 3.7946~\textrm{kg/m$^2$ s} \\
\hbar\vF\NF\kB T_c &= 3.9952\cdot10^{-8}~\textrm{J/m$^2$}\\
2m_3/\hbar &= 9.4979\cdot10^7~\textrm{kg/Js}. \nonumber
\end{split}
\end{equation}
For easier conversion between Joules and electron volts, I also add
the relation
\begin{equation}
\begin{split}
1 ~\textrm{eV} &= 1.6021773\cdot10^{-19}~\textrm{J}.
\nonumber
\end{split}
\end{equation}
%Now I have just about used up this

%\vspace{50mm}
%{
%\begin{center}
%\begin{tabular}{| p{12 cm}| }
%\hline %\\
%\noindent
%\textbf{Would you believe...?} \\
%During this master's thesis work I used 7 graphite pencils which were 
%labeled ``Staedtler Minerva 130 60, 
%2 HB''. They costed me FIM 1 apiece. \\%\\
%\hline
%\end{tabular}
%\end{center}
%}

\appendix
\newpage

\end{document}